\documentclass[structabstract]{aa} 
\usepackage[utf8]{inputenc}
\usepackage{txfonts} 
\usepackage{graphicx} 
\usepackage{natbib} 
\usepackage{url} 

\usepackage{graphicx}	% Including figure files
\usepackage{amsmath}	% Advanced maths commands
\usepackage{amssymb}	% Extra maths symbols
\usepackage[dvipsnames]{xcolor}

\usepackage{comment}

\usepackage[colorlinks = true,
            linkcolor = blue,
            urlcolor  = blue,
            citecolor = blue,
            anchorcolor = blue]{hyperref}

\usepackage{ulem}

\usepackage{longtable}

\usepackage{threeparttable}

\usepackage{colortbl}

\usepackage{subcaption}
\usepackage{mwe}

\usepackage{enumitem}

\usepackage{pifont}
\newcommand{\xmark}{\ding{55}}%

\begin{document}

   \title{A detailed spectroscopic study of tidal disruption events}

%   \subtitle{I. Overviewing the $\kappa$-mechanism}

   \author{P. Charalampopoulos\inst{1}\fnmsep\thanks{Contact e-mail: \href{mailto:pngchr@space.dtu.dk}{pngchr@space.dtu.dk}}\href{https://orcid.org/0000-0002-7706-5668}\
          \and
          G. Leloudas\inst{1}\href{https://orcid.org/0000-0002-8597-0756}
          \and
          D. B. Malesani\inst{1}\href{https://orcid.org/0000-0002-7517-326X}
          \and
          T. Wevers\inst{2}\href{https://orcid.org/0000-0002-4043-9400}
          \and
          I. Arcavi\inst{3,4}\href{https://orcid.org/0000-0001-7090-4898}
          \and
          M. Nicholl\inst{5,6}\href{https://orcid.org/0000-0002-2555-3192}
          \and
          M. Pursiainen\inst{1}\href{https://orcid.org/0000-0003-4663-4300}
          \and
          A. Lawrence\inst{7}
          \and
          J. P. Anderson\inst{2}\href{https://orcid.org/0000-0003-0227-3451}
          \and
          S. Benetti\inst{8}
          \and
          G. Cannizzaro\inst{9,10}\href{https://orcid.org/0000-0003-3623-4987}
          \and
          T.-W. Chen\inst{11}\href{https://orcid.org/0000-0002-1066-6098}
          \and
          L. Galbany\inst{12}\href{https://orcid.org/0000-0002-1296-6887}
          \and
          M. Gromadzki\inst{13}\href{https://orcid.org/0000-0002-1650-1518}
          \and
          C. P. Guti\'errez\inst{14,15}\href{https://orcid.org/0000-0003-2375-2064}
          \and
          C. Inserra\inst{16}\href{https://orcid.org/0000-0002-3968-4409}
          \and
          P. G. Jonker\inst{10,9}\href{https://orcid.org/0000-0001-5679-0695}
          \and
          T. E. M\"uller-Bravo\inst{17}\href{https://orcid.org/0000-0003-3939-7167}
          \and
          F. Onori\inst{18}
          \and
          P. Short\inst{7}\href{https://orcid.org/0000-0002-5096-9464}
          \and
          J. Sollerman\inst{11}\href{https://orcid.org/0000-0003-1546-661}
          \and
          D. R. Young\inst{19}\href{https://orcid.org/0000-0002-1229-2499} \    
          }

   \institute{DTU Space, National Space Institute, Technical University of Denmark, Elektrovej 327, DK-2800 Kgs. Lyngby, Denmark
            %  \email{wuchterl@amok.ast.univie.ac.at}
         \and
         European Southern Observatory, Alonso de C\'ordova 3107, Casilla 19, Santiago, Chile
         \and
         The School of Physics and Astronomy, Tel Aviv University, Tel Aviv 69978, Israel
         \and
         CIFAR Azrieli Global Scholars program, CIFAR, Toronto, ON M5G 1M1 Canada
         \and
         School of Physics and Astronomy, University of Birmingham, Birmingham B15 2TT, UK
         \and
         Institute for Gravitational Wave Astronomy, University of Birmingham, Birmingham B15 2TT, UK
         \and
         Institute for Astronomy, University of Edinburgh, Royal Observatory, Blackford Hill EH9 3HJ, UK
         \and
         INAF - Osservatorio Astronomico di Padova, Vicolo dell'Osservatorio 5, 35122 Padova, Italy
         \and
         Department of Astrophysics/IMAPP, Radboud University, P.O. Box 9010, 6500 GL Nijmegen, the Netherlands
         \and
         SRON, Netherlands Institute for Space Research, Sorbonnelaan, 2, NL-3584CA Utrecht, the Netherlands
         \and
         The Oskar Klein Centre, Department of Astronomy, Stockholm University, AlbaNova, SE-10691 Stockholm, Sweden
         \and
         Institute of Space Sciences (ICE, CSIC), Campus UAB, Carrer de Can Magrans, s/n, E-08193 Barcelona, Spain.
         \and
         Astronomical Observatory, University of Warsaw, Al. Ujazdowskie 4, 00-478 Warszawa, Poland
         \and
         Finnish Centre for Astronomy with ESO (FINCA), FI-20014 University of Turku, Finland
         \and
         Tuorla Observatory, Department of Physics and Astronomy, FI-20014 University of Turku, Finland
         \and
         School of Physics \& Astronomy, Cardiff University, Queens Buildings, The Parade, Cardiff, CF24 3AA, UK
         \and
         School of Physics and Astronomy, University of Southampton, Southampton, Hampshire, SO17 1BJ, UK
         \and
         INAF - Osservatorio Astronomico d'Abruzzo via M. Maggini snc, I-64100 Teramo, Italy
         \and
         Astrophysics Research Centre, School of Mathematics and Physics, Queen's University Belfast, Belfast BT7 1NN, UK  
             }

   \date{Received - ; accepted -}

% \abstract{}{}{}{}{} 
% 5 {} token are mandatory
 
%   \abstract
%   % context heading (optional)
%   % {} leave it empty if necessary  

%   % conclusions heading (optional), leave it empty if necessary 
%   {}

  \abstract
   {Spectroscopically, tidal disruption events (TDEs) are characterized by broad ($\sim 10^{4}$ km~s$^{-1}$) emission lines and show a large diversity as well as different line profiles. After carefully and consistently performing a series of data reduction tasks including host galaxy light subtraction, we present here the first detailed, spectroscopic population study of 16 optical and UV TDEs. We study a number of emission lines prominent among TDEs including Hydrogen, Helium, and Bowen lines and we quantify their evolution with time in terms of line luminosities, velocity widths, and velocity offsets. We report a time lag between the peaks of the optical light curves and the peak luminosity of H$\alpha$ spanning between $\sim$ 7 -- 45 days. If interpreted as light echoes, these lags correspond to distances of $\sim$~2~--~12~$\,\times\,10^{16}$~cm, which are one to two orders of magnitudes larger than the estimated blackbody radii (R$_{\rm BB}$) of the same TDEs and we discuss the possible origin of this surprisingly large discrepancy. We also report time lags for the peak luminosity of the \ion{He}{I} 5876 \AA\,line, which are smaller than the ones of H$\alpha$ for H TDEs and similar or larger for \ion{N}{III} Bowen TDEs. We report that \ion{N}{III} Bowen TDEs have lower H$\alpha$ velocity widths compared to the rest of the TDEs in our sample and we also find that a strong X-ray to optical ratio might imply weakening of the line widths. Furthermore, we study the evolution of line luminosities and ratios with respect to their radii (R$_{\rm BB}$) and temperatures (T$_{\rm BB}$). We find a linear relationship between H$\alpha$ luminosity and the R$_{\rm BB}$ (L$_{\rm line} \propto$ R$_{\rm BB}$) and potentially an inverse power-law relation with T$_{\rm BB}$ (L$_{\rm line} \propto$ T$_{\rm BB}^{-\beta}$), leading to weaker H$\alpha$ emission for T$_{\rm BB}$ $\geq$ 25\,000 K. The \ion{He}{II}/\ion{He}{I} ratio becomes large at the same temperatures, possibly pointing to an ionization effect. The \ion{He}{II}/H$\alpha$ ratio becomes larger as the photospheric radius recedes, implying a stratified photosphere where Helium lies deeper than Hydrogen. We suggest that the large diversity of the spectroscopic features seen in TDEs along with their X-ray properties can potentially be attributed to viewing angle effects.}
   

   \maketitle
%
%________________________________________________________________

\section{Introduction} \label{sec:intro}

Tidal disruption events (TDEs) occur when the trajectory of a star intersects the tidal radius (R$_{\rm t}$) of a supermassive black hole (SMBH) lurking in the nucleus of a galaxy, in a pericenter distance (R$_{\rm p}$) smaller than the R$_{\rm t}$ where R$_{\rm t}$~$\approx$~R$_{\rm *}$(M$_{\rm BH}$/M$_{\rm *}$)$^{1/3}$ with R$_{\rm *}$ and M$_{\rm *}$ being the radius and mass of the star and M$_{\rm BH}$ being the mass of the SMBH \citep{Hills1975}. The immense gravitational field of the SMBH causes a large spread in the specific orbital binding energy of the star (much greater than its mean binding energy) and the star gets ripped apart in a TDE \citep{Rees1988}. Self-gravity stretches the stellar debris into a long thin stream, around half of which remains bound to the SMBH, and the stream starts circularising around the SMBH into highly eccentric orbits \citep{Rees1988,Evans1989}. A strong, luminous, transient flare is eventually produced \citep{Lacy1982,Rees1988,Evans1989,Phinney1989} which can emit above the Eddington luminosity \citep{Strubbe2009,Lodato2011} with L$_{bol} \sim$ 10$^{41-45}$ erg~s$^{-1}$. TDEs are a unique tool for studying SMBHs with masses $\leq$ 10$^{8}$ M$_\odot$ as they conveniently evolve on ``human'' timescales of a few months. If M$_{\rm BH}$ $>$ 10$^{8}$, a solar-mass star is disrupted within the Schwarzschild radius hence no luminous flare occurs (see \citealt{Kesden2012}), unless the SMBH is rapidly spinning \citep{Leloudas2016}. The occurrence of TDEs was predicted by theorists almost four decades ago \citep{Hills1975}, however observations of such exotic events started a lot later, first in the X-ray regime \citep{Komossa1999}, followed by the ultraviolet (UV) \citep{Gezari2006}, and finally reached the optical wavelengths \citep{Gezari2012}. Furthermore, there are some TDEs discovered in the mid-infrared \citep{Mattila2018} and others that launch relativistic jets and outflows leading to bright radio emission (e.g., \citealt{Zauderer2011,VanVelzen2016,Alexander}). 

When the bound debris starts to orbit around the SMBH, relativistic precession effects leads to a self-intersection of the debris stream and dissipation of energy \citep{Strubbe2009,Shiokawa2015,Guillochon2015,Bonnerot2020}. There are different models that try to explain the rise of such a luminous flare; either the radiation is produced when the intersecting stellar debris streams circularize and form a viscous accretion disk around the SMBH \citep{Rees1988,Phinney1989} or earlier, if radiation is produced directly from the stream collisions \citep{Piran2015,Jiang2016}. TDEs were expected to peak at the X-ray wavelengths as they were considered to be accretion-powered events \citep{Komossa2002}. %? See Andy comment . 
However, the X-ray properties of TDEs turn out to be much more diverse; up to 50\% of TDE candidates show no X-ray emission, others emit primarily in the X-rays and some ``intermediate cases'' show both optical/UV and moderate X-ray emission with X-ray to optical ratios spanning between $\leq$ 10$^{-4}$ to $\geq$ 10$^{3}$ \citep{Auchettl2017}. This large diversity casts doubts on our understanding of the underlying emission mechanisms. 

There have been two main families of models trying to explain such strong optical/UV emission (luminosities $\sim 10^{44}$ erg~s$^{-1}$) without detectable X-rays. The first scenario proposes that there must be some material around the SMBH which reprocesses the accretion disk emission to less energetic wavelengths (e.g., \citealt{Loeb1997,Strubbe2009,Guillochon2014,Roth2016}). A unified TDE scenario has been proposed \citep{Dai2018} which describes the TDE geometry with a thick, super-Eddington accretion disk. Due to inefficient accretion (mainly at early times) there is a polar relativistic jet as well as outflows of material \citep{Metzger2016} that can be optically thick and thus reprocess radiation to longer wavelengths. Depending on the line of sight of the observer, a TDE can be perceived as ``optical'' if viewed edge-on (all X-rays are reprocessed) or as ``relativistic/X-ray'' if viewed face-on (X-rays escape from the outflow/jet/funnel). Intermediate angles can reveal both optical and X-ray emission. A second scenario suggests that the optical/UV emission is produced by shocks occurring by collision/self-intersection of the debris streams  \citep{Piran2015,Jiang2016}. In this scenario, since the stream collision happens off-center, fluctuations in the intersection point drive material to the center later to form an accretion disk and produce the (sometimes) observed X-rays (which are delayed compared to the optical/UV emission \citealt{Pasham2017}) and the collisions between the debris streams can launch material on unbound trajectories. In both scenarios the bound debris form a photosphere -- either around the SMBH or the intersection point -- and outflowing material can be traced -- either by inefficient accretion or as unbound material leaving the self-intersection point produced by the collision of the streams. In both cases, if this material is directed to the observer's line of sight, it can produce blueshifted emission lines \citep{Nicholl2019}. \citet{Lu2020} also find that unbound debris can be produced from the shock experienced by the self-crossing debris stream due to relativistic precession when it returns near the SMBH. In their model, termed ``collision-induced outflow'' (CIO), the optical/UV radiation is not produced by the shock between the debris but from accretion of infalling matter from the intersection point to the BH. The EUV/X-ray radiation from the accretion is reprocessed by the CIO and re-emitted to optical/UV wavelengths. In this picture, it is the position and the bulk movement of the line emitting region of the CIO that sets the emission lines profiles and offsets (blueshifted or redshifted depending on the outflowing direction of the CIO with respect to the observer's line of sight) as well as the X-ray properties of a TDE (blocking or not the view of the accretion disk for the observer). In order to test the different competing models for TDEs, it is vital to examine their spectra and quantify the evolution of the emission lines and their ratios.

Early spectroscopic work on optically selected TDE candidates \citep{Arcavi2014} suggested that TDEs show a range of spectral properties with broad He and/or H emission features. However, the increasing number of TDE discoveries revealed a larger diversity than previously thought. Nitrogen and Oxygen lines have been discovered \citep{Blagorodnova2018,Leloudas2019,Onori2019} and attributed to the Bowen fluorescence mechanism \citep{Bowen1934,Bowen1935}, narrow low-ionization Iron lines have been identified implying the ionization of high-density, optically-thick gas by the X-rays of an accretion disk \citep{Wevers2019,Cannizzaro2021} and highly blueshifted, broad Balmer absorption and emission lines which have been attributed to outflows \citep{Hung2019,Nicholl2020}.
The spectroscopic heterogeneity is enhanced by diversity in line profiles, including double-peaked Balmer emission lines possibly probing the accretion disk \citep{Short2020a,Hung2020}.

For the Bowen lines to emerge, there must be a source of X-ray/far-ultraviolet (FUV) photons that will trigger a cascade of transitions and eventually result to the high-ionization Nitrogen and Oxygen lines \citep{Leloudas2019}. If an orientation-dependent unification scenario where the reprocessing happens around the accretion disk is at work \citep[such as the one proposed by][]{Dai2018}, line-photons should undergo a lot of scattering \citep{Roth2017} for large viewing angles resulting in broader emission lines than in those TDEs with prominent X-ray emission (smaller inclination). Thus, by measuring the widths of emission lines it is possible to test whether they depend on the viewing angle (based on the X-ray properties of the studied TDE) apart from only the kinematics of virially bound gas and Doppler broadening.
Another way to test proposed TDE scenarios is to search for wavelength shifts from the central wavelengths of the emission lines. Blueshifts have already been observed in a number of TDEs \citep{Arcavi2014,Holoien2015,Nicholl2020}. In the \citet{Dai2018} unification scenario, we might expect blueshifts to be correlated with the X-ray emission since they are both detected for small viewing angles (close to the poles). If the photosphere becomes thinner with time and easier for X-rays to penetrate \citep{Jonker2020}, then the observer should start seeing blueshifted lines at late-times, accompanying the rise of the X-ray luminosity \citep{Nicholl2019}. In the collisions paradigm, blueshifts should be unrelated with changes in the X-ray flux while in the CIO scenario, lines could be blueshifted, redshifted or even a combination depending on the outflowing direction. 

Furthermore, emission line ratios also provide critical insights. The radiative transfer calculations of \citet{Roth2016} suggest that the strength of lines in TDEs are set by the wavelength-dependent optical depths and they find that the Helium photosphere should lie deeper than the Hydrogen one in a ``stratified'' TDE atmosphere. An increasing \ion{He}{II}/H$\alpha$ luminosity ratio can confirm this prediction; since the blackbody photosphere of TDEs are found to recede with time (e.g., \citealt{vanvelzen2021,Hinkle2021a}) Helium lines should become stronger compared to Hydrogen, since the former lies deeper and gets emitted over a larger volume while the latter is self-absorbed at most radii and becomes weaker for smaller photospheres \citep{Nicholl2019}.
In addition, the H$\alpha$/H$\beta$ ratio, \ion{He}{II}/\ion{He}{I} ratio and the strength of the Bowen lines (and their evolution) can place critical constraints on the dominant emission line mechanism and the ionization state of the debris. Despite this large potential, there has been no systematic and comparative study of the spectroscopic properties of TDEs so far.

\indent In this paper, we present the first systematic spectroscopic analysis of a sample of optical/UV TDEs. In Sect. \ref{sec:data} we introduce our sample and the spectroscopic data for the TDEs and their host galaxies. In Sect. \ref{sec:analysis} we describe in detail our data reduction and in Sect. \ref{sec:els}, the methodology we employ to quantify the properties of the spectral lines. In Sect. \ref{sec:results} we present our results and in Sect. \ref{sec:discussion} we discuss their implications and present our interpretations. Sect. \ref{sec:conclusion} contains our summary and conclusions. Throughout the paper we assume a $\Lambda$CDM cosmology with H$_{0}$ = 67.4 km~s$^{-1}$ Mpc$^{-1}$, $\Omega_{\rm m}$ = 0.315 and $\Omega_{\rm \Lambda}$ = 0.685 \citep{Aghanim2020}.

\section{TDE sample and observational data} \label{sec:data}
To select our spectroscopic analysis sample, we started with candidates that have been proposed in the literature to be optical/UV TDEs (see \citealt{VanVelzen2020} for a proposed definition of this class). The criteria that had to be met for a TDE to enter our sample were: i) to have at least least two spectra, and ii) a host galaxy spectrum in order to perform proper host galaxy subtraction (for details see Sect. \ref{subsub:hgs}). Concerning the first criterion, we made an exception for PTF09ge \citep{Arcavi2014}, which only has one spectrum but it is one of the few early TDEs with pre-maximum data. 
The second criterion essentially imposed a ``cut'' for events discovered after 2019, as transient light could still be contaminating the host galaxy spectrum. 
In fact, a low luminosity long-lasting UV plateau has been found in late-time TDE observations \citep{VanVelzen2019} but we assume that the optical radiation is undetectable at such late stages. 
Therefore, objects discovered after 2019 are deferred to a future analysis.

Our final sample of 16 TDEs is presented in Table \ref{tab:sample}. We list the discovery and IAU name of the TDEs (we hereafter refer to TDEs with their IAU name, if they exist), the redshift, the luminosity distance, the Galactic extinction \citep{Schlafly2010} and the time of peak and the respective light curve band, as reported in the literature. For those TDEs where the peak was not observed (discovered after peak), we quote instead the time of discovery.

\begin{table*}
\renewcommand{\arraystretch}{1.2}
\setlength\tabcolsep{0.36cm}
\fontsize{10}{11}\selectfont
\begin{center}
\caption{Sample presentation}\label{tab:sample}
\begin{tabular}{l l c c c c c c}
\hline
Discovery & IAU  & Redshift & D$_{\rm L}$ & A$_{\rm V_{MW}}$ & t$_{\rm peak/max}$$^{b}$ & Peak/max & Discovery \\Name & Name  & & (Mpc) & (mag) & (MJD) & filter$^{c}$ & Ref$^{a}$ \\
\hline
PTF09ge &     & 0.064  & 289.2   & 0.044  & 54995.0 &R (PTF) & 1\\
PTF09djl &      & 0.184  & 900.0   & 0.047 & 55066.4 &R (PTF) & 1\\
PS1-10jh &        & 0.170  & 822.3   & 0.036 & 55389.8 &g ( PS1) & 2\\
LSQ12dyw &           &  0.090     & 414.2   & 0.074 & 56143.3& V-like (LSQ) & 3\\
ASASSN-14ae &         & 0.044  & 194.1   & 0.047 & 56682.5&- & 4\\
ASASSN-14li &       & 0.021      & 90.2   & 0.068 & 56983.6&- & 5\\
iPTF15af &            & 0.079  & 360.8  & 0.090 & 57037.3&g (PS1) & 6\\
ASASSN-15oi &     &  0.048     & 216.3   & 0.182 & 57248.2&- & 7\\
iPTF16axa &    & 0.108      & 503.2  & 0.121 & 57537.3&- & 8\\
iPTF16fnl &     & 0.016      & 71.1   & 0.219 & 57632.1&g (iPTF) & 9\\
PS17dhz & AT2017eqx  & 0.109     & 507.7   & 0.169 & 57921.6& i \& o (PS1 \& ATLAS) & 10\\
PS18kh & AT2018zr       & 0.075  & 341.6   & 0.124 & 58195.1& g (ASASSN) & 11\\
ASASSN-18pg	 & AT2018dyb     & 0.018  & 78.6  & 0.616 & 58343.6& g (ASASSN) & 12\\
ASASSN-18ul & AT2018fyk  & 0.060  & 270.4  & 0.036 & 58369.0&- & 13\\
ASASSN-18zj & AT2018hyz  & 0.046  & 203.9  & 0.092 & 58429.0$^{d}$&g (ASASSN) & 14\\
ZTF19aabbnzo & AT2018lna   & 0.080 & 365.7  & 0.126 & 58505.0& g (ZTF) & 15\\
\hline
\end{tabular}
\\[-10pt]
\end{center}
$^{a}$The discovery paper for each source (first journal article to present a classification and observed properties): $^1$\citet{Arcavi2014}, $^2$\citet{Gezari2012}, $^3$This work, $^4$\citet{Holoien2014}, 
$^5$\citet{Miller2015}, $^6$\citet{Blagorodnova2018}, $^7$\citet{Holoien2015}, $^8$\citet{Hung2017}, $^{9}$\citet{Blagorodnova2017},
$^{10}$\citet{Nicholl2019},$^{11}$\citet{Holoien2018}, $^{12}$\citet{Leloudas2019}, $^{13}$\citet{Wevers2019}, $^{13}$\citet{Short2020a}, $^{15}$\citet{vanvelzen2021}. \\
$^{b}$ Time of the peak of each TDE as reported in the discovery paper. For the events that were discovered post-peak, we provide the date of the first detection. \\
$^{c}$ The light curve band that was used in the discovery paper in order to interpolate between gaps in the data and retrieve the peak date. The sources that were only detected post-peak are denoted with a dash\\
$^{d}$ Taken from \citet{gomez1925}
\end{table*}

The sources from which we obtained our spectra (for both TDEs and host galaxies) vary. The majority of published spectra were retrieved from the Weizmann Interactive Supernova data REPository \citep[WISeREP\footnote{\url{http://www.weizmann.ac.il/astrophysics/wiserep}};][]{Yaron2012} or provided directly to us from authors; ASASSN-15oi and AT2018lna (unpublished spectra) were observed by the (e)PESSTO survey (the (extended) Public ESO Spectroscopic Survey for Transient Objects Survey; \citealt{Smartt2014}) using the EFOSC2 on the New Technology Telescope (NTT) at the La Silla Observatory, Chile. The NTT spectra were reduced in a standard manner with the aid of the PESSTO pipeline \citep{Smartt2014}. LSQ12dyw is a TDE discovered in 2012 whose data have not been analyzed previously, although its possible nature had been discussed in several circulars \citep{Smartt2012,12dyw2,12dyw3}.
With the knowledge accumulated since 2012, it is possible to classify this event as a \textit{bona fide} optical TDE. Its reduced spectra are already publicly available through the PESSTO data release one (DR1) and a dedicated publication, including the light curve and host galaxy properties is in preparation. Here, we focus on the spectroscopic properties of this event as part of a larger sample.

For some targets the host galaxy spectrum was available and we performed the host subtraction ourselves while for others the spectra were already host subtracted in WISeREP (ASASSN-14li and ASASSN-14ae) or provided host subtracted to us by colleagues (AT2017eqx, AT2018hyz).
Finally, for a few events, we obtained new host galaxy spectroscopy. We obtained the host galaxy spectra of iPTF16axa and AT2018zr using the ALFOSC on the Nordic Optical Telescope (NOT) on La Palma, Spain. The NOT spectra were reduced using standard \texttt{iraf} reduction tasks \citep{Tody1986}. The host galaxy spectra of ASASSN-15oi, AT2018lna and AT2018dyb were obtained with the NTT as part of the ePESSTO survey.
The exact data with all spectral epochs, including the host galaxy, are presented in Table \ref{tab:sample2}.

\section{Data processing} \label{sec:analysis}

There were several processes that needed to be carried out consistently for all the reduced spectra of our sample before moving on to the study and accurate analysis of the emission lines. These were the following: i) scaling of the spectral fluxes with the photometry, ii) correction for Galactic extinction, iii) subtraction of the host galaxy spectrum and iv) fitting and removing the continuum from the host subtracted spectra. All these processes were performed using customized scripts in Python. This section describes those processes in detail. 
The homogeneously reduced and analyzed spectra will be made publicly available via WISeREP \citep{Yaron2012}.

\subsection{Scaling fluxes with photometry} \label{subsub:sfwp}
Despite the fact that the standard reduction processes of the raw spectra include flux calibration using a spectrophotometric standard star, there can still be slit and fiber losses that some times can be differential (i.e., depend on wavelength). %losses due to differences in slit/fiber widths. 
In order to account for this, we scaled all reduced spectra (including the host galaxy spectra) with the respective optical photometry of each transient/galaxy. The photometric data,
%(preferentially g and r bands or Swift V and B bands in order to coincide with the wavelengths of our optical spectra) 
were retrieved either from the literature (see Table \ref{tab:sample2}), the High Energy Astrophysics Science Archive Research Center (\textsc{HEASARC}) data archive\footnote{\url{https://heasarc.gsfc.nasa.gov/cgi-bin/W3Browse/w3browse.pl}} for Swift's \textsc{UVOT} light curves or the \textsc{Lasair} broker \footnote{\url{https://lasair.roe.ac.uk/}} \citep{Smith2019}. 
Since photometric data were not always available for the exact date of spectroscopic observations, we obtained the flux value via polynomial interpolation. 
%Furthermore, if there was a discrepancy between the flux in a specific wavelength after scaling with two (or more) different bands, we were fitting a linear polynomial to the photometric points and then scaled the flux of the spectrum using this polynomial (even more accurate process).
In order to account for differential slit losses, in a few cases we ``mangled" the spectrum to reproduce the colors in multiple photometric bands (by multiplying with a linear function interpolating between scaling factors in different wavelengths).

\subsection{Correction for Galactic extinction} \label{subsub:cfge}
In order to correct spectra for Galactic extinction, we used the \textsc{extinction}\footnote{\url{https://github.com/kbarbary/extinction}} package of Python and employed the extinction curve of \citet{Cardelli1989}. The $A_{V}$ value used for each TDE can be found in Table \ref{tab:sample}. Correction for host galaxy extinction was not applied due to lack of data to constrain its significance. In addition, most of our TDE sample are found in passive, quiescent galaxies with little evidence for dust \citep{Arcavi2014,French2020}. 

\subsection{Host galaxy subtraction} \label{subsub:hgs}
\begin{figure}
\centering
\includegraphics[width=0.47 \textwidth]{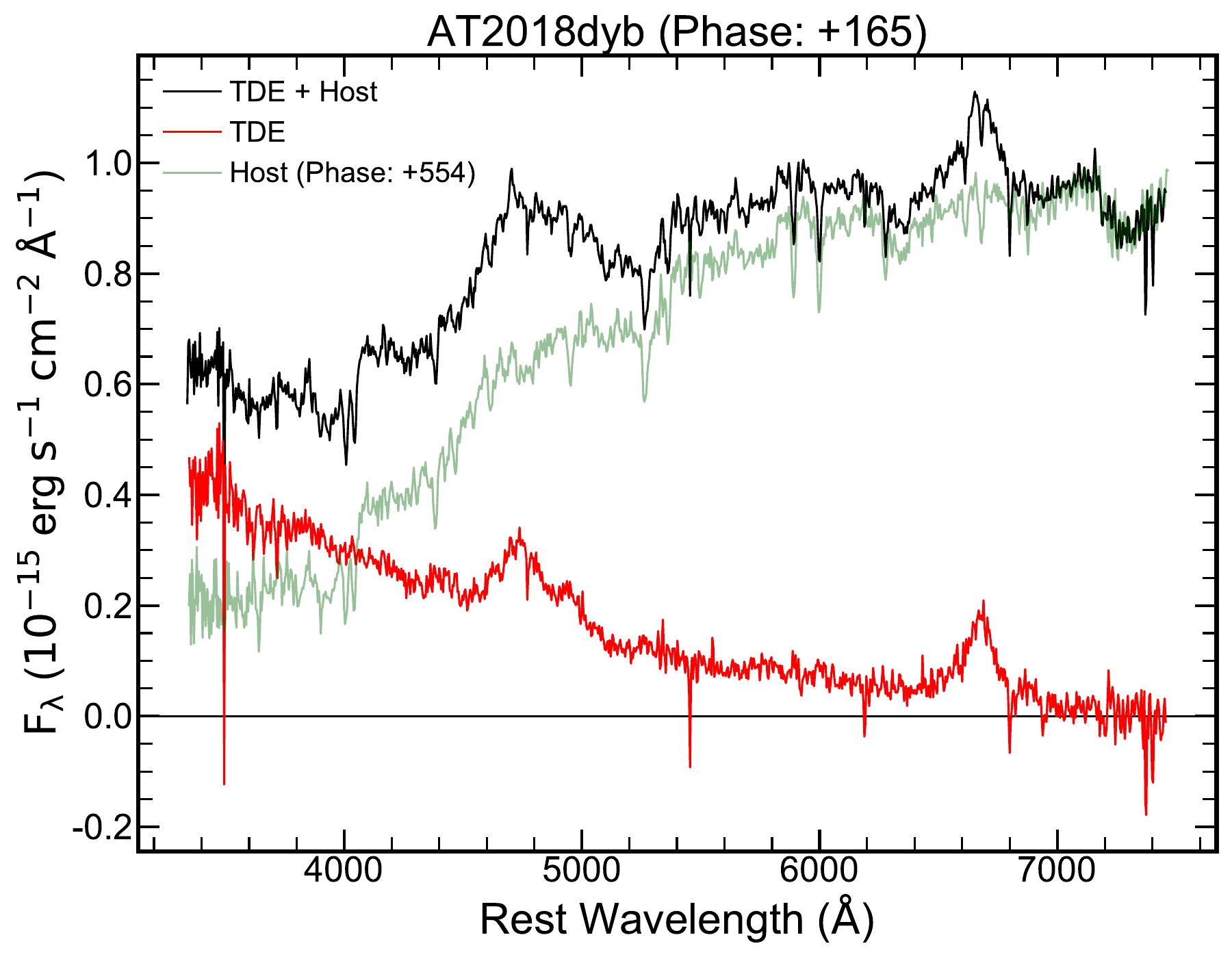}
\caption{
Example host subtraction using the spectrum of TDE AT2018dyb (black) at 165 days post peak ($+165$ days) and the host galaxy spectrum, obtained at $+$554 days  (green). Despite the significant host contamination, the resulting TDE spectrum (red) is of high quality, allowing for an accurate study of the emission lines even at late stages of the TDE evolution.
}\label{fig:hostgalsub}
\end{figure}

TDEs are embedded in their host galaxy and any obtained spectrum is a superposition of the TDE and host galaxy. The host contamination is especially significant at late phases and at redder wavelengths due to the dimming of the TDE; since TDEs are intrinsically blue and the host galaxies that host them are usually passive and quiescent (hence red), the contamination of the host at red wavelengths becomes quickly significant. For this reason, in order to study the TDE flare, the host light needs to be carefully removed. We have performed the host subtractions for all the spectra of our sample unless the spectra were provided to us already host subtracted (see Table \ref{apdx:tables}, ``\textit{Notes}'' column). In Fig. \ref{fig:hostgalsub}, we plot and visualize the result of such a procedure (for TDE AT2018dyb) where the host galaxy contribution is removed (along with its absorption lines) and the resulting spectrum is the TDE flare itself.

\subsection{Continuum removal} \label{subsub:cs}
\begin{figure}
\centering
\includegraphics[width=0.47 \textwidth]{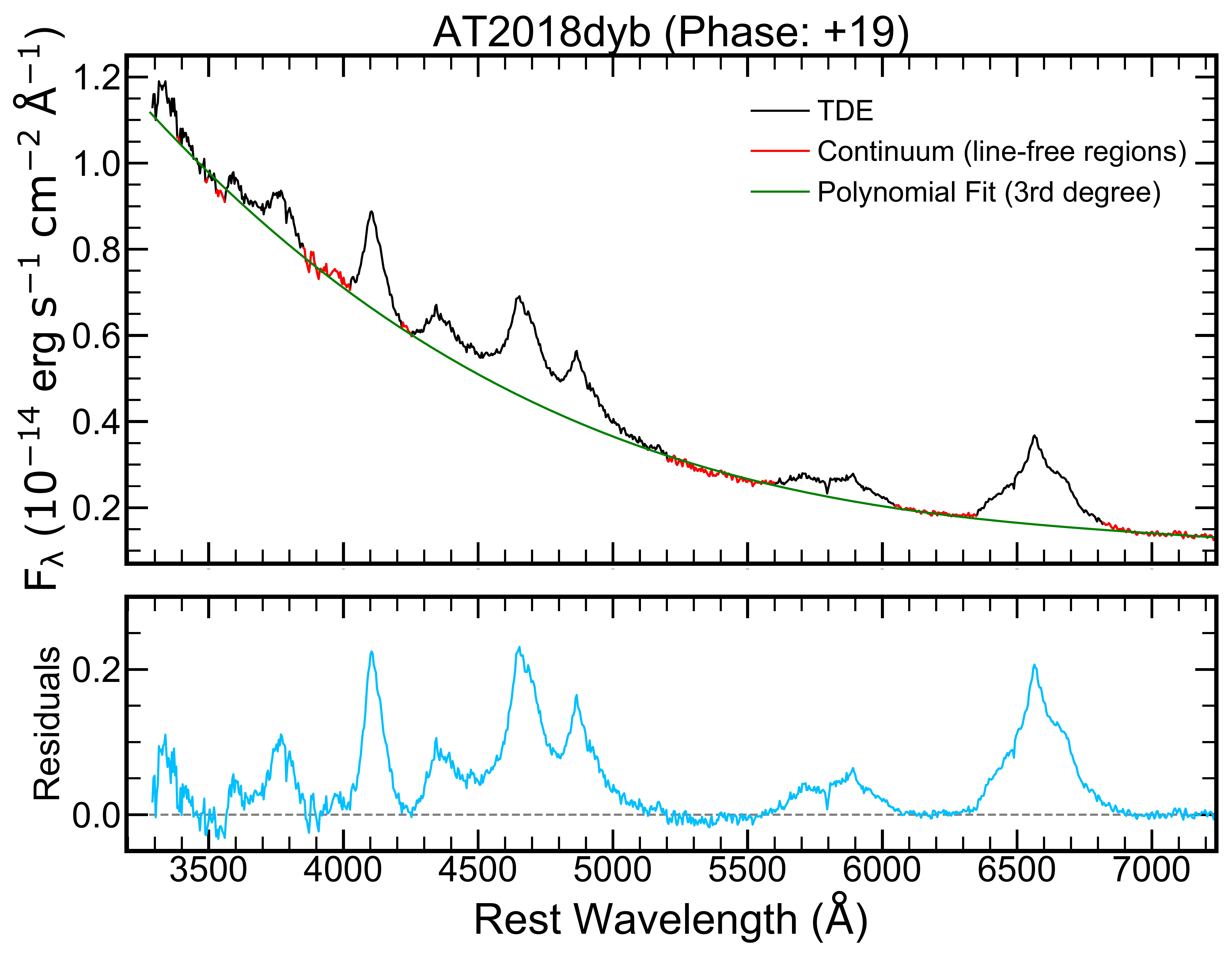}
\caption{Example continuum removal from the spectrum of TDE AT2018dyb at $+18$ days (black). The line-free regions are marked with red and the polynomial fit to the continuum with green. The lower panel, shows the residuals, i.e., the pure emission line spectrum of the TDE (blue).
}\label{fig:contsub}
\end{figure}

Since our focus is on the spectral lines of the TDE flare, the continuum needs to be removed. In order to fit a continuum to our host subtracted spectra we first carefully chose the line-free regions of each spectrum in order to fit the continuum in those areas. This procedure is often not trivial and sometimes subjective. The spectra of TDEs are typically dominated by broad emission lines and certain areas of the spectrum are occasionally heavily blended (e.g., the 4300 -- 4900 \AA\, area). This makes the choice of the line-free regions particularly challenging for some TDEs. The advantage with our study is that this procedure was performed consistently for the entire sample of TDEs, following the same criteria when selecting the line-free regions of each TDE and epoch as well as applying similar techniques for the whole sample in order to fit and remove it. Typical selections of the line-free regions involve: 3900 -- 4000 \AA, 4220 -- 4280 \AA, 5100 -- 5550 \AA, 6000 -- 6350 \AA\, and 6800 -- 7000+. Of course these ranges were adjusted to the specific features of each TDE and were modified for different TDEs and epochs. After marking the line-free regions we
fit them using a polynomial. 
We also experimented with power-law functions but we found that the polynomials often resulted in better reduced chi-square (${\rm \chi_{\rm \nu}}^{2}$) values. For each spectrum, we tried $3^{rd}$-$5^{th}$ order polynomials and used the one yielding the lowest ${\rm \chi_{\rm \nu}}^{2}$ value. In Fig. \ref{fig:contsub}, we plot and visualize these procedures (for TDE AT2018dyb). After obtaining the best fit continuum, we subtract it from the host subtracted TDE spectrum in order to study the emission lines and retrieve line luminosities, full width at half maxima (FWHM) and velocity offsets. Another possibility would be to use a blackbody to estimate the continuum level. However, using the blackbody temperatures estimated from the photometry, we found that this method was unsuitable for our purposes leading to either a continuum estimation that was not precise enough or even to systematic over- or under-estimation of the continuum level for different spectral ranges and epochs and much larger ${\rm \chi_{\rm \nu}}^{2}$ values. This was also pointed out by \citealt{Hung2019}. The continuum removal represents the largest systematic uncertainty in our analysis.

\section{Emission line analysis} \label{sec:els}

After we obtained the emission line spectra, we performed our spectroscopic line study using customized Python scripts employing the \textsc{lmfit}\footnote{\url{https://lmfit.github.io/lmfit-py/}} package \citep{Newville2016} where a Levenberg-Marquardt algorithm (i.e., least-squares method) was used for fitting. In this study, we focused on these following emission lines which are common in TDE spectra \citep[e.g.,][]{Arcavi2014,Leloudas2019,vanvelzen2021}: H$\alpha$, H$\beta$, \ion{He}{II} 4686 \AA, \ion{He}{I} 5876 \AA,\, and the Bowen fluorescence lines \ion{N}{III} \textasciitilde4640 \AA\, (doublet at 4634 \AA\, and 4641 \AA), \ion{N}{III} \textasciitilde4100 \AA\, (doublet at 4097 \AA\, and 4104 \AA) \citep{Osterbrock1974}. However, additional lines, such as H$\gamma$ and \ion{He}{I} 6678 \AA,\, are also included in our fits as part of the deblending.

\subsection{Line profiles}
\label{subsub:lp}
Most studies in the literature fit emission lines of TDEs with a single Gaussian in order to simplify the process of quantifying their properties \citep[e.g][]{Arcavi2014,Blagorodnova2017,Hung2017}. Nevertheless, the line profiles of TDEs can be very complicated and diverse (e.g., \citealt{Holoien2018,Short2020a,Nicholl2020}). The line profiles can be a result of many different physical processes; a large electron-scattering
optical depth, the kinematics of an accretion disk %(resulting in double-peak profiles) 
and/or outflows which induce asymmetries.

During our analysis, we tried different ways to obtain measurements on the emission lines, including fitting different line profiles (Gaussian and Lorentzian) and direct integration. The direct integration method is only useful for isolated lines, therefore it is meaningless to use it for the 4300 -- 4900 \AA\, region, especially for \ion{N}{III} Bowen TDEs. In reality, even the most isolated lines in TDEs are often blended with other emerging lines or demonstrate asymmetric bumps. For example, the \ion{He}{I} 6678 \AA\, line sometimes contributes significantly to the red side of H$\alpha$ for some TDEs (e.g., \citealt{Blagorodnova2017,Leloudas2019,Nicholl2019}). 
Another example is the occasional emergence of a line (attributed to N~II~5754 \AA\, in \citealt{Blagorodnova2017}) on the blue side of \ion{He}{I} 5876 \AA\, (or maybe the Na~I~5889 \AA\, line on the red side). In these 2-line blended cases one can either use multi-profile fitting in order to de-blend the lines or, in case the line profile deviates from the familiar spectral profiles (Gaussian or Lorentzian), one can use direct integration in order to measure the flux of the whole blend and then fit a spectral profile to the emerging line/asymmetric bump in order to subtract it from the total underlying flux. 

\begin{figure}
\centering
\includegraphics[trim={0 0cm 0 0},clip,width=0.47 \textwidth]{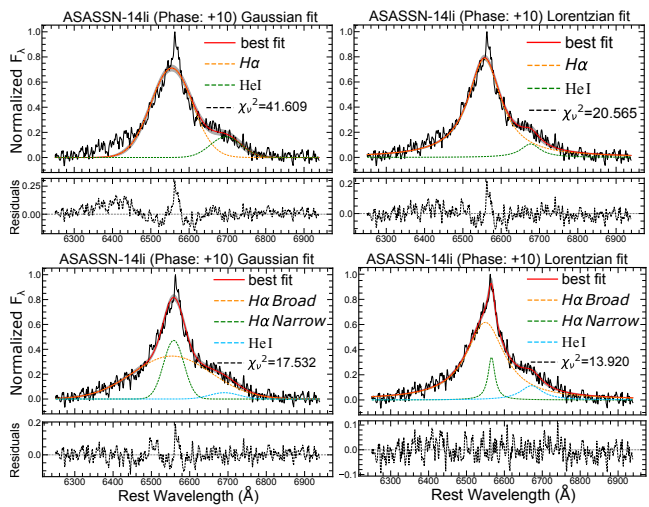}
\caption{We fit the H$\alpha$ profile of ASASSN-14li at $+14$ days with a single Gaussian (top left), a single Lorentzian (top right), a double Gaussian (bottom left) and a double Lorentzian (bottom right). Although a double Lorentzian (broad and narrow profile) provides the best fit, a single Lorentzian and a double Gaussian also capture well the line profile (while a single Gaussian results in a worse fit, particularly at the blue wing) and yield comparable results for the line luminosity. For simplicity, in our sample comparative analysis we use the single Lorentzian fit for ASSASN-14li.
}\label{fig:GL_fits0}
\end{figure}

It has been argued that the emission lines of some TDEs can be better modeled by two components. For example, \citet{Holoien2020} used a narrow and a broad component to fit the H$\alpha$ line of AT2018dyb and \citet{Nicholl2020} and \citet{Hinkle2021} did the same for AT2019qiz and AT2019azh respectively. 
Although this may yield more accurate results and may in fact be physically motivated for the study of individual events, it is not practical for the purpose of our comparative sample analysis. To keep the fitting simplified, we chose to make model fits with single components.
In Fig. \ref{fig:GL_fits0}, we explore the effect of this choice: 
we present four different fits for the H$\alpha$ line of TDE ASASSN-14li including single or double (broad and narrow) Gaussians or Lorentzians. In addition, we separately fit the \ion{He}{I} 6678 \AA\, line on the red side of H$\alpha$. 
The double Lorentzian fit results in the best ${\rm \chi_{\rm \nu}}^{2}$ value but the interesting fact here is that the double Gaussian has just a slightly better ${\rm \chi_{\rm \nu}}^{2}$ than the single Lorentzian (17.5 and 20.6) and the residuals look similar. The single Gaussian scores worse in terms of ${\rm \chi_{\rm \nu}}^{2}$ and fails to capture the blue wing of the line.
Furthermore, the line luminosities resulting from the three best fits are all within 1$\sigma$ from each other: the single Lorentzian fit results in L = 9.36 $\pm$ 0.13$\,\,\times\,$ 10$^{40}$~erg~s$^{-1}$, the double Gaussian fit results in L = 8.77 $\pm$ 0.55$\,\,\times\,$ 10$^{40}$~erg~s$^{-1}$ and the double Lorentzian fit results in L = 9.48 $\pm$ 0.33$\,\,\times\,$ 10$^{40}$~erg~s$^{-1}$. 
At the same time, the double component solution makes the comparison to other TDEs, as well as the definition of line widths and velocity offsets, significantly more complicated. 
For this reason, we chose to keep the single Lorentzian solution, which captures well the line profile, for our analysis. 

For all TDEs, we chose between a Gaussian or a Lorentzian profile depending on the ${\rm \chi_{\rm \nu}}^{2}$ of the fit. A Lorentzian profile is usually attributed to collisional (or pressure) broadening which could be in accordance with the electron scattering \citep{Roth2017} models. 
We used the Lorentzian fits for two TDEs of our sample, namely ASASSN-14li and AT2018dyb. These TDEs exhibited broad wings in their line profiles and the Lorentzian was a very good match. 

\begin{figure}
\centering
\includegraphics[trim={0 0 0 0},clip,width=0.47 \textwidth]{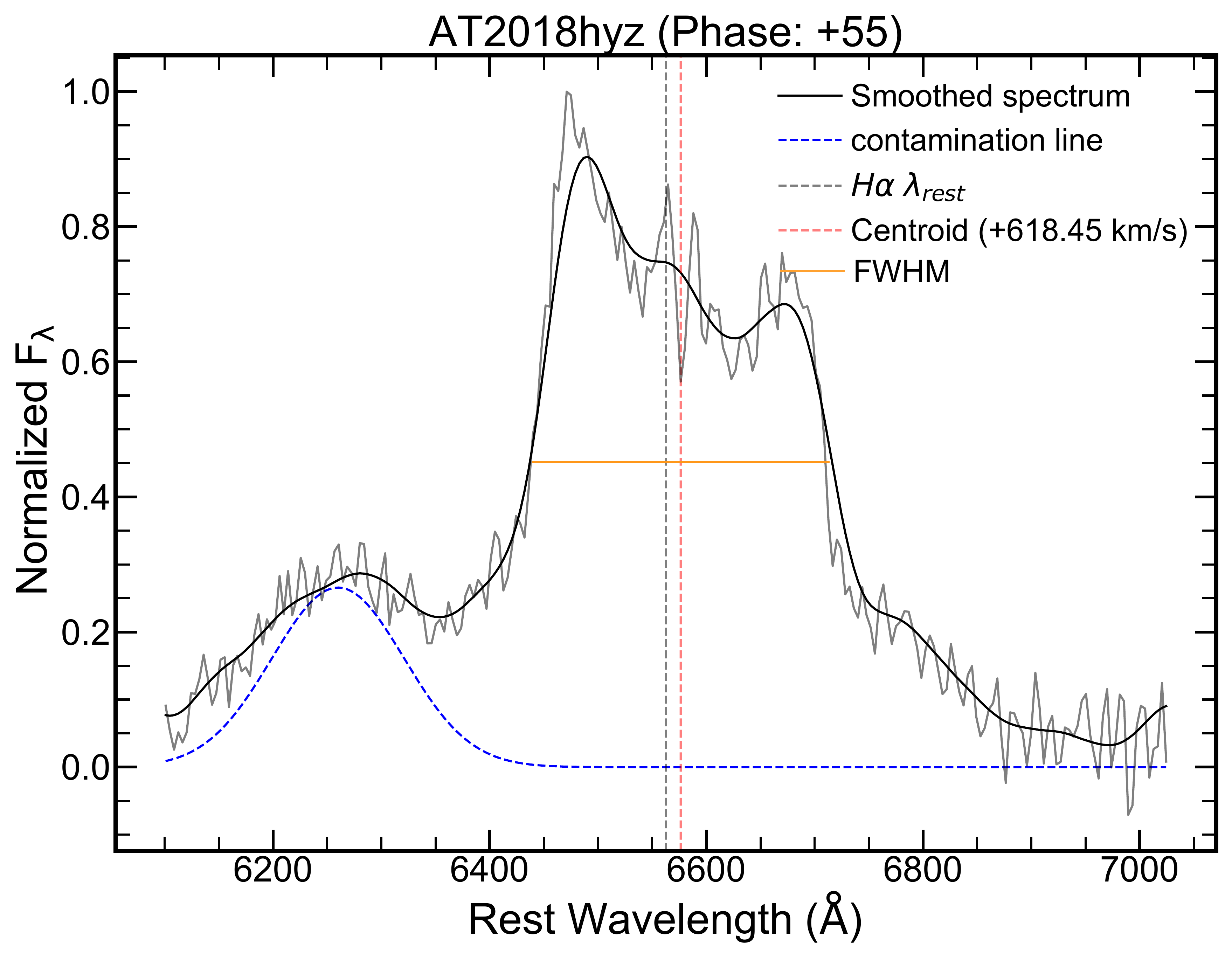}
\caption{Demonstration of our direct integration method for profiles where a Gaussian or Lorentzian fit was not possible. This is the H$\alpha$ line of TDE AT2018hyz. We employed direct integration to measure the flux of the whole blend and determine the line width and velocity offset (see text for details). A Gaussian profile was fit to the line on the blue side of H$\alpha$ in order to subtract it from the total underlying flux. }.
\label{fig:DG_fits}
\end{figure}

When the line profile deviated from the familiar spectral profiles, and neither a Lorentzian nor a Gaussian provided a good fit, we were forced to use direct integration in order to quantify the line properties. 
Such a case is illustrated in 
Figure \ref{fig:DG_fits} for TDE AT2018hyz, which demonstrated peculiar double-peaked Balmer lines \citep{Short2020a,Hung2020}.
One issue that the direct integration introduces (apart from being impossible to use in heavily blended areas) is the measuring of the FWHM and the offset of the studied line since it is not a free parameter of the fit anymore (as it is for example in a Gaussian or Lorentzian fit). In order to overcome this issue, we used a custom script in Python which first smooths the spectrum, then locates the data points on the left and right of the maximum that have flux values closest to the half of the maximum and then calculates the distance between them on the x-axis. In addition, it calculates the mean of the above length and measures its deviation from the rest wavelength of the studied line and finally it converts everything to velocity space. We use a custom Monte-Carlo method (10\,000 iterations of re-sampling the data assuming Gaussian error distribution) in order to calculate uncertainties for the flux (luminosity), FWHM and offset of the line. 
The TDEs that were fit with direct integration were PTF09ge, AT2018hyz, AT2018zr (double-peak profiles) and LSQ12dyw (boxy and broad red shoulder profiles). Fortunately these TDEs did not show Bowen features so we used direct integration for both H$\alpha$ and H$\beta$ as well as \ion{He}{II} (if present). 
%This allowed us to remain consistent within the measurements of e.g., line luminosity ratios (H$\alpha$/H$\beta$ or \ion{He}{II}/H$\alpha$) of each TDE.

\subsection{Fit parameters and constraints}\label{subsub:fpac}
The area around 4300 -- 4900 \AA\, is heavily blended and could contain in some cases as many as nine different lines blended together, namely the 4100 \AA\, line (\ion{N}{III} \& H$\delta$), H$\gamma$ 4340.5 \AA, \ion{N}{III} 4379 \AA, \ion{Fe}{II} $\sim$ 4550 \AA\, ($\lambda\lambda$ 4512,4568,4625), \ion{N}{III} 4640 \AA\,, \ion{He}{II} 4686 \AA\, and H$\beta$ 4861.3 \AA\, (some times an absorption trough around 4225 \AA\, is also seen but it may be a continuum removal artefact). 
The de-blending of such areas is not a trivial process and can be particularly complicated (especially for lower S/N spectra). Consequently, we fit the entire TDE spectrum simultaneously and, since this is a fit with many free parameters, we provide some physical information by imposing some reasonable constraints in order to reduce the number of possible solutions and help the fit converge. We require that lines of the same ion have a similar width; the FWHM of H$\beta$ and H$\gamma$ to be within $\pm$ 3000 km~s$^{-1}$ of that of H$\alpha$ and the FWHM of the \ion{N}{III} 4640 \AA\, line to be within $\pm 3000$ km~s$^{-1}$ of that of the \ion{N}{III} 4100 \AA. In some few cases where the \ion{N}{III} 4640 \AA\, and \ion{He}{II} 4686 \AA\, were completely unresolved (either because of low S/N or because the two lines were intrinsically very broad, or because they were blended with a mix of lines that emerge sometimes between H$\gamma$ and \ion{N}{III} 4640 \AA) and the fit did not converge, we set an extra constraint where we require that the amplitude of the \ion{N}{III} 4640 \AA\, line is similar (i.e., within a factor of 2) to the one of \ion{N}{III} 4100 \AA\, (in practice this helped the fit converge).

\begin{figure}
\centering
\includegraphics[width=0.47 \textwidth]{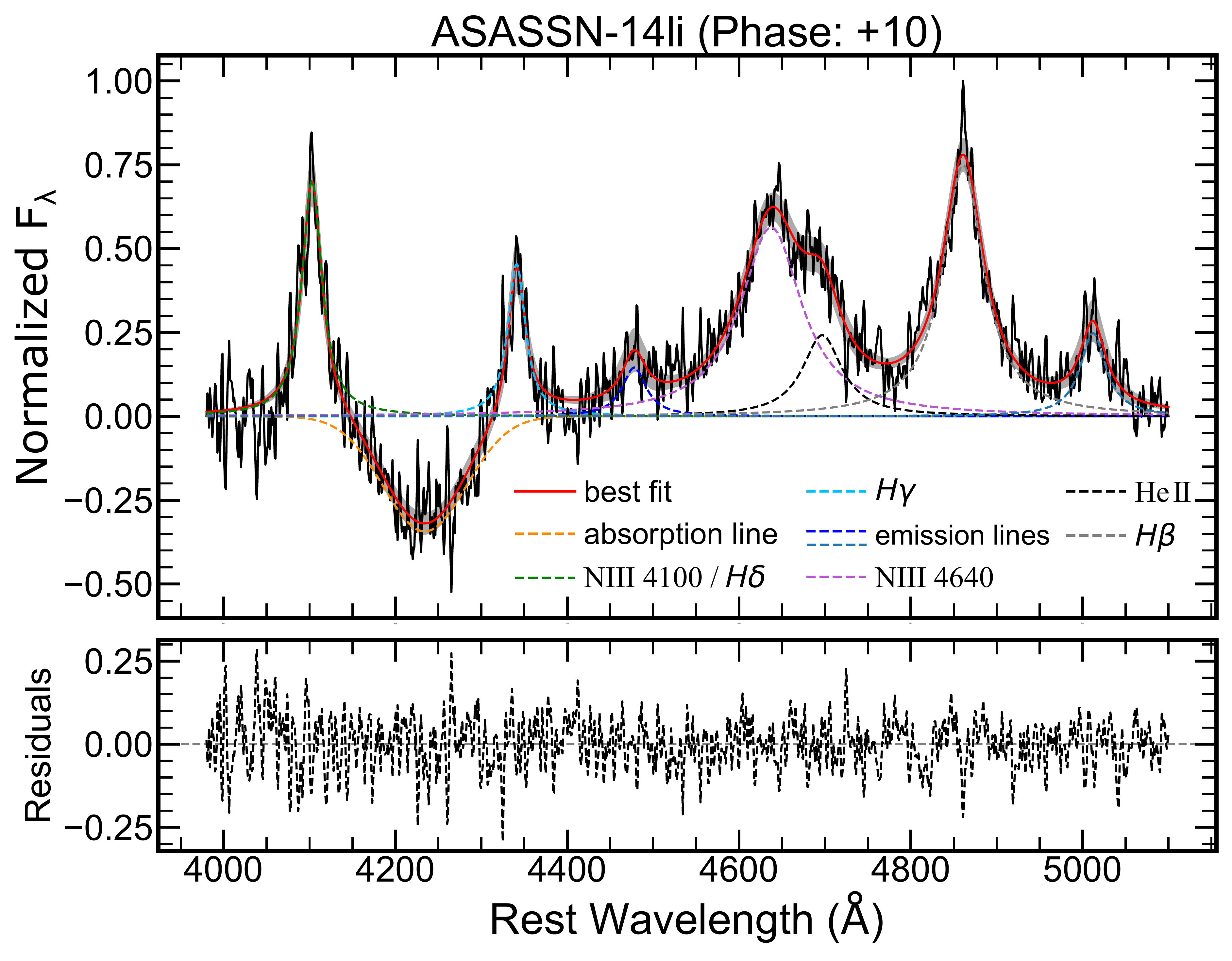}
\caption{Example fit for the blue part of the spectrum of ASASSN-14li at a phase of $+10$ days. This fit requires seven Lorentzians. We note the presence of an absorption trough on the blue side of H$\gamma$, which is fit here by an extra Gaussian. This absorption feature is present in more TDEs, probably an artefact resulting from the simplest possible continuum removal.
}\label{fig:debl_fit}
\end{figure}

In Fig. \ref{fig:debl_fit} an example case is presented (for TDE ASASSN-14li) where we fit seven Lorentzian and one Gaussian in order to achieve an acceptable fit and minimize the residuals. The Gaussian is used for an absorption feature at 4200 \AA,\, which is present in a few more TDEs. It is not clear if this corresponds to real absorption or whether it is an artefact of the simplest choice of removing the continuum. Setting the continuum at the minimum of this absorption trough, would result in unrealistic choices for the rest of the spectrum. Such complicated fits resulted in large correlations between pairs of fitted variables, especially between \ion{N}{III} 4640 \AA\, and \ion{He}{II} 4686 \AA, which consequently resulted in large errors. Another caveat of such complicated fits is that the offsets and widths of these blended lines might occasionally not be very trustworthy, something that is mirrored in the very large errors of these parameters. This is why we only study velocity offsets for the most isolated lines, such as H$\alpha$, \ion{N}{III} 4100 \AA\, and \ion{He}{I} 5876 \AA.

\subsection{Calculation of uncertainties}
\label{subsub:cou}
After a fit using the least-squares method has completed successfully, standard errors for the fitted variables and correlations between pairs of fitted variables are automatically calculated from the covariance matrix. In principle, the uncertainties in the parameters are closely tied to the goodness-of-fit statistics (chi-square). 
%as the standard errors or 1$\sigma$ uncertainties on a parameter is found where the chi-square has increased by unity from the minimum. 
The \textsc{lmfit} documentation argues that since it is often not the case that one has realistic estimates of the data uncertainties (error spectrum), the standard errors or 1$\sigma$ uncertainties reported by \textsc{lmfit}  
%are not those that increase chi-square by 1. 
are those that rescale the uncertainty in the data such that the reduced chi-square would be 1, assuming the underlying model is true. Consequently, if the reduced chi-square is far from 1, this re-scaling often makes the reported uncertainties large. 
In this work we have adopted the \textsc{lmfit} approach
since our error spectra do not include uncertainty estimates for a number of data analysis procedures.
In particular, the uncertainty in the continuum removal is hard to quantify and may be the most dominant source of uncertainty.

\section{Results} \label{sec:results}
In this section, we present our measurements of line luminosities, line widths and velocity offsets for our TDE sample. We study the evolution of these quantities with time as well as their dependency on the blackbody temperature (T$_{\rm BB}$) and radius (R$_{\rm BB}$) as retrieved from the literature (see Table \ref{tab:sample2}). A discussion on the implications of our results and their physical interpretation is presented in Sect. \ref{sec:discussion}.

\subsection{Line luminosities} \label{sub:ll}

\subsubsection{H$\alpha$ luminosity evolution} \label{subsub:hale}
\begin{figure*}
\centering
\includegraphics[width=0.75 \textwidth]{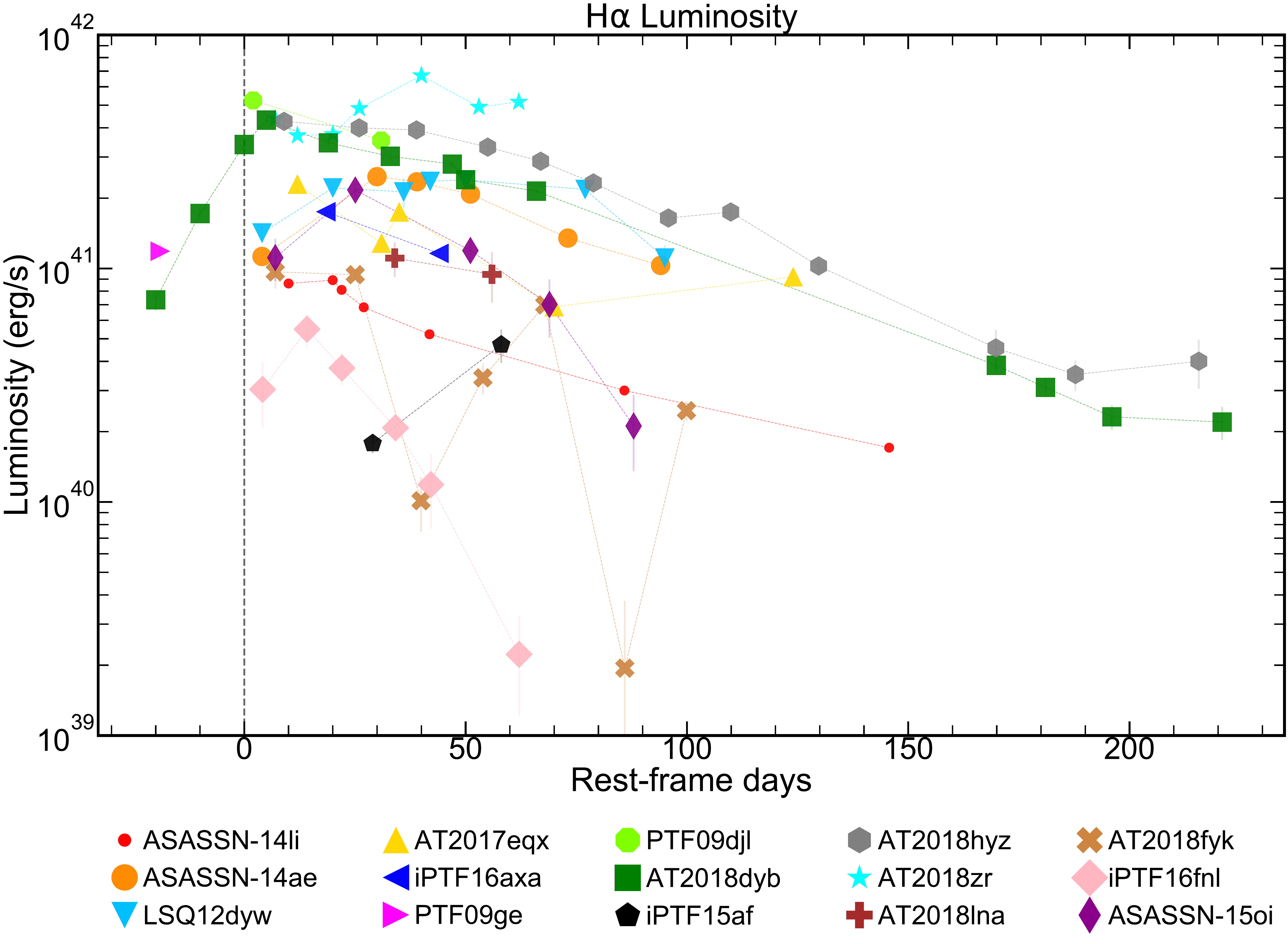}
\caption{Evolution of the H$\alpha$ line luminosity for all the TDEs of our sample. The dashed vertical line denotes the time of peak or discovery of each TDE. A number of TDEs show a lag between their H$\alpha$ luminosity peak and the time of the light curve peak. }
\label{fig:Ha_lums}
\end{figure*}

\begin{figure*}
        \centering
        \begin{subfigure}[b]{1\textwidth}
            \centering
            \includegraphics[trim={0 0cm 0 0},clip,width=\textwidth]{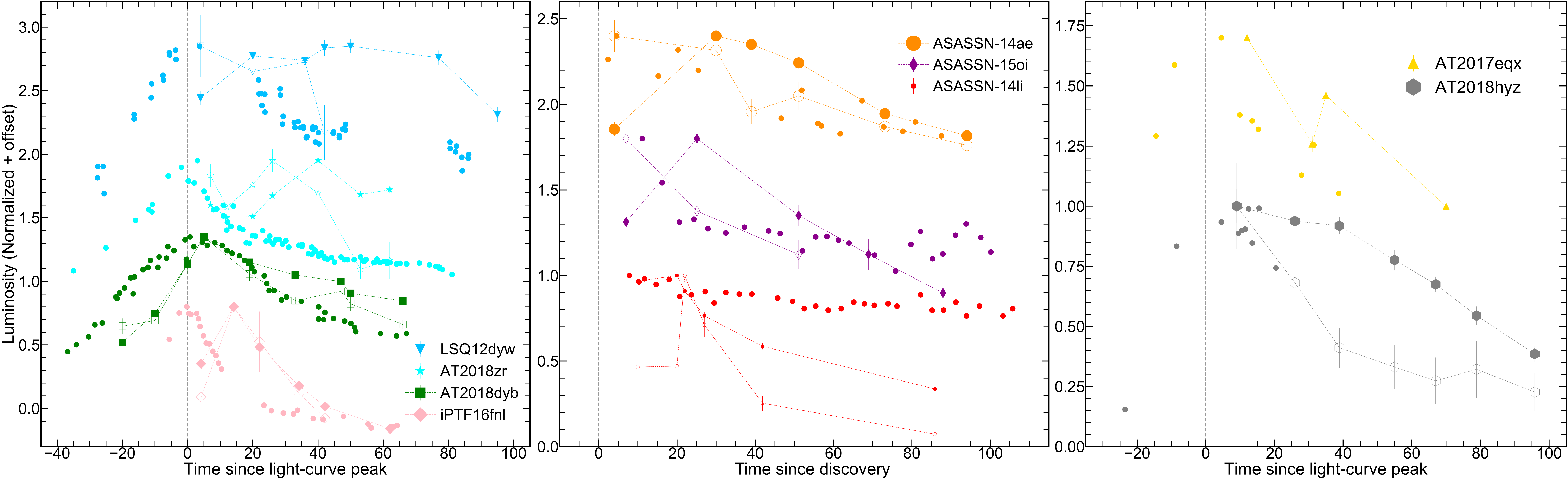}
            % \caption[Network2]%
            % {{\small Network 1}}    
            % \label{fig:off_sub}
        \end{subfigure}
        % \vskip\baselineskip
        \caption
        {Comparison of the H$\alpha$ (filled markers), \ion{He}{I} 5876 \AA\, (empty markers) and continuum light curves for the TDEs in our sample that have a ``determinable'' time lag between the H$\alpha$ and the optical light curve luminosities (i.e., events with $\leq$ two spectra are not plotted). Left panel: events that were observed pre-peak and the lag between the continuum and H$\alpha$ luminosities are obvious. The dashed vertical line denotes the time of peak (see Table \ref{tab:sample}). Middle panel: events discovered post-peak but for which the H$\alpha$ luminosity shows a delayed peak and for which  a lower-limit can be placed on the lag. The dashed vertical line denotes the time of discovery of these TDEs as reported in their discovery paper. The plotted light curves are in the Swift $V$ band. Right panel: events for which the existence of a time lag cannot be claimed. The dashed vertical line denotes the time of peak.} 
        \label{fig:ha_heI_lc}
    \end{figure*}

\begin{figure}
\centering
\includegraphics[width=0.45 \textwidth]{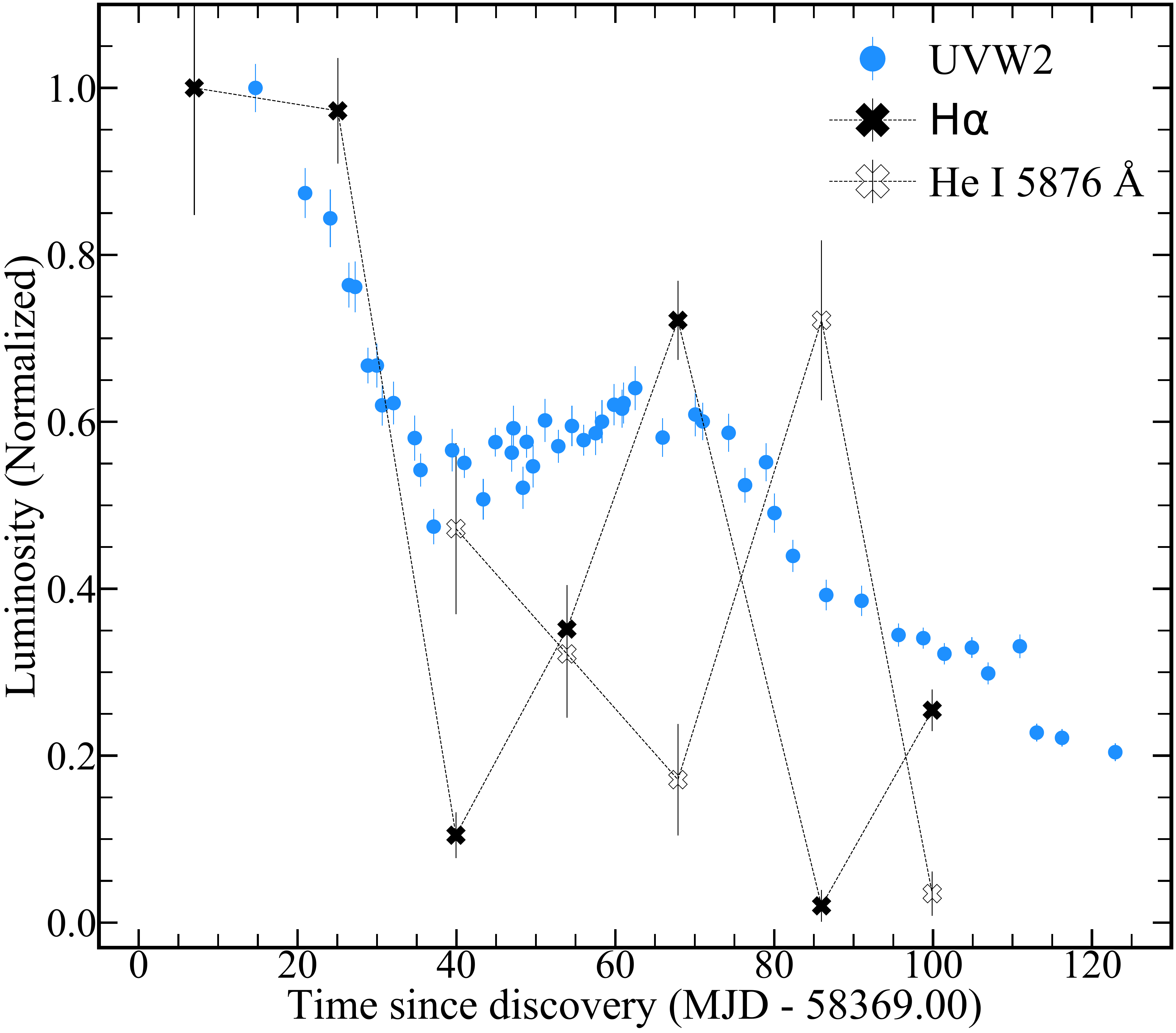}
\caption{The evolution of the H$\alpha$ and \ion{He}{I} 5876 \AA\, line luminosities of TDE AT2018fyk are plotted with black and empty X-marks respectively and the Swift UVW2 band light curve \citep{Wevers2019} is plotted in blue circles. All luminosity curves are normalized to one. The line luminosities respond to variations of the light curve with a small lag for H$\alpha$ and a larger one for \ion{He}{I}.}
\label{fig:18fyk_UVW2}
\end{figure}

Figure \ref{fig:Ha_lums} shows the evolution of the H$\alpha$ line for all the TDEs in our sample. 
At peak, the typical H$\alpha$ luminosities span a range of $\sim$ 0.5 -- 7 $\,\times\,10^{41}$ erg s$^{-1}$. 
Similar to the broad-band light curves, for most events the line shows a rise in the luminosity  until it starts to decline with time. 
Interestingly, for many TDEs in our sample %(LSQ12dyw, iPTF16fnl, AT2018zr and AT2018dyb), 
the H$\alpha$ luminosity is still rising after the respective optical light curve of the TDE has reached its peak (marked in Fig. \ref{fig:Ha_lums} by the dashed vertical line). In other words, there is a measurable time lag between the light curve and H$\alpha$ luminosity maxima. 
A comparison of the H$\alpha$ and continuum luminosity evolution of these events is shown in Fig. \ref{fig:ha_heI_lc}. In the left panel, four events are shown for which this time lag is obvious as the line luminosity peaks after the continuum light curve (LSQ12dyw, iPTF16fnl, AT2018zr and AT2018dyb). 
However, the existence of a delayed peak in H$\alpha$ can also be deduced, and a lower limit can be placed, even for events where the continuum peak has not been observed (the TDE was discovered after maximum). This is the case for three events of our sample (ASASSN-14li, ASASSN-14ae, ASASSN-15oi), which are shown in the middle panel of Fig. \ref{fig:ha_heI_lc}. On the other hand, only two events in our sample (AT2017eqx and AT2018hyz) do not show any evidence for a delayed peak in H$\alpha$ (Fig. \ref{fig:ha_heI_lc}; right panel), while a conclusion is not possible for events where we have $\leq$ two spectra (hence these events are not included in Fig. \ref{fig:ha_heI_lc}). A special case is AT2018fyk, which showed multiple maxima in its light curves (see \citealt{Wevers2019,Wevers2021} for more details).
However, AT2018fyk is the TDE that provides further evidence  that the line luminosity responds to variations in the continuum light curve.
%  and interestingly we see the same behavior in L$_{\rm H\alpha}$. 
This connection is visualized in Fig. \ref{fig:18fyk_UVW2}; the lag is small and a proper cross-correlation analysis is needed to robustly quantify it.
However, this is not possible to do in this work due to the small number of available spectra and the sparse coverage.
In Table \ref{tab:lag} we quantify the H$\alpha$ lag after simply fitting a $3^{rd}$ order polynomial around the peak of the H$\alpha$ luminosity in order to determine the time of its peak. The lags span $\sim7$ to $\sim45$ days. We note that the \ion{N}{III} Bowen TDEs show smaller lag values compared to the rest which ties nicely with the fact that \ion{N}{III} Bowen TDEs have consistently lower R$_{\rm BB}$ values \citep{vanvelzen2021}. These results are further discussed in Sect. \ref{subsec:dtl}.

\begin{table}
\renewcommand{\arraystretch}{1.2}
\setlength\tabcolsep{0.06cm}
\fontsize{8.8}{11}\selectfont
% \begin{center}
\caption{Time lag and inferred distances between line and continuum luminosities. %and the optical light curves for the TDEs that show a peak in their H$\alpha$ luminosity evolution.
}\label{tab:lag}
% \begin{center}
\begin{tabular}{c c c c c }
\hline
TDE  & $\tau_{\rm lag,\ion{He}{I}}$ &$\tau_{\rm lag,H\alpha}$ & r$_{\rm lag,H\alpha}$ & $\sim$ R$_{\rm BB}$ peak  \\
& (days) & (days)& ($10^{16}$ cm) & ($10^{14}$ cm)\\
\hline
% PTF09ge & - \\
% PTF09djl & $\leq$ 2 \\
ASASSN-14li* (\ion{N}{III} )& 22.95 $\pm 4.6$  &14.97 $\pm$ 0.4& 3.88 &2.57\\
LSQ12dyw & $\leq$ 4 &45.94 $\pm$ 5.7& 12.22&17.8 \\
ASASSN-14ae* & $\leq$ 4 &32.20 $\pm$ 1.1& 8.34 &7.92\\
% iPTF15af &- \\
ASASSN-15oi* & $\leq$ 7 & 26.83 $\pm$ 1.9& 6.95&10.0\\
% iPTF16axa &$\leq$ 12 \\
iPTF16fnl (\ion{N}{III} )& 15.97 $\pm 2.3$  & 13.66 $\pm$ 2.7& 3.54&1.28\\
% AT2017eqx & $\leq$ 12 \\
AT2018zr & 26.73 $\pm 3.8$ & 42.08 $\pm$ 3.6& 10.8&14.1\\
AT2018dyb (\ion{N}{III} )& 7.52 $\pm 1.8$  &6.92 $\pm$ 1.1& 1.79 &8.12\\
% AT2018fyk & $\leq$ 7 \\
% AT2018hyz & $\leq$ 9\\
% AT2018lna & $\leq$ 34 \\
\hline
\end{tabular}
\\[-0pt]
First column: 
TDEs for which a time lag between H$\alpha$ and continuum is measurable. TDEs for which the light curve peak has not been observed are indicated with an asterisk, and in these cases the tabulated H$\alpha$ and \ion{He}{I} is a lower limit.
%If the TDE has photometric data only after the light curve peak, we indicate this with an asterisk symbol next to its name. In this cases the discovered time lag is a lower-limit.
If the TDE shows \ion{N}{III} Bowen lines we note it in a parenthesis next to their name. 
Second column: Lag between peak \ion{He}{I} luminosity and the optical light curve peak. If the line does not peak we show the upper-limit. Third column: Lag between peak H$\alpha$ luminosity and the optical light curve peak. %The H$\alpha$ peak has been determined by fitting a $3^{rd}$ order polynomial.  
Fourth column: Deduced distances extracted from equation: r$_{\rm lag,H\alpha}$ = $\tau_{\rm lag,H\alpha} ~c$ (see Sect. \ref{subsec:dtl}) where $\tau_{\rm lag,H\alpha}$ = t$_{\rm peak,H\alpha}$ - t$_{\rm peak,LC}$. Fifth column: The blackbody radius value at peak for each TDE (see Table \ref{tab:sample2}).
% \end{center}
\end{table}

\subsubsection{Other emission lines} \label{subsub:oel}
The line luminosities of H$\beta$, \ion{He}{II} 4686 \AA, \ion{He}{I} 5876 \AA\, and \ion{N}{III} 4100 \AA\, and 4640 \AA\, can be found in Fig. \ref{fig:lums_sub} of the Appendix. Similar to H$\alpha$, we observe a time lag in the line luminosities of H$\beta$ for all the TDEs of Table \ref{tab:lag} except for iPTF16fnl. For \ion{He}{I} 5876 \AA\, we also observe a delayed peak with respect to the continuum and we try to quantify this in a way similar to H$\alpha$. 
More specifically, there is clearly a lag for ASASSN-14li, AT2018dyb, iPTF16fnl and AT2018zr while for ASASSN-14ae and LSQ12dyw a peak may exist (within uncertainties) however, since we cannot be certain, we place an upper-limit. A lag is not detected for ASASSN-15oi and AT2018hyz. The \ion{He}{I} lags are tabulated in Table \ref{tab:lag} and visually presented in Fig. \ref{fig:ha_heI_lc}. 
For most events the time lag in \ion{He}{I} appears smaller than for H$\alpha$, although this difference is minimized and inverted for \ion{N}{III} Bowen TDEs. 
The behavior of \ion{He}{I} in AT2018fyk is especially interesting: it seems to show an anticorrelation with the H$\alpha$ (Fig. \ref{fig:18fyk_UVW2}).
The sample is smaller for TDEs with N III lines, but for \ion{N}{III} 4100 \AA\, a lag is observed for ASASSN-14li and AT2018dyb. All the aforementioned lags are detected compared to the same light curve peaks that we used in order to study the H$\alpha$ lag.

In the case of the Bowen TDEs, the error-bars on \ion{He}{II} and \ion{N}{III} 4640 \AA\, are very large, especially for late times or for low S/N spectra in general. This is very reasonable because by adding specific constraints (see Sect. \ref{subsub:fpac}), we ``force'' the fitting of an extra line where most of the times the peaks of the two lines are not resolved. This creates a large correlation between pairs of fit parameters (in this case the two aforementioned lines) and this is consequently mirrored in the error-bars (see Sect. \ref{subsub:cou}). Sequentially, the line ratios which contain these lines (which are presented in Sect. \ref{subsub:lr}) also have very large error-bars due to error propagation. Because of this, we have removed the error-bars of TDE AT2018dyb from every plot that contains a ratio which includes \ion{He}{II} or \ion{N}{III} 4640 \AA\, for visual purposes (the two line peaks are not resolved in this one so the fitted parameters are highly correlated).

\subsubsection{Line ratios} \label{subsub:lr}

\begin{figure}
\centering
\includegraphics[trim={0 0cm 0 0},clip,width=0.47 \textwidth]{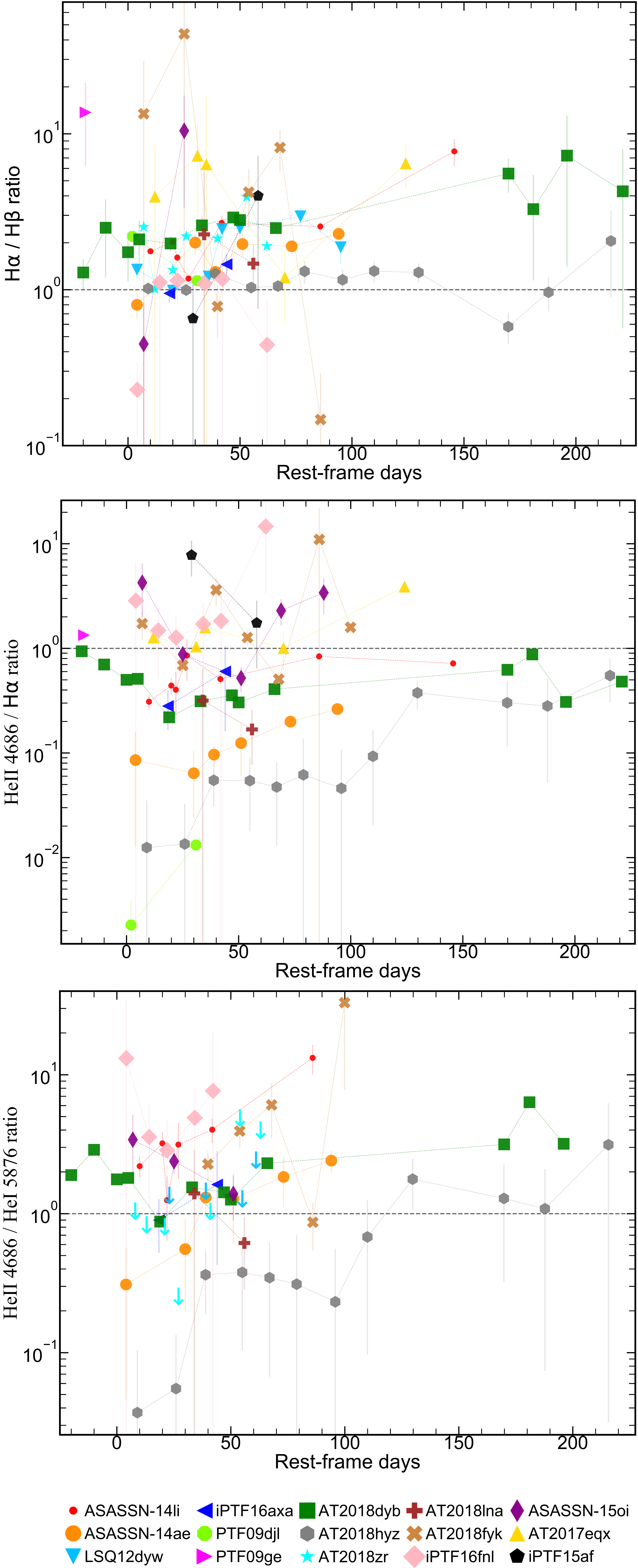}
\caption{Evolution of line luminosity ratios with time: H$\alpha$/H$\beta$ (top), \ion{He}{II}/H$\alpha$ (middle) and \ion{He}{II}/\ion{He}{I} (bottom). The vertical axes in all panels are logarithmic. The H$\alpha$/H$\beta$ ratio shows a large diversity within the sample and, occasionally, a significant time evolution for individual events. \ion{He}{II} seems to be persistent in TDEs as the \ion{He}{II}/H$\alpha$ and \ion{He}{II}/\ion{He}{I} ratios do not drop with time. AT2018zr and LSQ12dyw do not show \ion{He}{II} in their spectra hence we place upper-limits for those in the \ion{He}{II}/\ion{He}{I} ratio plot. For visual purposes, the large error bars in the \ion{He}{II} ratios of AT2018dyb have been removed (see Sect. \ref{subsub:oel}).}
\label{fig:ratios}
\end{figure}

In Fig. \ref{fig:ratios} we present the time evolution of three different line luminosity ratios; H$\alpha$/H$\beta$ (top panel), \ion{He}{II}/H$\alpha$ (middle panel) and \ion{He}{II}/\ion{He}{I} (bottom panel). The H$\alpha$/H$\beta$ ratio highly varies for TDEs going from below 1 for a few events to values as high as 10. The ones that show values higher than 4 at the early times (i.e., during the first 100 days after the TDE peaked) are either TDEs that strangely show prominent H$\alpha$ emission but weak to almost non existent H$\beta$ (AT2017eqx and AT2018fyk) or those whose Hydrogen emission is weak and suppressed relative
to Helium (PTF09ge and ASASSN-15oi) and this makes the Balmer lines harder to fit. Furthermore, some events show a flat Balmer decrement throughout their evolution while others have one that highly varies. In a typical AGN broad line region (BLR) the H$\alpha$/H$\beta$ ratio is $\sim$ 3 -- 4 which is consistent with case B recombination \citep{Osterbrock1974} and it has been a topic of discussion whether TDEs do also show this value. \citet{Short2020a} showed that AT2018hyz has a flat Balmer decrement ($\sim$ 1.5) which could imply that the Balmer emission is dominated by collisional excitation rather than photoionization. However, our sample shows a large range of values for individual events as well as for the whole sample. This may point to a variety of physical conditions responsible for the emergence of the Balmer lines that differ from TDE to TDE and occasionally change during the evolution of a single event. We caution that, as explained in Sect. \ref{subsub:cfge}, we have ignored the effect of host galaxy reddening, as we believe it to be negligible in these galaxies. We cannot exclude that some reddening might be present, which would affect the value of these ratios. Reddening, however, cannot be a dominant source of diversity accounting for the large range of H$\alpha$/H$\beta$ values found here. This  would require negative extinction for $\sim$80\% of the TDE hosts studied here (assuming case B recombination). We conclude that TDEs have a preference for relatively low values for this ratio.

The \ion{He}{II}/H$\alpha$ ratio shows a general rising trend from early to late-times for the ``well-sampled'' ($\geq$ two spectra) TDEs in our sample (see also Sect. \ref{subsub:eitobtar} where we examine this ratio as a function of the blackbody radius). It is interesting that for ASASSN-14ae, ASASSN-15oi, AT2018dyb, iPTF16fnl, this ratio initially drops and then consistently rises. For AT2017eqx the ratio shows a plateau and then rises at late-times. The ratio shows a rising trend throughout the evolution of ASASSN-14li and AT2018hyz.
The \ion{He}{II}/\ion{He}{I} ratio is presented, for the first time in the literature. It shows a general rising trend with time for the well-sampled events (see also Sect. \ref{subsub:eitobtar} where we examine this ratio as a function of the blackbody temperature). The ratio shows a rising trend throughout the evolution of ASASSN-14li, ASASSN-14e and AT2018hyz (and AT2018fyk with the exception of one epoch) while it shows an initial drop and then rise for AT2018dyb and iPTF16fnl. However the ratio drops for ASASSN-15oi (for first three epochs, after that \ion{He}{I} disappears). The general rising trend of these two luminosity ratios indicates that \ion{He}{II} is very persistent as TDEs evolve, and fades slower (or increases later) than other emission lines.

The \ion{He}{II}/\ion{N}{III} 4640 \AA\, and the \ion{N}{III} 4100/4640 \AA\, can be found in Fig. \ref{fig:HeII_NIIIs} of the Appendix. 
%(although their large uncertainties make them hard to trust). 
The \ion{He}{II}/\ion{N}{III} 4640 \AA\, seems to have a rising trend with time which is not surprising since, as discussed above, \ion{He}{II} fades slower (or increases later) compared to other TDE emission lines. This is an important ratio in order to understand how the Bowen blend evolves in TDEs but, as discussed already, the blending makes the fitting hard and introduces large uncertainties which makes it difficult to draw conclusions from. The \ion{N}{III} 4100/4640 \AA\, ratio shows values close to one (within the uncertainties).

\subsubsection{Evolution in terms of bolometric temperature and radius} \label{subsub:eitobtar}

\begin{figure}
\centering
\includegraphics[trim={0 9cm 0 0},clip,width=0.47 \textwidth]{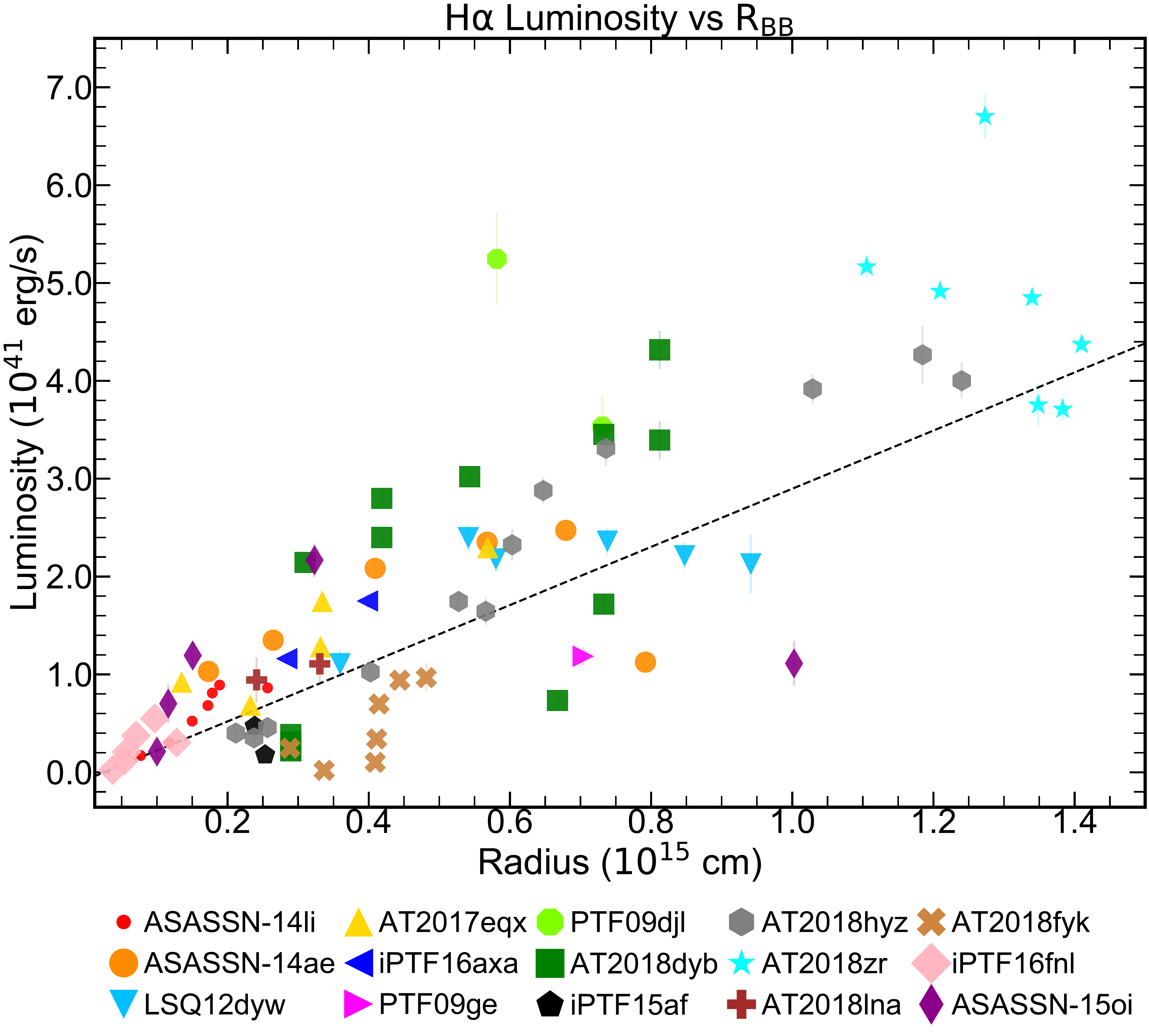}
\includegraphics[trim={0 0 0 0},clip,width=0.47 \textwidth]{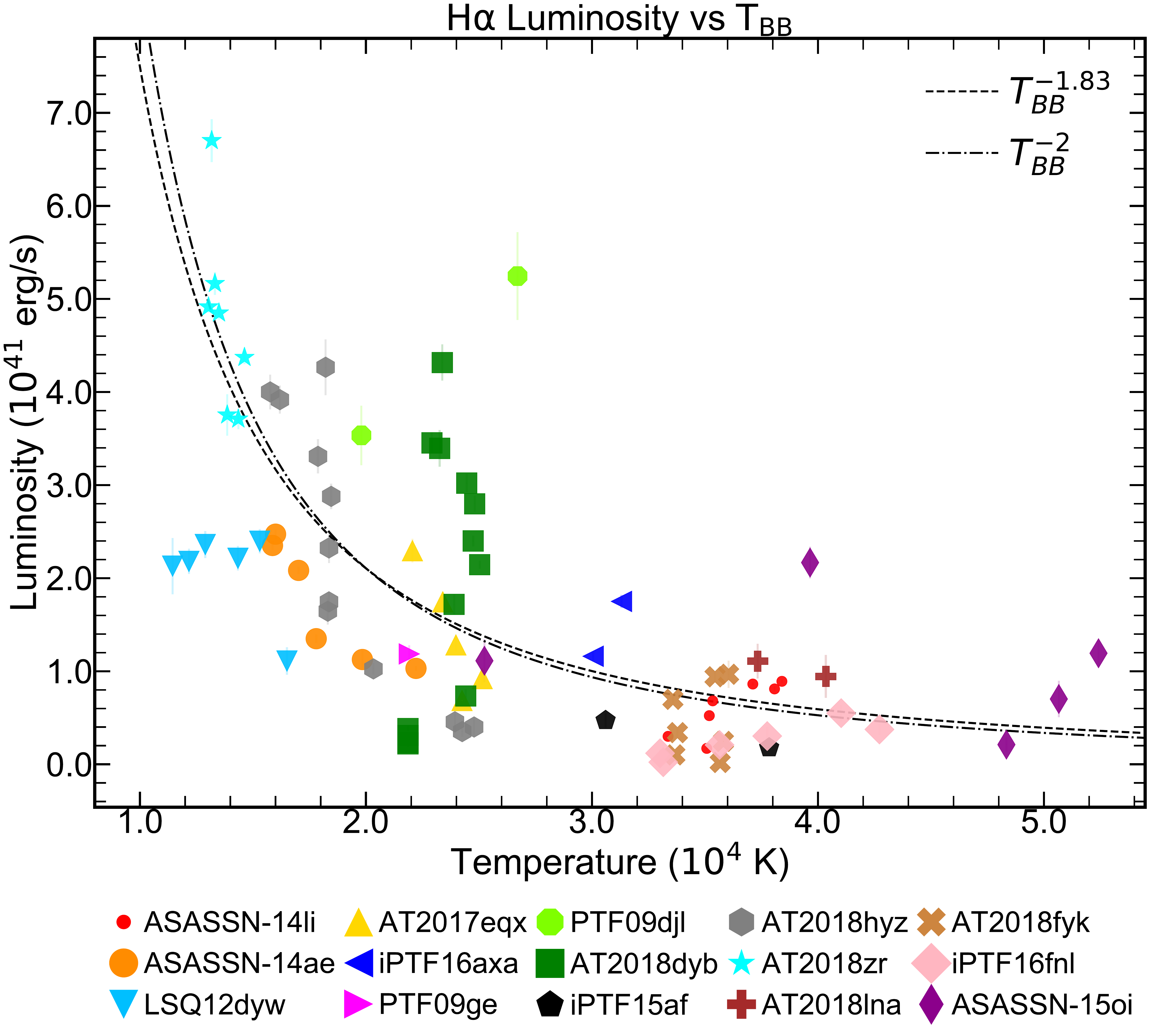}
\caption{The H$\alpha$ line luminosity as a function of R$_{\rm BB}$ (top panel) and T$_{\rm BB}$ (bottom panel). The dashed lines are the best fits for the data; linear for R$_{\rm BB}$ and inverse power-law for T$_{\rm BB}$. The dotted-dashed line in the bottom panel is an inverse power-law fit with the exponent fixed at 2. H$\alpha$ luminosity seems to follow a linear relationship with the blackbody radius (L$_{\rm H\alpha} \propto$ R$_{\rm BB}$) and potentially an inverse power-law relationship with blackbody temperature (L$_{\rm H\alpha} \propto$ T$_{\rm BB}^{-\beta}$).}
\label{fig:Ha_T_R}
\end{figure}

It is common in the literature to use a blackbody model in order to describe the optical photosphere of TDEs, yielding characteristic temperatures and radii (e.g., \citealt{vanvelzen2021,Hinkle2021a}). Here we investigate the evolution of spectroscopic properties in relation to the blackbody radius (R$_{\rm BB}$) and temperature (T$_{\rm BB}$). 
These quantities evolution with time were retrieved from the literature (see Table \ref{tab:sample2} for details) and we linearly interpolated between two data points of the time evolution curves, in order to get the T$_{\rm BB}$ and R$_{\rm BB}$ for the times that match our spectra. In Fig. \ref{fig:Ha_T_R} we present the evolution of the H$\alpha$ luminosity in relation to the R$_{\rm BB}$ (top panel) and T$_{\rm BB}$ (bottom panel). It is clear that as R$_{\rm BB}$ increases (both for individual events as well as for the statistical sample) the H$\alpha$ luminosity is rising as well, seemingly following a linear trend (L$_{\rm H\alpha} \propto$ R$_{\rm BB}$). Furthermore the luminosity of H$\alpha$ seems to have an inverse power-law relation with T$_{\rm BB}$ (L$_{\rm H\alpha} \propto$ T$_{\rm BB}^{-\beta}$).

\begin{figure*}
        \centering
        \begin{subfigure}[b]{1\textwidth}
            \centering
            \includegraphics[trim={0 0cm 0 0},clip,width=\textwidth]{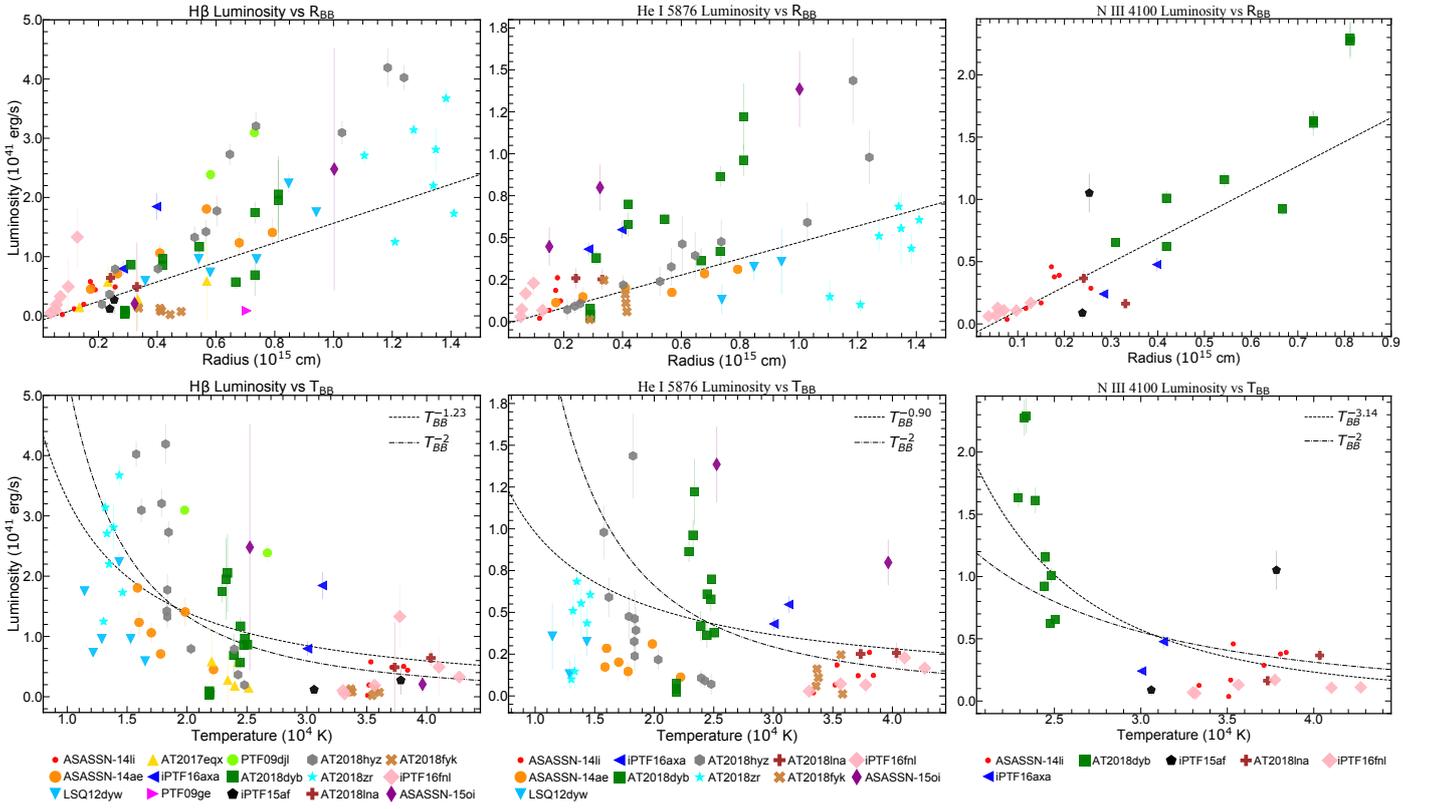}
            % \caption[Network2]%
            % {{\small Network 1}}    
            % \label{fig:lums_T_R_subplot}
        \end{subfigure}
        \vskip\baselineskip
        \caption
        {\small H$\beta$ (left panels), \ion{He}{I} 5876 \AA\, (middle panels) and \ion{N}{III} 4100 \AA\, (right panels) line luminosities as a function of R$_{\rm BB}$ (upper panels) and T$_{\rm BB}$ (lower panels). The dashed lines are the best fits for the data; linear for R$_{\rm BB}$ and inverse power-law for T$_{\rm BB}$. The dotted-dashed line in the bottom panels is an inverse power-law fit with the exponent fixed at 2.} 
        \label{fig:lums_T_R_subplot}
    \end{figure*}

In order to verify this result, in Fig. \ref{fig:lums_T_R_subplot} we plot the luminosities of H$\beta$, \ion{He}{I} 5876 \AA\, and \ion{N}{III} 4100 \AA\, (these lines were chosen as they are not blended and hence, are more trustworthy to measure) against the R$_{\rm BB}$ and T$_{\rm BB}$ of their respective TDEs. Interestingly, all these lines follow similar relations (i.e., L$_{\rm line} \propto$ R$_{\rm BB}$ and L$_{\rm line} \propto$ T$_{\rm BB}^{-\beta}$). In order to quantify these trends, we fit the line luminosities against R$_{\rm BB}$ with a linear regression and the line luminosities against T$_{\rm BB}$ with two different power-laws; one with the exponent fixed at $-$2 and one for which the exponent is free. The best fit results for the free exponent fit are: $\beta$=1.83 $\pm$ 0.14 for H$\alpha$, $\beta$=1.23 $\pm$ 0.18 for H$\beta$, $\beta$=0.9 $\pm$ 0.3 for \ion{He}{I} and $\beta$=3.14 $\pm$ 0.52 for \ion{N}{III}.

We ran statistical tests in order to look for an ordinal association between the aforementioned quantities. We used two rank correlation coefficient tests which provide a measure of the correspondence between two rankings, the Kendall's tau test and the Spearman's rho test. Both tests assess how well the relationship between two variables can be described using a monotonic function, in other words, if the resulting coefficients are 1 (or $-$1), X and Y are perfectly monotonically dependent random variables, while a coefficient of 0 would indicate that X and Y are monotonically uncorrelated random variables. Spearman's rho is considered more sensitive to outliers compared to Kendall's tau. %\citep{Spearman1904,Kendall1938}. 
In order to further investigate the relationship between the line luminosities and R$_{\rm BB}$ and T$_{\rm BB}$ we also ran a Pearson's correlation coefficient test which measures the linear correlation between two sets of data. Again, a result of 1 (or $-$1) indicates that X and Y are perfectly linearly dependent random variables while a result of 0 would indicate that they are linearly uncorrelated. Pearson's r is sensitive to outliers, more than the aforementioned non-parametric tests. We choose a significance of $\alpha$=0.05 (i.e., 95\% significance) for rejecting the null hypothesis (i.e., rejecting that there is a correlation). The results can be found in Table \ref{tab:ktsp}. We provide the scores of the tests and their associated p-values. 
We note here that statistical significance assesses whether a correlation has arisen (or not) by chance. By squaring the test scores, one can probe how ``strong'' a correlation is.
%The practical importance of a correlation (i.e., how much of the variation in one of the two variables is associated with variation in the other) can be assessed by squaring the score of a test (coefficient of determination). 
There is a significant (p<0.001) monotonic and linear correlation between the luminosities of all studied lines and R$_{\rm BB}$. There is a significant monotonic and linear correlation between the luminosities of H$\alpha$ and H$\beta$ and T$_{\rm BB}$. There is also a weak correlation for \ion{N}{III} 4100 \AA\, but not for \ion{He}{I} 5876 \AA\, which has a p-value lower than 0.05 only for the Spearman's rho test. Since the data showed a linear correlation with temperature in most of the cases, we also tried a linear regression fit but the ${\rm \chi_{\rm \nu}}^{2}$ was always indicating a worse fit than the power-laws.

\begin{table}
\renewcommand{\arraystretch}{1.4}
\setlength\tabcolsep{0.04cm}
\fontsize{9}{11}\selectfont
% \begin{center}
\caption{Statistical test scores and p-values for different emission line luminosities and luminosity ratios as a function of R$_{\rm BB}$ and T$_{\rm BB}$.}\label{tab:ktsp}
% \begin{center}
\begin{tabular}{l | c c c }
\hline
 & Kendall's & Spearman's & Pearson's \\
 & $\tau$ & $\rho$ & r\\
\hline
R$_{\rm BB}$ & & & \\
\hline
H$\alpha$ (N=83)&0.61 (<0.001) &0.81 (<0.001)&0.83 (<0.001) \\
H$\beta$ (N=79)&0.53 (<0.001) &0.72 (<0.001)&0.80 (<0.001) \\
\ion{He}{I} 5876 \AA\, (N=63)&0.44 (<0.001)&0.63 (<0.001)&0.56 (<0.001) \\
\ion{N}{III} 4100 \AA\, (N=28)&0.71 (<0.001)&0.88 (<0.001)&0.92 (<0.001) \\
\ion{He}{II}/ H$\alpha$ (N=70)&$-$0.36 (<0.001)&$-$0.50 (<0.001)&$-$0.25 (0.033)\\
\noalign{\global\arrayrulewidth=0.7mm}\hline
\noalign{\global\arrayrulewidth=0.4pt}
\hline
T$_{\rm BB}$ & & & \\
\hline
H$\alpha$ (N=83)&$-$0.45 (<0.001)&$-$0.65 (<0.001)& $-$0.61 (<0.001) \\
H$\beta$ (N=79)&$-$0.38 (<0.001) &$-$0.59 (<0.001)&$-$0.57 (<0.001) \\
\ion{He}{I} 5876 \AA\, (N=63)&$-$0.16 (0.070\xmark) &$-$0.26 (0.043)&$-$0.23 (0.066\xmark) \\
\ion{N}{III} 4100 \AA\, (N=28)&$-$0.41 (0.002)&$-$0.57 (0.002)&$-$0.72 (<0.001)\\
\ion{He}{II}/ \ion{He}{I} (N=53)&0.35 (<0.001)&0.51 (<0.001)&0.30 (0.027)\\
\noalign{\global\arrayrulewidth=0.7mm}\hline
\noalign{\global\arrayrulewidth=0.4pt}
\end{tabular}
\\[-0pt]
Kendall's Tau, Spearman's Rho and Pearson's r scores and equivalent p-values in the parentheses. The results are significant for p<0.05. If the result is not significant, the p-value is accompanied by an x mark. The number of data points used for each line are provided in the parentheses next to the line name.
% \end{center}
\end{table}

\begin{figure}
\centering
\includegraphics[trim={0 0cm 0 0},clip,width=0.47 \textwidth]{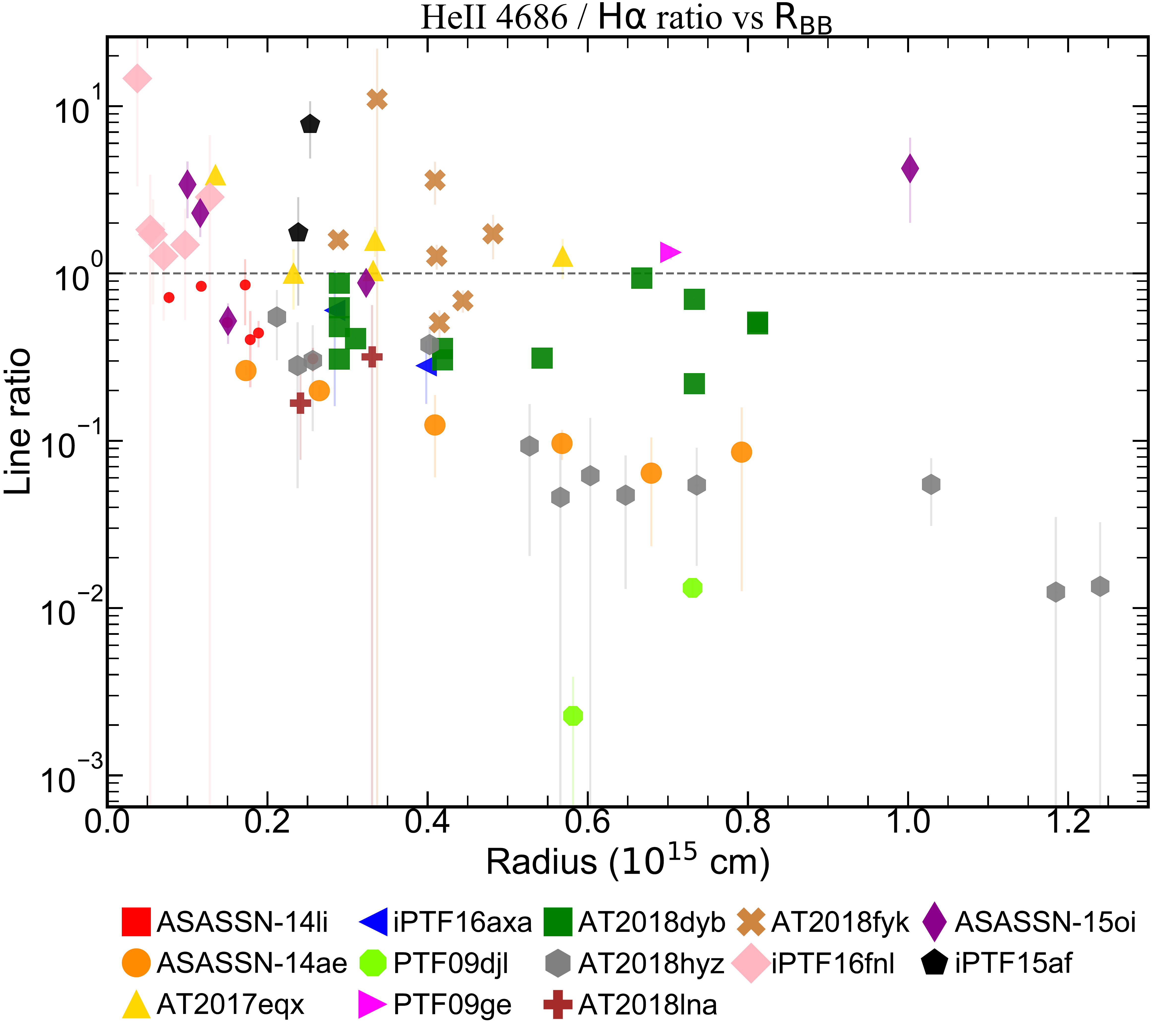}
\caption{\ion{He}{II}/H$\alpha$ luminosity ratio against R$_{\rm BB}$. The ratio increases for most TDEs as the R$_{\rm BB}$ decreases. For visual purposes, the large error bars in the \ion{He}{II} ratios of AT2018dyb have been removed (see Sect. \ref{subsub:oel}).}
\label{fig:HeII_Ha_R}
\end{figure}

\begin{figure}
\centering
\includegraphics[trim={0 0cm 0 0},clip,width=0.47 \textwidth]{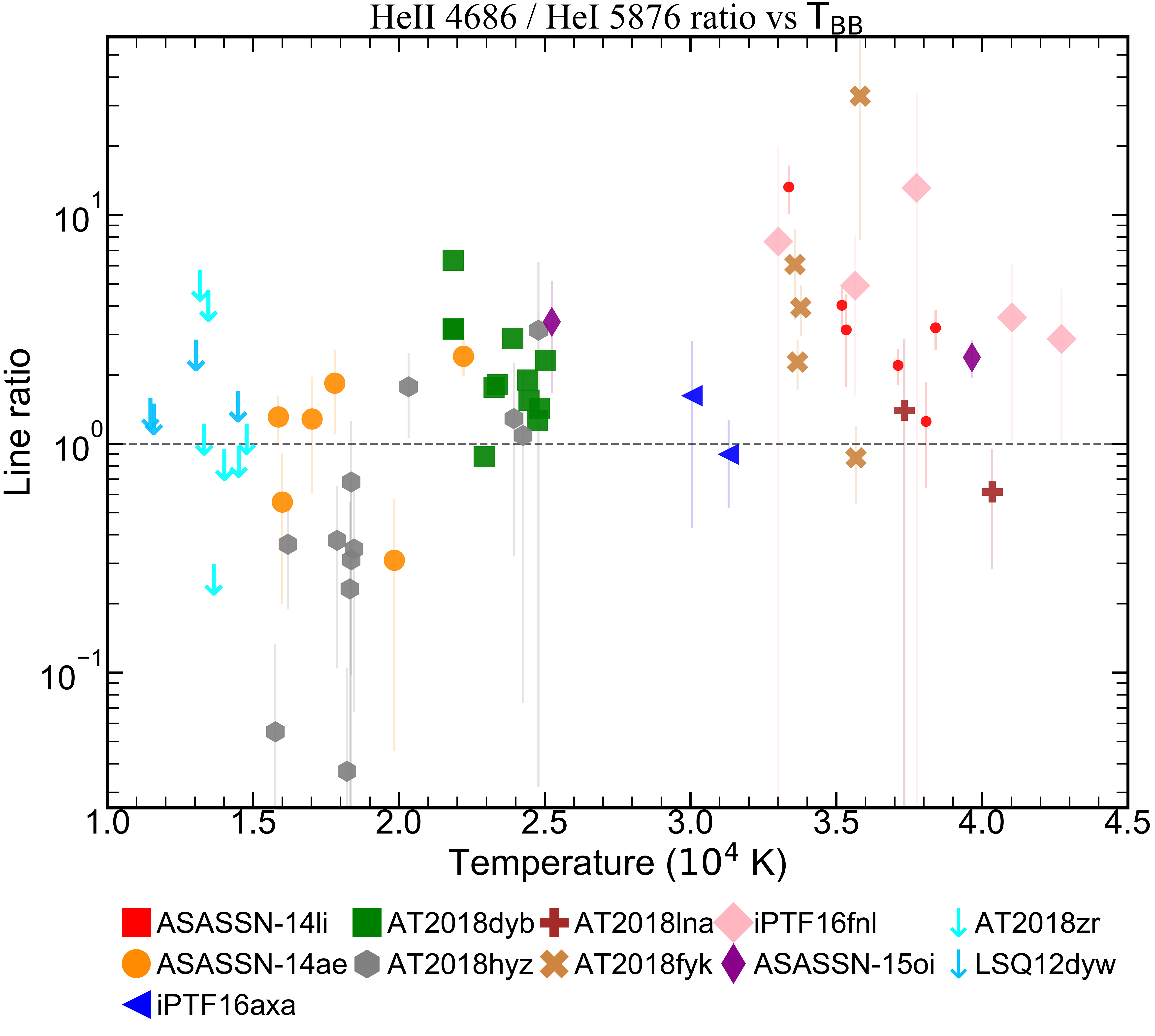}
\caption{\ion{He}{II}/\ion{He}{I} luminosity ratio against T$_{\rm BB}$. The ratio reaches values smaller than one for those TDEs that have low temperatures. In addition, TDEs AT2018zr and LSQ12dyw, which have the lowest T$_{\rm BB}$ in our sample, do not show \ion{He}{II}, only \ion{He}{I} hence we place upper-limits for those. An indication that probably He (in the line emitting region) is not yet ionized for such photospheric temperatures (~$\leq$ 20\,000 K).}
\label{fig:HeII_HeI_T}
\end{figure}

We visually examined all possible combinations of line ratios as a function of R$_{\rm BB}$ and T$_{\rm BB}$. We present here the two results that we consider most noteworthy as they have important physical implications; in Fig. \ref{fig:HeII_Ha_R} we present the \ion{He}{II}/H$\alpha$ luminosity ratio as a function of R$_{\rm BB}$ which shows a general rising trend for our sample as the R$_{\rm BB}$ becomes smaller. In Fig. \ref{fig:HeII_HeI_T} we present the \ion{He}{II}/\ion{He}{I} luminosity ratio as a function of T$_{\rm BB}$. We find that this ratio has values smaller than one for those TDEs that have low temperatures (i.e., AT2018hyz and ASASSN-14ae with temperatures $\leq$ 20\,000 K).

We ran the same correlation coefficient tests in order to look for statistically significant correlations in those two graphs and the results are presented in Table \ref{tab:ktsp} as well. Although the scores do not suggest a very strong correlation, all the results are statistically significant for monotonicity (p<0.001) and linearity (0.02<p<0.04).

\subsection{Line widths} \label{sub:lw}

\subsubsection{H$\alpha$ FWHM time evolution} \label{subsub:hafe}
\begin{figure}
\centering
\includegraphics[trim={0 0cm 0 0},clip,width=0.47 \textwidth]{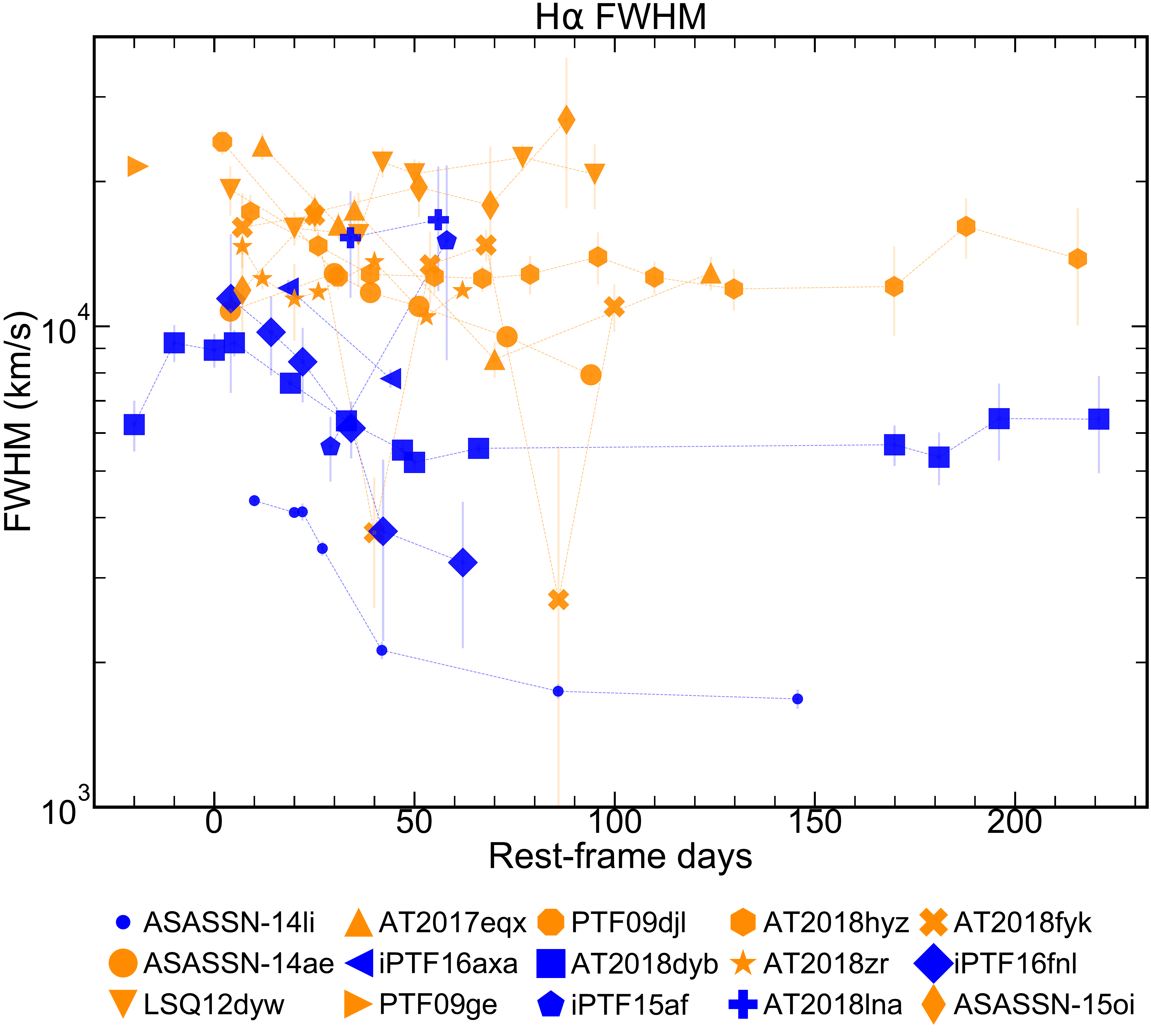}
\caption{Evolution of the H$\alpha$ FWHM with time. The graph is color coded for Bowen \ion{N}{III} TDEs (blue) and not Bowen (orange). The former seem to consistently have broader line widths than the latter.
}\label{fig:Ha_FWHMs_BCC}
\end{figure}

In Fig. \ref{fig:Ha_FWHMs_BCC} we present the evolution of the H$\alpha$ FWHM with time for all the TDEs of our sample. The FWHM slowly drop with time but remain relatively broad even at late times (several months after peak), in contrast with reverberation mapped AGNs where a decrease in luminosity is accompanied by an increase in line widths (e.g., \citealt{Peterson2004, Denney2008}). This has already been pointed out in studies of individual events (e.g., \citealt{Holoien2016}) and here we further strengthen this conclusion using a sample of TDEs. The Bowen \ion{N}{III} TDEs are found to have systematically lower line widths than the rest of the sample. Although this may be partly due to a selection bias, it is unlikely that this is the sole explanation (see Sect. \ref{subsec:dbsc}).
Plots for the FWHM evolution of other emission lines can be found in Fig. \ref{fig:fwhms_sub} of the Appendix.

\subsubsection{Dependencies of the H$\alpha$ FWHM}
\label{subsub:habhlp}

\begin{figure*}
        \centering
        \begin{subfigure}[b]{1\textwidth}
            \centering
            \includegraphics[trim={0 0cm 0 0},clip,width=\textwidth]{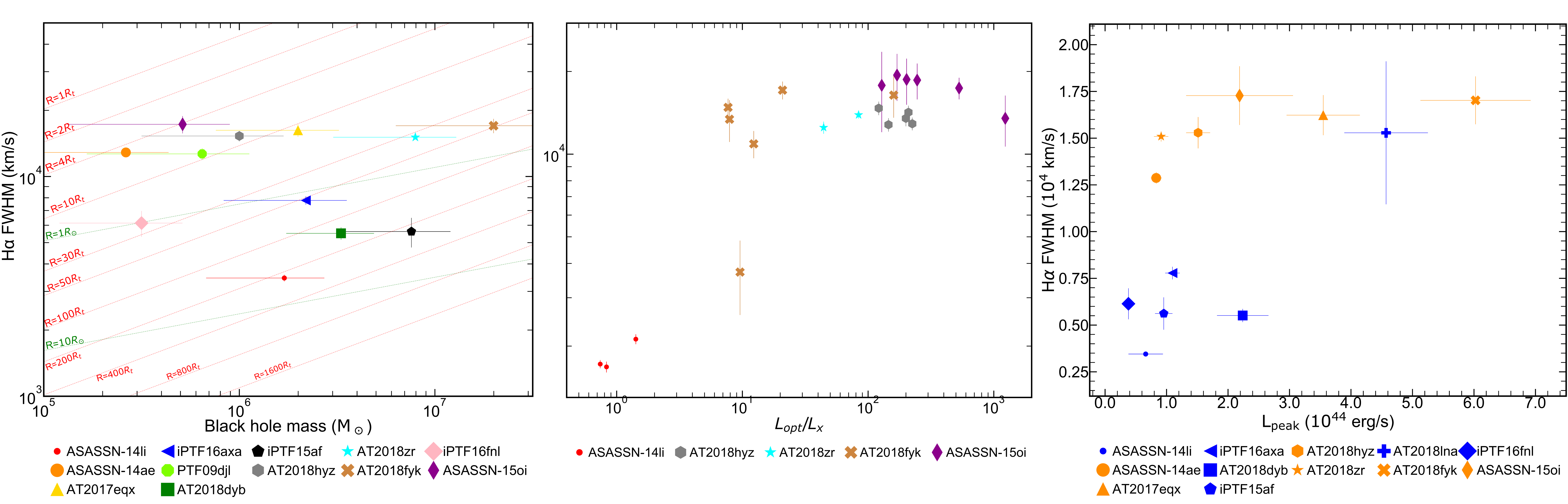}
            % \caption[Network2]%
            % {{\small Network 1}}    
        \end{subfigure}
        % \vskip\baselineskip
        \caption{Left panel: FWHM of H$\alpha$ around 30 days after peak/discovery against the black hole mass of each respective TDE. The dashed red lines represent the expected keplerian velocity correlations for bound material at different radii (assuming a sun-like star). The solid green lines are the velocities for the outflowing material assuming R$_{\rm p}$ = R$_{\rm t}$ (pericenter and tidal radii respectively) for a sun-like star (1 R$_{\rm *}$) or red giant (10 R$_{\rm *}$) \citep{Arcavi2014}. No evidence for correlation is found between the line widths and the BH masses. Middle panel: FWHM of H$\alpha$ against the L$_{\rm opt}$/L$_{\rm X}$ of X-ray TDEs (taken from \citealt{Wevers2019}) at the epochs for which we have spectra. The only TDE that exhibits strong X-ray emission during these epochs is ASASSN-14li and it is the one with the lowest FWHM. Right panel: FWHM of H$\alpha$ around 30 days after peak/discovery (same FWHM as left panel) against the peak bolometric optical/UV luminosities of each TDE taken from \citet{Hinkle2020}. The graph is color coded for Bowen \ion{N}{III} TDEs (blue) and not Bowen (orange). Low FWHM TDEs (\ion{N}{III} Bowen TDEs) seem to have low L$_{\rm peak}$ values.} 
        \label{fig:Ha_FWHM_MBH_Lp_Lx}
    \end{figure*}

\citet{Arcavi2014} examined the dependency of the line width of TDEs to their BH mass and looked for correlations based on two different scenarios; i) the velocities of the lines are attributed to circulation of bound material around the BH in keplerian orbits or ii) attributed to outflowing material. They found that there is no apparent correlation with either of these scenarios. We are now able to check their results with a larger sample. 
In the left panel of Fig. \ref{fig:Ha_FWHM_MBH_Lp_Lx}, %, inspired by Fig. 17 of \citet{Arcavi2014}, 
we plot the FWHM of H$\alpha$ (at the closest available epoch to +30 days after peak/discovery in order to include as many TDEs as possible) against the black hole mass of each respective TDE (taken from \citealt{Wevers2019a}). The dashed red lines represent the expected keplerian velocity correlations for bound material at different radii (assuming a sun-like star). The solid green lines are the \citet{Strubbe2009} velocities for outflowing material assuming R$_{\rm p}$ = R$_{\rm t}$ for a sun-like star (1 R$_{\rm *}$) or red giant (10 R$_{\rm *}$) \citep[see][Eqs. 2 \& 4]{Arcavi2014}. Consistent with previous studies, we do not find evidence for any correlation between the line widths and SMBH masses. In the middle panel, we plot the FWHM of H$\alpha$ against the L$_{\rm opt}$/L$_{\rm X}$ of X-ray TDEs (taken from \citet{Wevers2019a} who calculated BH masses using the M -- $\sigma$ relation by measuring bulge velocity dispersions using absorption lines) at the epochs for which we have available spectra. ASASSN-14li is the only TDE with high L$_{\rm opt}$/L$_{\rm X}$ ($<$2) and shows much lower velocities than the rest. Interestingly, AT2018fyk has one epoch where the FWHM significantly drops compared to the rest and this is when its optical/UV light curves showed a dip while its X-ray light curve was rising \citep{Wevers2019}.
In the right panel, we plot the same FWHM (around 30 days after peak/discovery) against the peak bolometric optical/UV luminosities of each TDE taken from \citet{Hinkle2021a}. We see that TDEs with low FWHM (which are the \ion{N}{III} Bowen TDEs except for AT2018lna) do not have high peak luminosities while the rest of the TDEs show a wide range of values. We ran the same statistical correlation tests as in Sect. \ref{subsub:eitobtar} in order to look for monotonicity and linearity. All results are statistically significant (p-values lower than 0.05) with the following scores and p-values: Kendall's tau 0.52 with p=0.021, Spearman's rho 0.66 with p=0.020 and Pearson's r 0.64 with p=0.025.

\subsection{Velocity offsets} \label{sub:loff}

Velocity offsets from the rest wavelength of a spectral line can probe kinematics of the line forming region and can determine if it approaches (blueshift) or recedes (redshift) from the observer. Blueshifts in TDE velocities, seen in certain spectral lines, can possibly determine whether there is an outflow of material along our line of sight -- this could arise from either a disk wind or from unbound debris streams \citep{Nicholl2019,Hung2019,Nicholl2020}. 
\begin{figure}
\centering
\includegraphics[trim={0 0cm 0 0},clip,width=0.46 \textwidth]{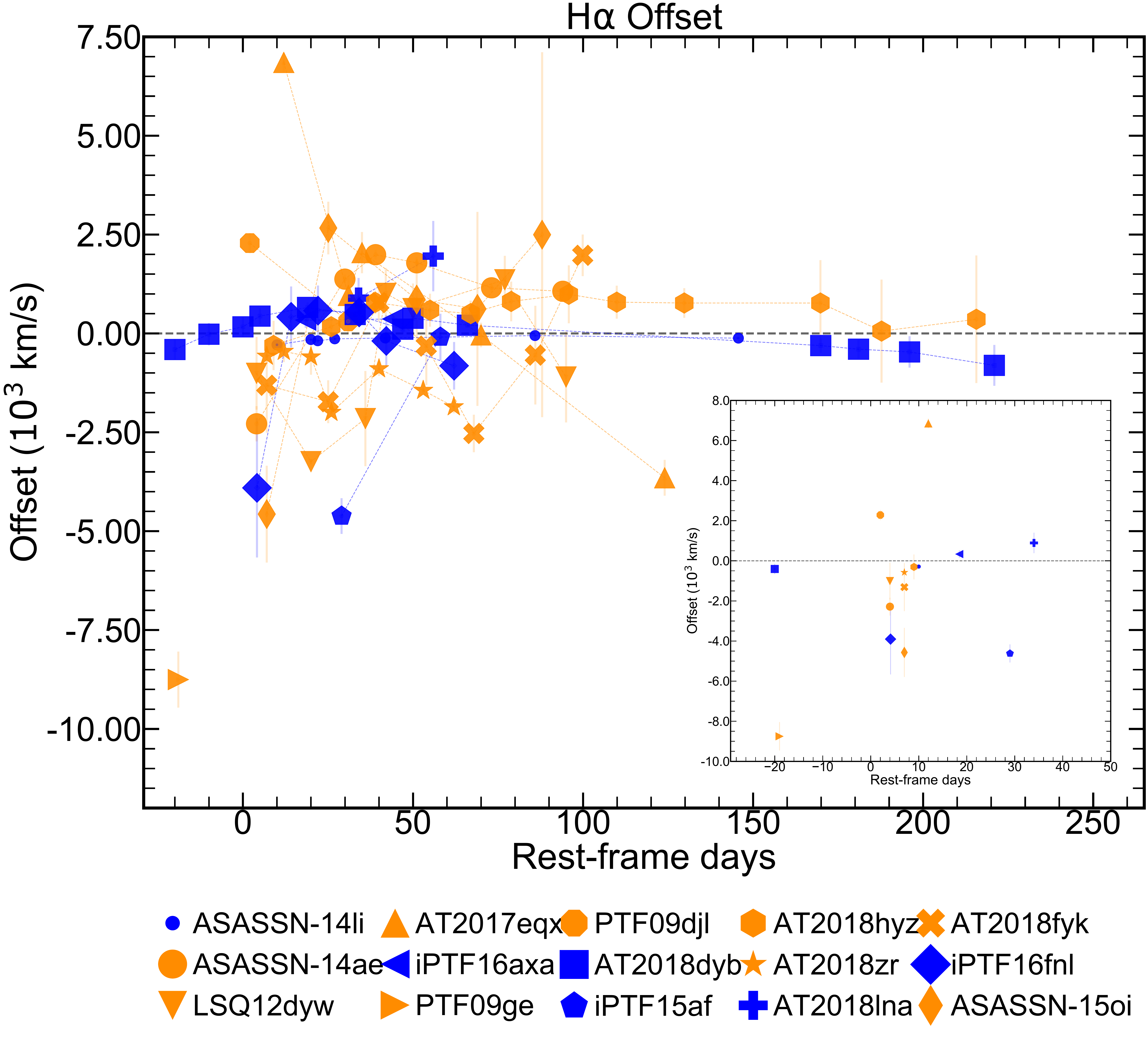}
\caption{Evolution of the H$\alpha$ velocity offset with time. The graph is color coded for Bowen \ion{N}{III} TDEs (blue) and not Bowen (orange). The embedded panel contains the H$\alpha$ offset of the first available epoch of each TDE. The H$\alpha$ line at the first epoch for the majority of the TDEs is blueshifted. The \ion{N}{III} Bowen TDEs show lower offsets compared to the rest of the TDEs which show more extreme values.}
\label{fig:ha_off}
\end{figure}

Figure \ref{fig:ha_off} shows the evolution of the H$\alpha$ line offset with time for the TDEs in our sample. The Fig. is color coded with blue being the Bowen \ion{N}{III} TDEs and orange those that do not show any signs of \ion{N}{III}. An interesting fact here is that for the majority of our sample (11 out of 15), the H$\alpha$ line is blueshifted for the first spectrum of each respective TDE with blueshifts varying from $\sim$ $-$200 to $\sim$ $-$8700 km~s$^{-1}$
%(and the extreme case of PTF09ge with $-$8700 km~s$^{-1}$).
A simple binomial test yields a $\sim$ 4$\%$ chance that this would occur by chance. Another interesting fact is that for eight TDEs, the lines change between being blueshifted and redshifted from their central wavelength at least once throughout the evolution of the event. Velocity offsets for other emission lines %(color coded for Bowen - not Bowen) 
can be found in Fig. \ref{fig:offs_sub} of the Appendix.

\begin{figure}
\centering
\includegraphics[trim={0 0cm 0 0},clip,width=0.47 \textwidth]{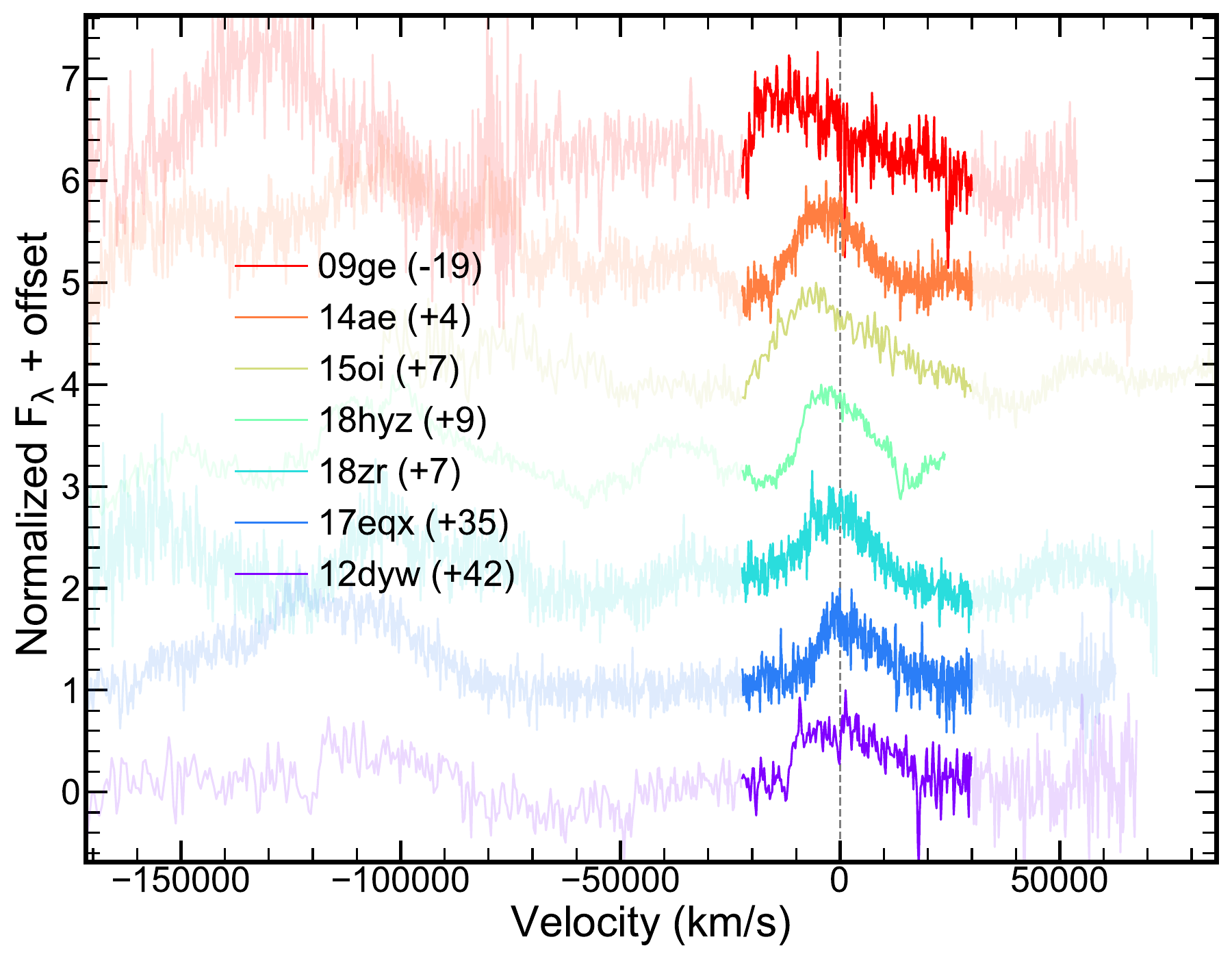}
\caption{H$\alpha$ line profiles (and \ion{He}{II} for ASASSN-15oi) for a single epoch (denoted in the parentheses is the phase of the spectrum) seen in seven TDEs of our sample. These profiles show a blueshifted peak or/and a broad and smooth red shoulder. \citet{Roth2017} predicted that profiles with such characteristics should arise due to the effects of electron scattering above a hot photosphere in an outflowing gas.}
\label{fig:roth_out3}
\end{figure}

Several TDEs have shown early time blueshifted emission lines \citep{Arcavi2014,Holoien2015,Nicholl2020} or absorption lines \citep{Hung2019} attributed to outflows. The TDE AT2019qiz \citep{Nicholl2020} showed some complicated H$\alpha$ and \ion{He}{II} profiles which exhibited a peak blueshifted from the rest wavelength as well as a broad and smooth red shoulder. \citet{Roth2017} calculated line profiles including the effects of electron scattering above a hot photosphere in an outflowing gas and their model resulted in similar profiles (namely, blueshifted peaks and broad red shoulders). They also suggest that as the outflow expands and the photospheric radius increases, these ``\textit{outflow profiles}'' will move from blueshifted to lower velocity offsets; closer to the rest wavelength. In our study, we encountered several such profiles that exhibit one or both of these characteristics (blueshifted peaks or/and broad red shoulders) and we plot them in Fig. \ref{fig:roth_out3}. We find that when this profile is encountered before or around peak it is indeed blueshifted while for the two TDEs that exhibit it later, it is more centered on the rest wavelength as predicted by \citealt{Roth2017}. This highlights how crucial it is to characterize and subsequently observe more TDEs prior to peak. Spectroscopy during the rise, peak and fall of the light curve as well as multi-wavelength coverage of TDEs, is essential in order to further investigate the nature of these emerging outflows.

\section{Discussion} \label{sec:discussion}

TDEs provide a unique tool for studying physics of accretion onto supermassive black holes. However, crucial details such as the geometry and the emission mechanism
are not fully understood, which has driven the development of many theoretical scenarios. These involve the reprocessing of X-rays of a promptly formed accretion disk into optical/UV wavelengths, either from a) an optically thick wind \citep{Dai2018} or b) from a CIO \citep{Lu2020} else c) that the optical emission is produced by collisions between the debris streams \citep{Piran2015,Jiang2016}. Consequently, the question of where and how the optical emission lines are formed is still open. In this section, we try to put the results of our spectroscopic study in a wider context in order to aid our understanding of these processes. 
We thus focus on key spectroscopic properties of TDEs and discuss whether they are compatible with current models and/or how they can help inform the development of future models.
In particular, we discuss the observed time lag between the continuum and emission lines, as well as the dependency of the emission line properties on the blackbody radius and temperature. Finally, we present a phenomenological discussion on the optical spectra of TDEs, we attempt a subgrouping based on their spectroscopic features and we examine these subgroups under the context of their X-ray emission and potential viewing angle effects.

\subsection{A time lag between the continuum and the spectral lines} \label{subsec:dtl}

As shown in Sect. \ref{subsub:hale}, there is a measurable time lag between the peaks of the light curves and the luminosities of H$\alpha$ for at least seven TDEs of our sample spanning from $\sim7$ to $\sim45$ days. A lag of H$\alpha$ with respect to the continuum has been found for AT2018dyb \citep{Holoien2020} and AT2019azh \citep{Hinkle2021}, %events more recent than those in our sample .
but we have shown in this work that this may be a common property of TDEs, as the presence a large delay can safely be excluded for only two events in a large sample (Fig. \ref{fig:ha_heI_lc}). In addition, Fig. \ref{fig:18fyk_UVW2} suggests that the line luminosity in TDEs seems to follow closely the variations of the broad band light curves, where the line luminosity of AT2018fyk clearly follows the variations of the light curves. Motivated by reverberation mapping studies which investigated the structure and radial extent of the broad line regions of AGNs \citep{Peterson1993}, it is possible to interpret the aforementioned lags as light echos, in other words responses to the continuum pulse. 
We consider a very simple model, a spherical shell of gas orbiting around the black hole at a distance $r$ (see Figs. 1 and 2 of \citealt{Peterson1993}). 
The travel time for continuum photons to reach the spherical shell is therefore $r/c$. In addition, assuming that the gas gets ionized and recombines instantaneously (emitting a line-photon toward the observer), it can be shown that the observed lag between the continuum and the line response is $\tau = (1 + \mathrm{cos}\theta )\cdot r/c$ \citep{Peterson1993} where $\theta$ is the angle in polar coordinates with respect to the observer, with which a photon leaves the central source. The most rapid response to the continuum source (i.e., $\tau=0$) would come from photons with an angle of $\theta=180^{\circ}$ that is directly toward the observer. The most delayed response would come from photons with an angle of $\theta=0^{\circ}$ that is heading to the far side of the shell; directly away from the observer.
The mean lag is shown to be $<\tau>=r/c$. 
It can easily be shown with simple geometrical calculations that a circular ring of gas orbiting around the BH would also have a $<\tau> \sim r/c$ no matter the inclination angle the observer is looking at the system.
It is also possible to imagine more complicated geometries but order of magnitude arguments will always result in $<\tau> \sim r/c$. The measuring of those lags is done in the most simple way; we measure the time difference of the H$\alpha$ peak from the optical light curve peak.
This is the equivalent of using Equation 9 of \citet{Peterson1993}, where the transfer function $\Psi(\tau)$ is a dirac delta function at $r/c$. However this is a good approximation for the scope of this work and the provided radii are indicative of the distances under discussion.
The distances that are therefore equivalent to the observed time lags are on the order of 
1.8$\,\times\,10^{16}$ -- 1.2$\,\times\,10^{17}$cm (7 -- 45 light days) (Table \ref{tab:lag}).
What is especially puzzling is that these values are one to two orders of magnitudes larger than the estimated blackbody radii for the same TDEs, where the continuum supposedly originates (R$_{\rm BB}$ = 1.3$\,\times\,10^{14}$ -- 1.8$\,\times\,10^{15}$ or 0.05 -- 0.7 light days; Table \ref{tab:lag}). It has been argued that the line emitting region in TDEs must be further out than the optical continuum photosphere \citep{Roth2016} however the distances inferred from the lags are much larger than R$_{\rm BB}$ at peak ($\sim$ ten to hundred times) and seem improbable. An interpretation of these results is crucial in order to probe the nature and the geometrical structure of the line emitting region in TDEs. 

% \begin{figure}
% \centering
% \includegraphics[width=0.45 \textwidth]{t_lag_vs_R_BB_Bowen_CC.pdf}
% \caption{The H$\alpha$ (filled markers) and \ion{He}{I} 5876 \AA\, (empty markers) luminosity time lag (see Table \ref{tab:lag}) against the blackbody radius at peak. The graph is color coded for Bowen \ion{N}{III} TDEs (blue) and not Bowen (orange). The dashed line is a linear regression fit for the H$\alpha$ lag only. \ion{N}{III} Bowen TDEs show lower H$\alpha$ lag values than the rest of the TDEs in the sample.}
% \label{fig:tlagvsRBB}
% \end{figure}

\begin{figure}
\centering
\includegraphics[width=0.45 \textwidth]{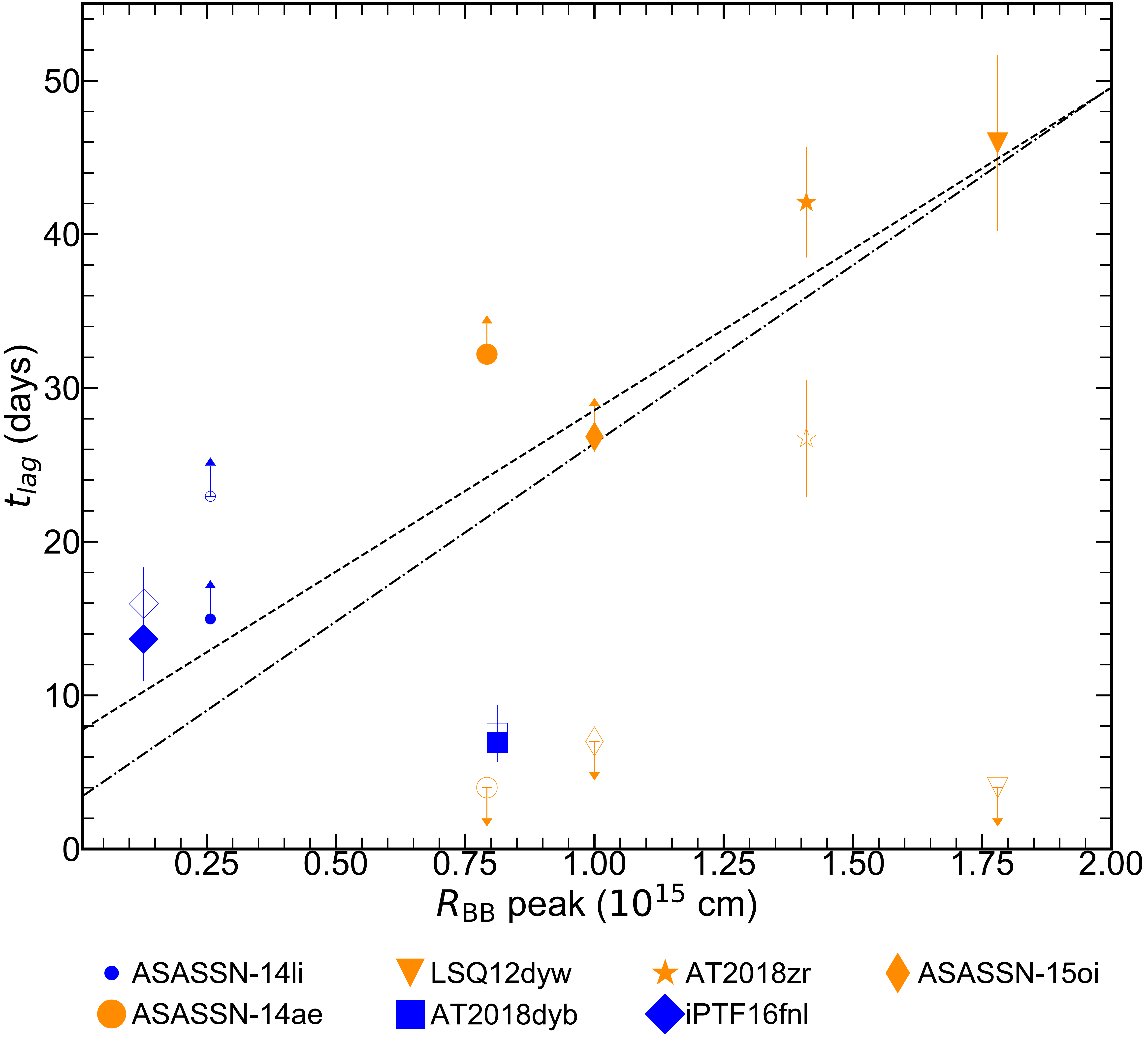}
\caption{The H$\alpha$ (filled markers) and \ion{He}{I} 5876 \AA\, (empty markers) luminosity time lag (see Table \ref{tab:lag}) against the blackbody radius at peak. The graph is color coded for Bowen \ion{N}{III} TDEs (blue) and not Bowen (orange). The dashed line is a linear regression fit for the H$\alpha$ lag only while the dot-dashed is the same fit without fitting the three lower-limits. \ion{N}{III} Bowen TDEs show lower H$\alpha$ lag values than the rest of the TDEs in the sample.}
\label{fig:tlagvsRBB}
\end{figure}

Outflows could be an option for the origin and nature of the material responsible for the aforementioned time lags. Outflows in TDEs have been predicted by several theoretical models \citep{Strubbe2009,Metzger2016,Roth2016,Roth2017,Dai2018} and have been confirmed by observations in some TDEs \citep[e.g.,][]{Alexander2016,Hung2019,Nicholl2020}. \citet{Nicholl2020} discovered an outflow for TDE AT2019qiz, and based on photometric, spectroscopic and radio data, propose a possible geometry for the evolution of the event. Outflows initially expand together with the blackbody photosphere until the latter stops expanding and starts to contract while the former keeps expanding. The authors suggest that the line emitting region lies somewhere between the outer edge of the (retracting) blackbody photosphere (maximum value of $\sim 7\,\times\,10^{14}$ cm) and the leading edge of the outflow (probed by observations in the radio). 
These observations imply a distance up to $10^{16}$~cm at 50 -- 100 days after the light curve peak. Therefore, the line emitting area has a characteristic size of $\sim$$10^{15}$ to $\sim$$10^{16}$~cm for a scale (maximum) velocity of $10^{4}$~km~s$^{-1}$ derived from fitting the spectral lines. If this scale velocity is in the order of $2\,\times\,10^{4}$ km~s$^{-1}$ (seen in many TDEs as we have shown in our work) it would imply a maximum radius for the line emitting region on the order of $\sim 5\,\times\,10^{16}$~cm, still unable to explain the largest distances derived from the H$\alpha$ lags. 
Furthermore, it is reasonable to expect that the lines would be produced closer to the inner edge of this region (i.e., closer to the outer edge of the blackbody photosphere) and not close to the leading edge of the outflow. Therefore, there is still a big discrepancy between this inner edge and the measured r$_{\rm lags}$ in this work.
Finally, the time lags that we measure are in the order of 7 -- 45 days after the light curve peak while the largest distances inferred by \citet{Nicholl2020} would occur around 50 -- 100 days after peak.
Hence, we conclude that it is improbable to explain the r$_{\rm lags}$ found in this work based on the AT2019qiz outflow scenario.
\citet{Metzger2016} also calculate the outer radius of the ejecta to be at $3.5\,\times\,10^{15}$~cm (t/t$_{\rm fb}$) meaning that it rapidly evolves to values greater than $10^{16}$~cm after the light curve's maximum (several fallback times). However, they also suggest that the emission lines are potentially produced somewhere between the optical photosphere and the outer shell of the ejecta.

\citet{Guillochon2014} suggest that the broad lines seen in TDEs are produced in a BLR structure (analogous to the one seen in AGNs) which lies above and below the forming accretion disk. Furthermore, they comment that if this is true, it would be reasonable to expect that these two astrophysical phenomena (i.e., AGNs and TDEs) should have many similarities in terms of velocity structures, as well as the components of the structure that play a key role in the production of the radiated light and the emergent emission lines. If their prediction is valid, it is not surprising that TDEs show the observed time lags pointed out in this work, as time lags (caused by the response of line luminosities to variations in the continuum) are commonly observed in AGNs. 
It has been  suggested that the line emitting region must be stratified, where Helium is closer to the black hole than Hydrogen \citep{Guillochon2014,Roth2016}. Regular AGNs are well known to be stratified as well with \ion{He}{II}, \ion{He}{I}, H$\beta$ and H$\alpha$ being emitted from closest to furthest from the black hole (e.g., see \citealt{Clavel1991,Peterson1999}). 
To test this, we investigated how lagged the Helium lines are compared to Hydrogen (see Sect. \ref{subsub:hale} - \ref{subsub:oel}).
We chose \ion{He}{I} 5876 \AA\, 
as this line is more isolated and easier to measure than \ion{He}{II}. We find a difference, depending on the spectroscopic type: H TDEs show larger lags in H$\alpha$ than \ion{He}{I}. On the other hand, for \ion{N}{III} Bowen TDEs,  \ion{He}{I} has similar lag values to H$\alpha$. 
%ASASSN-15oi does not show a peak for the \ion{He}{I} line luminosity. 
The average (lower-limit) lag value for H$\alpha$ is 40 days for the H TDEs and 12 days for the \ion{N}{III} Bowen TDEs (without including ASASSN-15oi because it does not belong in either of these categories as it is a He TDE with weak Balmer lines in some epochs). The average (lower-limit) lag value for \ion{He}{I} is 15.48 light days for \ion{N}{III} Bowen TDEs.
For the H TDEs, only AT2018zr shows a clear lag in \ion{He}{I}  since for ASASSN-14ae and LSQ12dyw we can only place upper-limits). 
These results are visualized in Fig. \ref{fig:tlagvsRBB} where we plot the time lags against the blackbody radii values at peak for each TDE. 
The H$\alpha$ lags (filled symbols) increase with increasing R$_{\rm BB}$ (the dashed line is a fit to H$\alpha$ and the dot-dashed one is the same without fitting the  three lower-limit points).
On the other hand, \ion{He}{I} (empty symbols) does not show such a correlation and the lags remain small (mostly upper-limits) even for the larger radii.
This picture can be understood by a combination of two factors: 
i) \ion{N}{III} Bowen TDEs have smaller radii than H TDEs \citep{vanvelzen2021};
ii) Helium lies deeper in the photosphere than Hydrogen \citep{Roth2016}. The ionization energy of neutral Hydrogen is 13.6 eV while the one of neutral Helium is 24.6 eV hence Helium would need a larger temperature in order to get ionized, hinting that it is indeed lying closer to the black hole than Hydrogen -- consistent with a stratified photosphere. 
We find here, however, that these differences are minimized in the more compact Bowen TDEs. \citet{Guillochon2014} focus their study on PS1-10jh, the only TDE in our sample that shows no Hydrogen at all, but only strong \ion{He}{II}. They suggest that the absence of Hydrogen lines indicates that the accretion disk had not yet extended to the distances required to produce the Hydrogen lines. Their calculations imply that \ion{He}{II} 4686 \AA\, should be produced at a distance of $\sim$2.1 light days from the SMBH (see their Fig. 7). Such small lags are beyond the precision that can be attained with the present data, as daily spectroscopic observations would be required.

\citet{Lu2020} find in their study that unbound debris can be produced from the shock of the self-crossing debris stream which occurs due to relativistic precession when it comes back near the SMBH (CIO). They suggest that this CIO is the reprocessing matter responsible for the emergence of the emission lines in TDEs. In this case, and assuming a strong shock, the mass can reach a few 10\% of the star's mass, the speed about 10\% of the speed of light and the covering factor can represent up to half of the sky since this gas is launched inside a cone with a large opening angle. For example, the distance travelled at 20\% of the speed of light during one month (about the time of light curve peak) would be about 1.5 $\times 10^{16}$~cm (C. Bonnerot; private communication with the authors). More detailed calculations are needed to determine whether this matter can actually result in the emission lines observed but this is outside the scope of this work. What is important here is that the $1.5 \times 10^{16}$ cm distance is again ``on the small side'' of the inferred r$_{\rm lags}$ found here (1.8$\,\times\,10^{16}$ -- 1.2$\,\times\,10^{17}$cm).

\begin{figure}
\centering
\includegraphics[width=0.47 \textwidth]{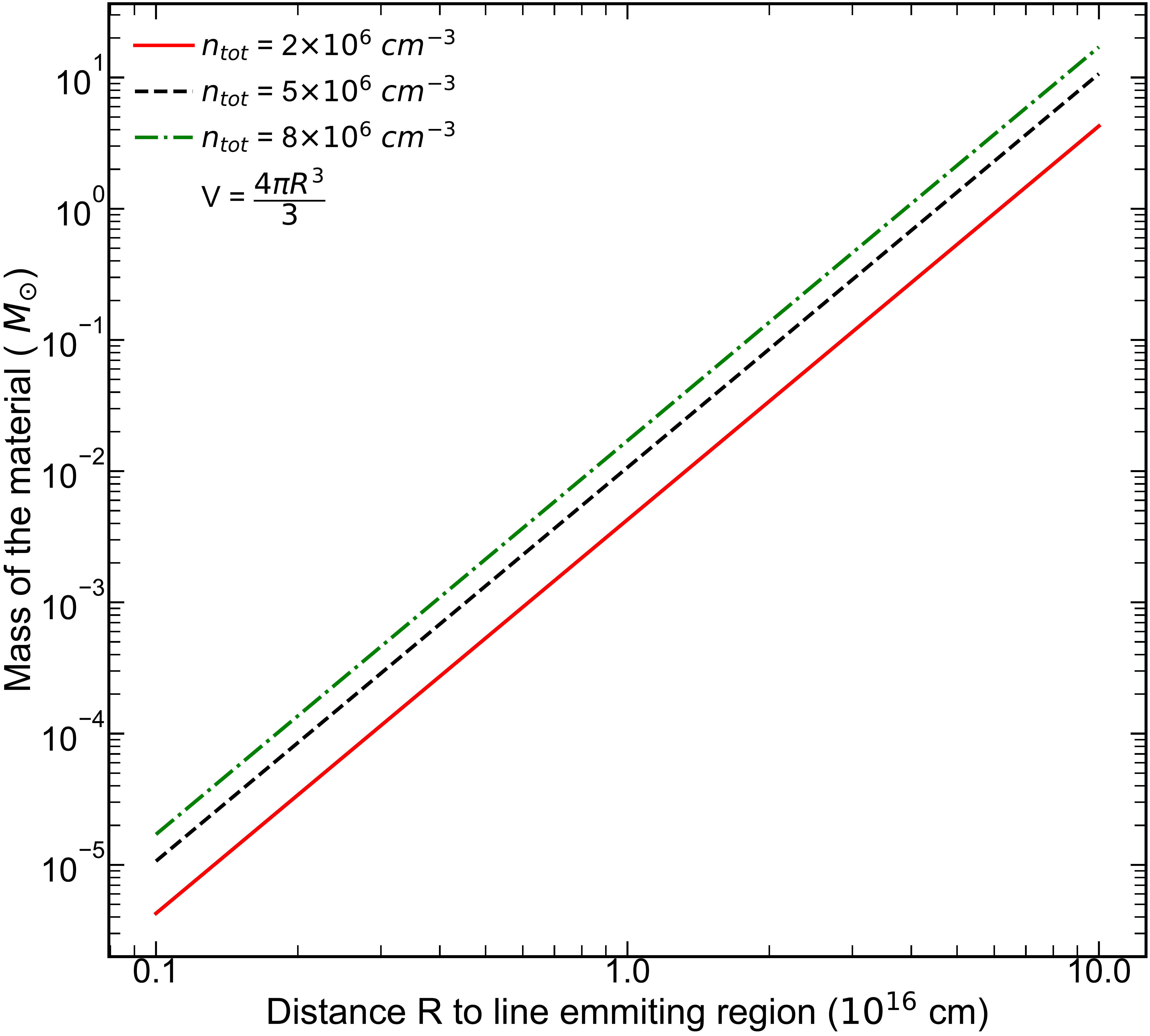}
\caption{Estimation of the mass of the reprocessing gas responsible for the emission of the spectral lines in TDEs in order to investigate if the recombination time ($\tau_{\rm rec}$ $\approx$ (n$_{\rm e}\alpha_{\rm B})^{-1}$, where n$_{\rm e}$ is the electron density and $\alpha_{\rm B}$ is the case B recombination coefficient) could be responsible for the lags that we discover. We assume its geometrical shape is a sphere (as the simplest approximation) and that the material's composition is solar-like and fully ionized. We perform three calculations with varying electron densities (which could result to the observed t$_{\rm lags}$) that in turn yield three total number densities n$_{\rm tot}$. For all the cases we get a sensible fraction of a solar mass (2.5 -- 10 \%) assuming the radius of the line emitting region is on the order of 10$^{16}$ cm. }
\label{fig:3_n_tots}
\end{figure}

Alternatively, but within a reprocessing scenario (optically thick wind, BLR-type or CIO), we examine the possibility that the recombination time $\tau_{\rm rec}$ $\approx$ (n$_{\rm e}\alpha_{\rm B})^{-1}$ (where n$_{\rm e}$ is the electron density and $\alpha_{\rm B}$ is the case B recombination coefficient) of the material responsible for the emission of the ``lagged'' photons is not as short as it is expected to be in the BLR of AGNs. In AGN studies, the BLR is expected to be dense enough (n$_{\rm e} \geq 10^{10}$~cm$^{-3}$) so that the cloud response to the continuum variations is nearly instantaneous which makes the light travel time across the BLR the most important timescale \citep{Peterson1993}. If we consider a spherical shell not as dense as the BLR of AGNs, it is possible to get a recombination time of $\sim$~7~--~45 days for distances shorter -- and more probable -- than the $\sim$ 10$^{17}$ cm r$_{\rm lags}$. For n$_{\rm e}$~=~10$^{6}$~cm$^{-3}$, $\tau_{\rm rec} \sim 41$ days while for n$_{\rm e}$~=~$5\,\times\,$10$^{6}$~cm$^{-3}$, $\tau_{\rm rec} \sim 8$ days. 
As this is promising, we attempt to estimate the mass of the reprocessing gas in this scenario and check whether it is consistent with originating from the debris of a disrupted star. In this case, we would expect a sensible fraction of a solar mass (e.g 1 -- 10 \% M$_{\odot}$). 
%If it is much higher then it must be pre-existing gas rather than TDE debris. 
In order to make this calculation we consider electron densities of 10$^{6}$ -- $5\,\times\,$10$^{6}$~cm$^{-3}$ for a solar-like composition material (71\% Hydrogen, 27\% Helium, 2\% metals) which is fully ionized (hence the resulting mass will technically be an upper-limit). This yields mass densities $\sim$~10$^{-24}$~n$_{\rm tot}$~g~cm$^{-3}$, where n$_{\rm tot}$ (which is the total number density) is $\sim$~1.7$\,\times\,$10$^{6}$ --~ 8.6$\,\times\,$10$^{6}$~cm$^{-3}$. In order to calculate the mass we need to make some reasonable assumption about the size of the line emitting region. For the simplest approximation we consider a sphere that would also be an upper-limit for the mass since every spherical shell of smaller thickness would end up in smaller masses (or larger distances to the line emmiting region). We visualize these calculations (for three different n$_{\rm tot}$ values) in Figure \ref{fig:3_n_tots}. We find that, for distances around $1.5 \times 10^{16}$~cm (which for example is the calculated distance of the CIO for 10 \% of the star's mass), we would get masses spanning $\sim$ 2.5 -- 10 \% M$_{\odot}$ for the different n$_{\rm tot}$ values. Hence we suggest that a low density stellar material could be the reprocessing layer responsible for the emission lines in TDEs. Even for asymmetric geometries of the reprocessing layer (such as the CIO which supposedly covers $\sim$ 50\% of the sky so not exactly a sphere or spherical shell), the approximation we made provides an upper-limit hence is consistent with the reprocessing layer being part of the TDE debris.

In conclusion, the very large inferred distances (r$_{\rm lags}$) that we calculate if we try to explain the measured time lags in TDEs as light echoes (similar to those measured at the BLR of AGNs) are hard to justify. We examine different possible scenarios of how a reproccessing material could have reached distances on the order of 10$^{17}$ cm; outflows, winds and unbound debris (CIO) do not seem able to reach that large distances at the time of the optical light curve peak. The possibility of a low -- compared to the BLR of AGNs -- electron density ($\sim$ 10$^{6}$ cm$^{-3}$) reprocessing material is promising as it could result in recombination times large enough to lead to the measured time lags. More TDEs with better sampled spectral coverage as well as photometric and spectroscopic pre-peak detections will be needed in order to perform a detailed cross-correlation analysis and acquire more robust results.

\subsection{Blackbody radius and temperature effects on the spectra} \label{subsec:dbb}
In Sect. \ref{subsub:eitobtar}, for the first time in the literature, the evolution of line luminosities as a function of blackbody radii and temperatures is presented. We find a clear trend between the luminosity of H$\alpha$ (and other lines as well with stronger or weaker correlation, as presented in Sect. \ref{subsub:eitobtar}) and the blackbody radius of TDEs. The larger the R$_{\rm BB}$ is, the stronger the H$\alpha$ emission. \citet{vanvelzen2021} show a T$_{\rm BB} \propto$ R$_{\rm BB}^{-1/2}$ scaling in TDEs. Motivated by this, we tried to fit the H$\alpha$ luminosity of TDEs as a function of their T$_{\rm BB}$ with an inverse power-law function in order to check if the T$_{\rm BB} \propto$ R$_{\rm BB}^{-1/2}$ relation can be confirmed from our results. The rationale is the following: since L$_{\rm H\alpha} \propto$ R$_{\rm BB}$ and T$_{\rm BB}~\propto$~R$_{\rm BB}^{-1/2}$, we would expect L$_{\rm H\alpha} \propto$ T$_{\rm BB}^{-1/2}$. The luminosities seem to indeed follow an inverse power-law relation and in order to further investigate this, we also fit a power-law with the exponent as a free parameter in order to account for the effects that L$_{BB}$ could have in the correlation (since L$_{BB}=4\pi\sigma$R$_{\rm BB}^{2}$T$_{\rm BB}^{4}$). We ran the same statistical tests as before in order to quantify the correlation that the blackbody luminosity could show with the H$\alpha$ luminosity and  we find the following scores: Kendall's tau 0.17 , Spearman's rho 0.22  and Pearson's $r$ 0.03. The correlation with L$_{\rm BB}$ is much weaker (and not significant) in comparison with the ones with R$_{\rm BB}$ and T$_{\rm BB}$ (see Table \ref{tab:ktsp}). Same applies for the correlations of the rest of the lines presented in Table \ref{tab:ktsp}. Therefore, we rule out the fact that the blackbody luminosity plays an important role in this scaling. The important question here is what drives the relation between line luminosity and R$_{\rm BB}$ (or T$_{\rm BB}$).

The fact that line luminosities and R$_{\rm BB}$ follow a monotonic and potentially linear trend for the whole sample is important. \citet{Roth2016} calculate that the smaller the outer radius of the reprocessing envelope is, the fainter the H$\alpha$ will be and vice versa. They actually find that for sufficiently small R$_{\rm BB}$ values, H$\alpha$ could even be seen in absorption (see their Figs. 7 and 8). This is not the case here but what we see in our work is that as the photospheric radius decreases, indeed, the H$\alpha$ emission fades. Additionally, the H$\alpha$ emission is weaker at high temperatures potentially following an inverse power-law relation (see Fig. \ref{fig:Ha_T_R}). A possible explanation for this could be that for photospheric temperatures greater than $3\,\times\,10^{4}$ K, Hydrogen starts getting ionized in the line emitting region hence the Balmer lines fade/get suppressed. Therefore, this could be an ionization effect. However, that would require that the temperature at the line emitting region is tightly correlated with the photospheric temperature which does not match with the surprisingly large distances inferred from the light echo scenario. Instead these tight relations of H$\alpha$ with the photospheric temperature and radius indeed argue that the line should form relatively close to the photosphere. 

The evolution of \ion{He}{II}/H$\alpha$ luminosity ratio in TDEs has also been a topic of discussion. The increase of this ratio could confirm something that we know from photometric analyses; namely that TDEs show a roughly constant (or small increasing) blackbody temperature evolution, while their blackbody radius shrinks with time (e.g., \citealt{Holoien2018}). If line strengths are mostly affected by the wavelength-dependent optical depth \citep{Roth2016}, a transition from Hydrogen-strong to Helium-strong spectra can be explained by a receding photosphere. If Helium lies deeper in the ``stratified'' photosphere then it will be emitted even for smaller photosphere sizes. On the other hand, Hydrogen is self-absorbed at most radii so a small/receding photosphere will suppress its emission \citep{Nicholl2019}. The general rising trend that this ratio shows with time seems to confirm the predictions of \citet{Roth2016}. Furthermore, the \ion{He}{II}/H$\alpha$ luminosity as a function of R$_{\rm BB}$ appears to rise as R$_{\rm BB}$ becomes smaller for our sample. This is also in agreement with \citet{Roth2016} who predict that Helium lies deeper in the reprocessing photosphere than Hydrogen so, as the photospheric radius shrinks, the more this ratio should increase. 

The \ion{He}{II}/\ion{He}{I} luminosity ratio in TDEs is an indicator of the ionization state of the debris. This ratio shows a general rising trend with time which implies that He is getting ionized as the photospheric radius shrinks something that is also in agreement with the predictions of \citet{Roth2016}. The \ion{He}{II}/\ion{He}{I} luminosity ratio is $<$ 1 for TDEs with low temperatures. An explanation for this is that He (in the line emitting region) has not been ionized for such low blackbody temperatures ($\leq$ 20\,000 K). AT2018zr and LSQ12dyw, with the lowest temperatures in our sample, do not show \ion{He}{II} 4686 \AA\, (we place upper-limits for those in Fig. \ref{fig:HeII_HeI_T}), suggesting that Helium has not been ionized at all. The fact that both these ratios grow stronger with time indicates that \ion{He}{II} indeed is stronger as the photospheric envelope shrinks \citep{Roth2016} and indeed fades slower (or increases later) than other emission lines.

In conclusion, the spectral properties of TDEs as a function of their blackbody radius and temperatures, are compatible with the picture proposed by \citet{Roth2016} where a larger photospheric radius would lead to stronger Hydrogen emission and a shrinking photospheric radius would lead to stronger/lasting \ion{He}{II} emission. Our results suggest that the emission lines are affected by the photospheric properties. This is additional evidence that the line forming region cannot be completely detached from the continuum photosphere (as the large distances deduced by the time lags might imply).

\subsection{Spectral classes of TDEs and connection to photometric properties} \label{subsec:dbsc}

\citet{vanvelzen2021} presented a basic spectroscopic classification scheme for TDEs (based on a single spectroscopic epoch) where they divide TDEs into three main subgroups (i.e., H, He or H+He TDEs). The H+He notation describe TDEs that show Hydrogen Balmer lines and the, characteristic in TDEs, \ion{He}{II} 4686 \AA\, line in their spectra. The H notation describes those that lack this \ion{He}{II} line while the He notation describes those that lack the Hydrogen balmer lines. In their work they studied the photometric properties of a sample of TDEs and investigated possible correlations with their spectroscopic classification. Their study showed that TDEs with Bowen lines have significantly lower blackbody radii than other TDEs. In Fig. \ref{fig:Ha_FWHMs_BCC} we show that \ion{N}{III} Bowen TDEs seem to systematically have lower H$\alpha$ widths than the rest of the TDEs in our sample. Since Bowen TDEs seem to have lower R$_{\rm BB}$ than the rest (see \citealt{vanvelzen2021}), the orbital velocity set by keplerian dynamics ($v=\sqrt{GM_{\rm BH}/R}$) should lead to higher velocities for the \ion{N}{III} Bowen TDEs. We have already shown that there is no correlation between the line widths and the BH masses in Section \ref{subsub:habhlp} (see left panel of Fig. \ref{fig:Ha_FWHM_MBH_Lp_Lx}). The fact that there is no correlation either between the blackbody radii and the line widths, hints that the line widths are not set by keplerian kinematics. It could therefore be electron scattering that dictates the widths of spectral lines in TDEs. If this is the case then we would expect that in Bowen TDEs, photons would undergo multiple scatterings before escaping as these scatterings can greatly boost the resonance of the Bowen lines. More scatterings would result in larger line widths, however we see the opposite result here. Another explanation could be that this might be a selection effect as broader lines result in higher blending, making it more difficult to distinguish Bowen features (e.g., also \ion{N}{III} 4100 \AA\, is blended with H$\delta$). 

\citet{VanVelzen2020} extended the basic classification scheme by adding the subtype of each TDE (i.e., whether they show Nitrogen, Oxygen or Iron lines) and by denoting the spectroscopic evolution of the type of each TDE (if any) with an arrow (e.g., H+He --> He). For example, ASASSN-14ae changes from an H to an H+He TDE while AT2017eqx shows the opposite behavior. In Table \ref{tab:sample_spec}, we have included  
the spectroscopic classification for the TDEs in our sample following \citet{VanVelzen2020}, but with some minor changes and additions. 
Changes include the spectroscopic type of ASASSN-15oi and PTF09ge, which have been presented as He TDEs so far. By host subtraction and careful continuum removal, it was possible to uncover that those TDEs exhibit the H$\alpha$ Balmer line (although \ion{He}{II} is the dominant line in their spectra) and we therefore classify them here as H+He.
In addition, except the N, O and Fe notations, we also note for each TDE: i) the presence of \ion{He}{I} 5876~\AA\, by \ion{He}{I} and ii) the presence of double-peaked H$\alpha$ profiles (e.g., \citealt{Holoien2016,Short2020a,Hung2020}) by \texttt{DP}.

In the following, we attempt a further subgrouping of our TDEs based on their spectroscopic features (see also Sec. \ref{subsec:view}, where these subgroups are associated with X-ray properties and possible viewing angles). 
In addition, we sort and group (separated by horizontal lines) the events in Table \ref{tab:sample_spec} based on their spectroscopic type  and features during their early times ($\sim$ 30 days post-peak, left side of the arrow). 
H+He type TDEs are divided into two subcategories: the Iron-rich ones (with the exception of PTF09ge they also contain the tentatively identified \ion{O}{III} line \citep{Wevers2019}) and the N-rich ones. H type TDEs are also divided into two subcategories; the ones that show double-peaked Balmer lines and those that do not. AT2017eqx does not match with any of these subcategories because although it is a H+He TDE, it does not show any signs of Nitrogen neither Iron or Oxygen lines. In addition it evolves on quick timescales to a He TDE so it is not included in these subcategories. Finally, PS1-10jh is the only He TDE in our sample.

\begin{table}
\renewcommand{\arraystretch}{1.2}
\setlength\tabcolsep{5pt}
\fontsize{8.8}{11}\selectfont
% \begin{center}
\caption{Spectroscopic classification of the TDEs in our sample.}\label{tab:sample_spec}
\begin{tabular}{c c c c}
\hline
Discovery & IAU & Spectroscopic & X-rays \\Name & Name & Type (features) & \\
\hline
\hline
PTF09ge & & H+He (\textbf{Fe}) & ?  \\
ASASSN-15oi & & H+He (\textbf{Fe}+O+\ion{He}{I}) --> He & Yes\\
ASASSN-18ul & AT2018fyk & H+He (\textbf{Fe}+O+\ion{He}{I}) & Yes\\
\hline
ASASSN-14li &  & H+He (\textbf{N}+\ion{He}{I}) & Yes \\
iPTF15af & & H+He (\textbf{N}+O) & -  \\
iPTF16axa & & H+He (\textbf{N}+O+\ion{He}{I}) & - \\
iPTF16fnl & &  H+He (\textbf{N}+\ion{He}{I}) & -  \\
ASASSN-18pg	 & AT2018dyb & H+He (\textbf{N}+O+\ion{He}{I}) & -\\
ZTF19aabbnzo & AT2018lna & H+He (\textbf{N}+\ion{He}{I}) & - \\
\hline
\hline
PTF09djl & & H (+\textbf{DP}) & ?  \\
PS18kh & AT2018zr & H (\ion{He}{I} +\textbf{DP}) & Yes\\
ASASSN-18zj & AT2018hyz & H (\ion{He}{I} +\textbf{DP}) --> H+He & Yes \\
\hline
LSQ12dyw & & H (\ion{He}{I}) & -  \\
ASASSN-14ae & & H (\ion{He}{I}) --> H+He & -  \\
\hline
\hline
PS1-10jh & & He & - \\
PS17dhz & AT2017eqx$^{*}$ & H+He--> He & -\\
\hline
\end{tabular}
% \\[-10pt]
The five different subcategories of TDEs based on their spectroscopic features (discussed in Sect. \ref{subsec:dbsc}) are divided by the vertical lines. We denote the specific feature that defines a subcategory with bold letters. From top to bottom: H+He Iron-rich, H+He Nitrogen strong, H double-peaked (DP), plain H and He TDEs. \\
$^{*}$AT2017eqx starts as a H+He however it does not show any signs of Nitrogen neither Iron or Oxygen lines. In addition it evolves on quick timescales to a He TDE hence it is not grouped with the other H+He TDEs.
% \end{center}
\end{table}

\begin{figure*}
        \centering
        \begin{subfigure}[b]{1\textwidth}
            \centering
            \includegraphics[trim={0 0cm 0 0},clip,width=\textwidth]{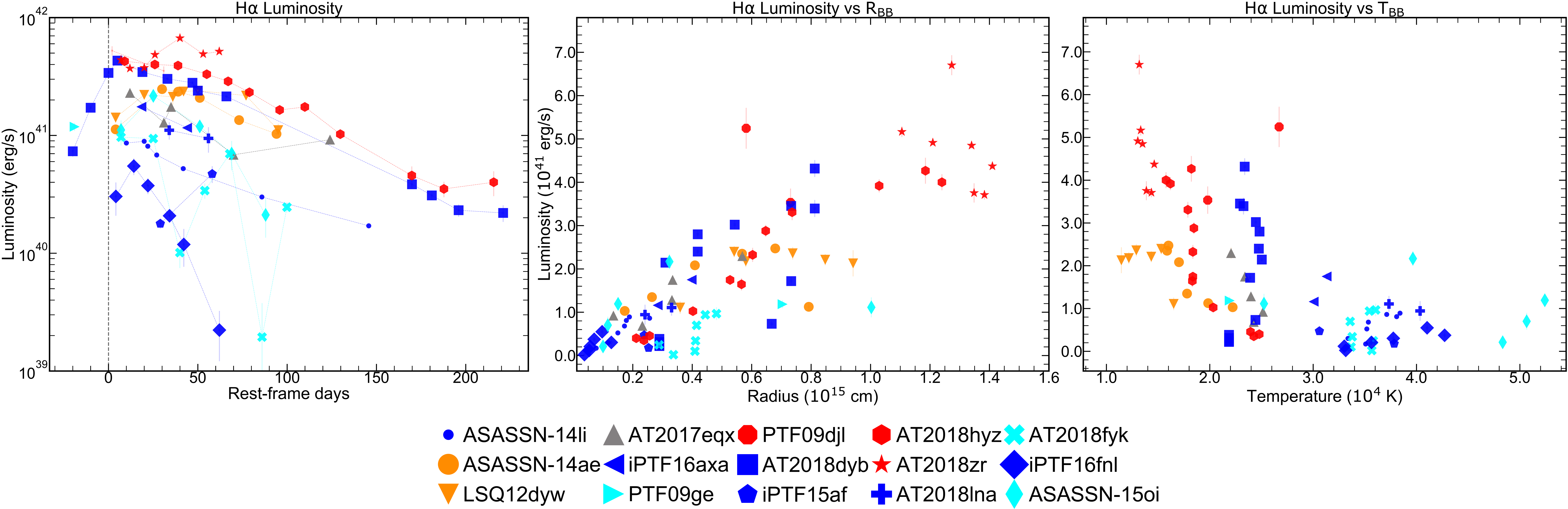}
            % \caption[Network2]%
            % {{\small Network 1}}    
            % \label{fig:off_sub}
        \end{subfigure}
        % \vskip\baselineskip
        \caption
        {Evolution of the H$\alpha$ line luminosity as a function of time (left panel), R$_{\rm BB}$ (middle panel) and T$_{\rm BB}$ (right panel). The plot is color-coded as follows: The H+He Iron-rich (with potentially \ion{O}{III} lines) events are plotted with cyan, the H+He Nitrogen Bowen events are plotted with blue, the double-peaked H events are plotted with red and the plain H events are plotted in orange. AT2017eqx does not meet the criteria of any of these four subcategories and is plotted in gray. The last two H subcategories show very high H$\alpha$ luminosity compared to the rest as well as low blackbody temperatures. This subcategory grouping is summarized in Table \ref{tab:4groups}.} 
        \label{fig:ha_three}
    \end{figure*}

In Fig. \ref{fig:ha_three} we present, once again, the H$\alpha$ line luminosity evolution with time, against R$_{\rm BB}$ and T$_{\rm BB}$ but this time with a different color coding based on the four aforementioned subcategories. The Iron-rich are plotted with cyan, the Nitrogen Bowen TDEs are plotted with blue, the double-peaked H events are plotted with red and the plain H events are plotted in orange. Interestingly, the two H subcategories have very high H$\alpha$ luminosity as well as low blackbody temperatures, in contrast with the H+He events that have low H$\alpha$ luminosities and high temperatures. The double-peaked events show the highest H$\alpha$ luminosities as well as the highest blackbody radii while the H+He events show the smallest radii (see also \citealt{vanvelzen2021}).

\subsection{Possible viewing angle effects} \label{subsec:view}

It has been suggested within the reprocessing scenario \citep{Dai2018} that the presence of X-ray emission in TDEs might depend on the viewing angle; those seen more face-on should be bright in X-rays while those seen more edge-on would be ``veiled'' in X-rays as they would all be reprocessed into optical/UV wavelengths as they pass through a reprocessing photosphere \citep[e.g.,][]{Auchettl2017}. If this is true, those TDEs that show strong X-ray emission should have narrower line widths as the escaping photons would undergo less scatterings before reaching the observer since they are viewed more face-on \citep{Leloudas2019}. In Fig. \ref{fig:Ha_FWHM_MBH_Lp_Lx} we see that the only TDE that showed strong X-ray emission (ASASSN-14li) has by far the lowest FWHM compared to the rest. Interestingly, the only spectrum for which AT2018fyk shows a lower FWHM value than the rest of its epochs is when its optical/UV light curves had a dip right before re-brightening again for a secondary maximum while the X-ray luminosity kept rising (something attributed to a rapid formation of an accretion disk). The fact that the line narrows when the optical/UV emission drops and the X-rays rise, further supports our point. So it seems that line widths are correlated with the X-ray intensity within X-ray bright TDEs favoring the reprocessing scenario \citep{Dai2018}. 

However X-ray TDEs do not seem to have systematically lower line widths than those TDEs that do not show X-rays in our sample. Maybe this points to reprocessing of the X-rays in the CIO \citep{Lu2020}. In the CIO scenario, the observer detects X-rays if the CIO is not between the observer and the accretion disk or does not if the CIO completely covers the accretion disk with respect to the observer. Hence the TDE pr  operties depend on the viewing angle and in this sense is similar to the one of \citet{Dai2018} but where they differ is that in the latter, the X-ray to optical luminosity ratio is dependant on the observer’s viewing angle with respect to the rotational axis of the accretion disk (instead of the CIO’s direction of expansion). This could explain the large diversity range of observed X-ray to optical flux ratios in TDEs. If the inner disk is completely veiled the observer sees no X-rays, if it partially veiled the observer sees weak X-rays and if it is not veiled the observer sees strong X-rays.
If both scenarios take place and the CIO does not cover the accretion disk for the observer -- which also happens to observe the accretion disk from small inclination angles -- then this could explain the correlation of X-ray strength with the optical line widths for X-ray bright TDEs.

In the previous section we attempted a subgrouping of TDEs based on their spectral features. Here we want to build on this subgrouping by examining each subcategory's X-ray properties. We present this further subgrouping in Table \ref{tab:4groups}.

\begin{table}
% \centering
%\def\arraystretch{1.3}
\setlength\tabcolsep{4pt}
\fontsize{9.4}{11}\selectfont
\caption{Grouping of the TDEs in our sample in four subcategories (two of the H+He type and two of the H type) based on their spectroscopic features at early times (pre-peak to a $\sim$ month after peak).}
\label{tab:4groups}
\begin{tabular}{c | cc p{0.2em} cc}
        %p{0.2em}
        \hline
        \hline
       \bf Type & \multicolumn{2}{c}{\bf H+He} & & \multicolumn{2}{c}{\bf H} \\
        \cline{1-2} \cline{2-3}  \cline{5-6}
       \bf subtype & \bf Fe (+O)  & \bf N  & & \bf DP  &\bf Plain  \\
         %\cline{2-2}  \cline{3-3} \cline{4-4}  \cline{5-5} \cline{6-6}
         \cline{1-2} \cline{2-3}  \cline{5-6}
      X-rays  & 2/2$^{\dagger}$ & 1/6$^{*}$ & & 2/2$^{\dagger}$ & 0/2 \\
      \cline{1-2} \cline{2-3}  \cline{5-6}
      Potential & Small & Large & & (Small to) & Intermediate \\
      viewing angles &  &  & & intermediate & (to large) \\
    %   Viewing angle & 0 -- 30$^{\circ}$ & 60 -- 90$^{\circ}$  & & 30 -- 60$^{\circ}$ & 45 -- 90$^{\circ}$ \\
        \hline
        \hline
\end{tabular}
%\\[-0pt]
$^{*}$ASASSN-14li is the only TDE of the H+He Nitrogen-rich subcategory that shows X-ray emission and in the same time has the lowest FWHM of all the TDEs in our sample. It is potentially viewed from an intermidiate angle rather than a large one.\\
$^{\dagger}$PTF09ge which is the third event of the H+He Iron-rich subcategory as well as PTF09djl which is the third event of the H double-peaked subcategory, were not monitored by any X-ray telescope while they were bright in the optical hence they could potentially be bright in X-rays as well during that time.
\end{table}

Concerning the subcategories of the H+He TDEs: the Iron-rich ones, ASASSN-15oi and AT2018fyk showed mild X-ray emission at early times becoming considerately stronger with time (L$_{\rm opt}$/L$_{\rm X}$ of $\sim$ 100 in early times to $\sim$ 1 and $\sim$ 10 respectively in later times) during their evolution while PTF09ge was not monitored by any X-ray telescope while it was bright in the optical hence we can not conclude whether it emitted in the X-rays or not. Concerning the N-rich ones, ASAASN-14li was the only one that showed strong X-rays (L$_{\rm opt}$/L$_{\rm X}$ $\sim$ 1 throughout the evolution). Concerning the H TDEs: The plain H TDEs were not detected in X-rays while two of the double-peaked ones (AT2018zr and AT2018hyz) showed mild X-ray emission (L$_{\rm opt}$/L$_{\rm X}$ of $\sim$ 100 for both) while the third (PTF09djl) was not monitored by any X-ray telescope while it was bright in the optical hence we can not conclude whether it emitted in the X-rays or not. 

We examine here whether the subcategories of Table \ref{tab:4groups} can be explained by potential viewing angle effects within the reprocessing scenario \citep{Dai2018} in which TDE properties are dependant on viewing angle. The events that show double-peaked profiles should be viewed from small to intermidiate angles (i $\sim$ 30$^{\circ}$ -- 60$^{\circ}$) since the double-peak is an attribute that would be ``lost'' if viewing the accretion disk face-on. These inclination angles are consistent with the accretion disk modeling performed in the individual papers of PTF09djl \citep{Arcavi2014}, AT2018zr \citep{Holoien2016} and AT2018hyz \citep{Short2020a,Hung2020}. AT2018zr and AT2018hyz both had mild X-ray emission (and about PTF09djl we do not know). The fact that these events show the largest R$_{\rm BB}$ makes their photospheres less dense hence easier to penetrate for X-rays. These events are color-coded in red in Fig. \ref{fig:ha_three}. The plain H TDEs without double-peak features which are plotted in orange, do not show X-ray emission and no Bowen lines either. They could be viewed from intermediate and large angles (i $\sim$ 45$^{\circ}$ -- 90$^{\circ}$) but the smaller R$_{\rm BB}$ they show compared to the Hydrogen-strong/double-peak ones (red) may point to a denser environment hence no escaping X-rays neither a clear view to the disk. Another possibility %(rather than the different R$_{\rm BB}$ values) 
explaining the potential differences in the photospheric density of the two Hydrogen-strong subclasses could be %the fact 
that double-peaked TDEs might come from partial disruptions of stars. The partial disruption would lead to a lower photospheric density, hence the escaping mild X-rays and double-peak features. AT2018hyz was indeed suggested to be a partial disruption, based on light curve modeling \citep{gomez1925}.  

\begin{figure}
\centering
\includegraphics[width=0.47 \textwidth]{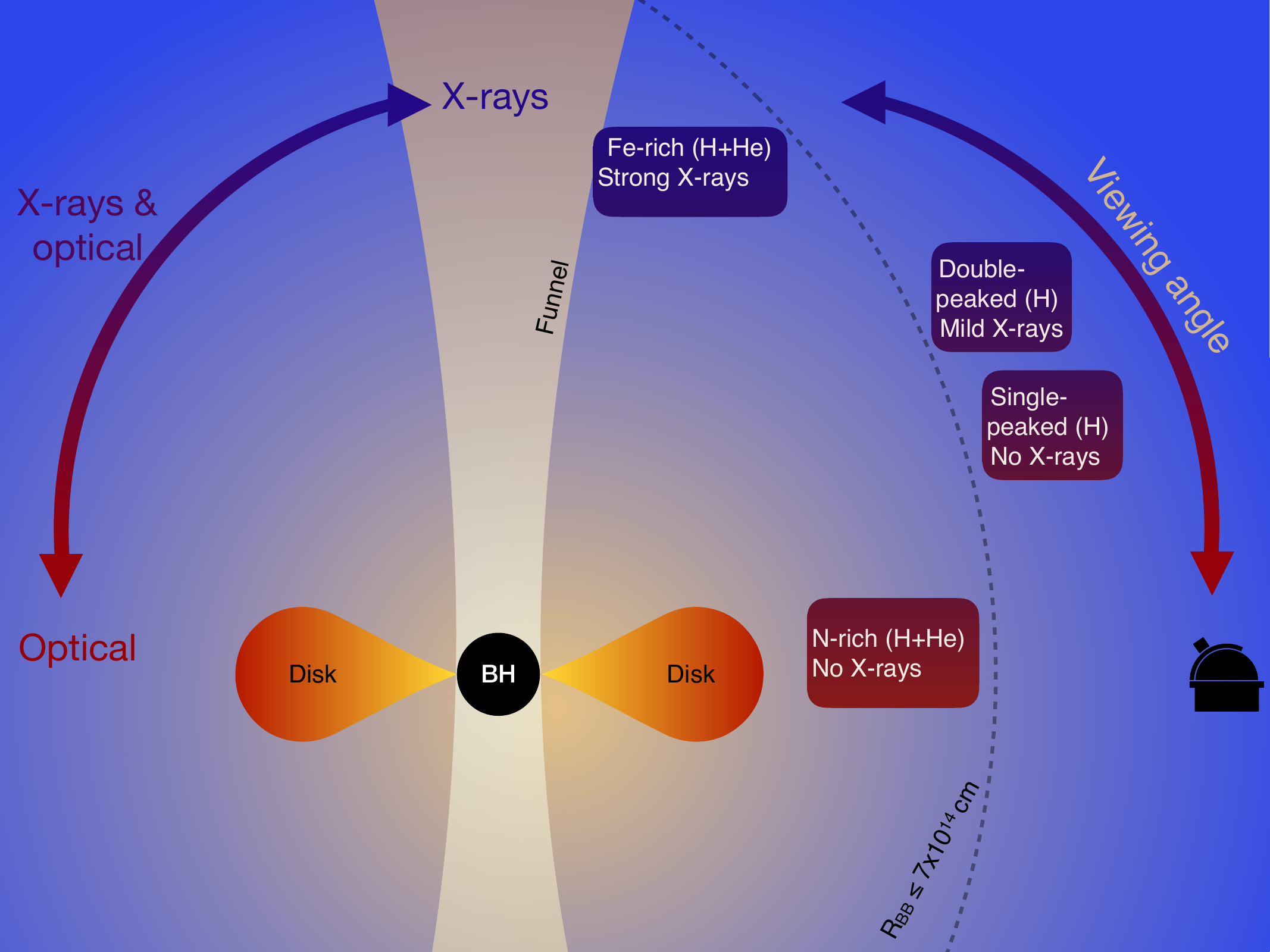}
\caption{A schematic  illustration (not to scale) showing the possible viewing angle dependence for the observed spectroscopic and photometric properties of TDEs. The four subclasses presented in Table \ref{tab:4groups} are distributed according to their potential viewing angle dependence, X-ray properties and R$_{\rm BB}$.}
\label{fig:schematic}
\end{figure}

The $\sim$ 4500 \AA, \ion{Fe}{II} lines are viewed in Type-1 AGNs and narrow-line Seyfert galaxies which are by definition viewed face on \citep{Antonucci1993}. It makes sense the same to apply for TDEs as ASASSN-15oi and AT2018fyk did show mild and later strong X-ray emission (and about PTF09ge we do not know). The fact that this subcategory shows the highest T$_{\rm BB}$ compared to the rest of the TDEs further supports this argument. These events are potentially viewed from very small angles (i $\sim$ 0$^{\circ}$ -- 30$^{\circ}$) and are plotted with cyan. On the other hand, Bowen \ion{N}{III} TDEs has been suggested that could be viewed edge-on in order for their X-ray emission to be reprocessed, a process needed in order to trigger the Bowen Fluorescence mechanism \citep{Leloudas2019}. The \ion{N}{III} Bowen TDEs of our sample (plotted in blue) do not show X-rays, apart from ASASSN-14li that had strong X-ray emission. These events could be viewed from large angles (i $\sim$ 60$^{\circ}$ -- 90$^{\circ}$). We show a schematic illustration of such a viewing angle dependent model in Fig. \ref{fig:schematic}.

One caveat here is that we would expect \ion{N}{III} Bowen TDEs to show larger line widths due to the scatterings however we find the opposite in this work. Potentially their small R$_{\rm BB}$ (and preference for smaller t$_{\rm lags}$) would make the line emitting photons travel through less material hence the smaller line widths. ASASSN-14li could be viewed from an intermidiate angle making it transparent to X-rays (since it also has the lower FWHM values). Else, the CIO scenario could also explain this peculiarity  as well as all the aforementioned viewing angle effects as it includes the many possible alignments of three points (namely i) observer ii) stream intersection point and iii) accretion disk).

\section{Summary and Conclusions} \label{sec:conclusion}

We present the first spectroscopic population study of optical/UV TDEs with a sample of 16 events.
We performed a careful and consistent series of data reduction tasks including host galaxy subtraction and continuum removal. This essentially imposed a ``cut'' for events discovered after 2019, as transient light could still be contaminating the host galaxy spectrum.
We study a number of emission lines prominent among TDEs and focus on H$\alpha$, H$\beta$, \ion{He}{II} 4686 \AA, \ion{He}{I} 5876 \AA\, and two \ion{N}{III} Bowen lines at 4100 \AA\, and 4640 \AA. We quantify their evolution in terms of line luminosities, velocity widths and velocity offsets by fitting Gaussians, Lorentzians and for some cases by direct integration and custom width and offset measurements.
Our study yielded the following findings:

\begin{enumerate}[label={\arabic*.}]

\item We find that H$\alpha$ line luminosity systematically peaks after the continuum light curves with time lags of $\sim$ 7 -- 45 days. \ion{N}{III} Bowen TDEs show small lag values compared to the rest. We discuss the possible origin of these lags in terms of outflows, BLR regions and unbound debris. We investigate the possibility of low electron density reprocessing material ($\sim$ 10$^{6}$ cm$^{-3}$) yielding recombination times that could explain the large inferred distances to the line emitting region.
\item We find that the \ion{He}{I} 5876 \AA\, line luminosity also shows lags with respect to the continuum light curves; smaller lags compared to those of H$\alpha$ are reported for the H TDEs and similar (or larger) for the \ion{N}{III} Bowen TDEs. This could point to a stratified TDE photosphere.
\item The widths of TDE lines drop slowly with time. We find that \ion{N}{III} Bowen TDEs have lower H$\alpha$ widths compared to the rest of the TDEs in our sample and we also find that a strong X-ray to optical ratio might lead to weakening of the line widths.
\item We find no apparent correlation between the line widths and the SMBH masses. Combined with the fact that \ion{N}{III} Bowen TDEs have lower R$_{\rm BB}$ values than H TDEs and in the same time have lower H$\alpha$ widths, points to the conclusion that line widths in TDEs are not set by the kinematics and maybe are dictated by electron scattering. 
\item H$\alpha$ velocity offsets are blueshifted in the earliest available spectrum for 11 TDEs of our sample.
\item We study the evolution of line luminosities and ratios with respect to their radii (R$_{\rm BB}$) and temperatures (T$_{\rm BB}$) and find a linear relationship between H$\alpha$ luminosity and the R$_{\rm BB}$ (L$_{\rm line} \propto$ R$_{\rm BB}$) consistent with theoretical predictions. We also report a potential inverse power-law relation with T$_{\rm BB}$ (L$_{\rm line} \propto$ T$_{\rm BB}^{-\beta}$) which could be driven by the ionization of Hydrogen at the line emitting region, when T$_{\rm BB}$ $\geq$ 25\,000 K in the blackbody photosphere. This would require that the temperature at the line emitting region is tightly correlated with the photospheric temperature which makes the surprisingly large distances inferred from the light echo scenario less likely.
\item We report that the \ion{He}{II}/H$\alpha$ ratio becomes stronger as the R$_{\rm BB}$ recedes, probably an optical depth effect which implies a stratified photosphere, consistent with theoretical predictions \citep{Roth2016}. The \ion{He}{II}/\ion{He}{I} ratio is decreasing for T$_{\rm BB}$ $\leq$ 20\,000 K pointing to a lack of ionization of Helium for these temperatures.
\item As part of our discussion, we attempt a grouping of TDEs into four subcategories based on their unique spectroscopic features and try to connect different spectral classes with their photometric properties. H TDEs divided in those that show double-peaked or Balmer profiles and those who do not (the first show X-rays while the second do not) and H+He TDEs divided in those that are Iron-rich and those that are \ion{N}{III} rich (the first show X-rays while the second do not). H TDEs have very high H$\alpha$ luminosity compared to the rest as well as low blackbody temperatures. 
\item Finally we discuss whether the large spectroscopic diversity of TDEs can be a result of viewing angle effects. We suggest a possible scheme (within the reprocessing scenario) where we attribute the spectroscopic features of each of the aforementioned subcategories to the observed viewing angle. 
\end{enumerate}
As the library of TDEs will keep getting larger in the era of the wide-field optical transient surveys, future work will aim to extend this sample and build upon the results of this work. A larger sample will put tighter constrains on the physical mechanisms responsible for the large diversity in the spectra of TDEs.

\begin{acknowledgements}

We thank Tom Holoien for providing us with the spectra of AT2018zr. P.C, G.L, D.B.M and M.P are supported by a research grant (19054)
from VILLUM FONDEN. IA is a CIFAR Azrieli Global Scholar in the Gravity and the Extreme Universe Program and acknowledges support from that program, from the European Research Council (ERC) under the European Union’s Horizon 2020 research and innovation program (grant agreement number 852097), from the Israel Science Foundation (grant number 2752/19), from the United States - Israel Binational Science Foundation (BSF), and from the Israeli Council for Higher Education Alon Fellowship. MN is supported by a Royal Astronomical Society Research Fellowship and by the European Research Council (ERC) under the European Union’s Horizon 2020 research and innovation programme (grant agreement No.~948381). T.-W.C. acknowledges the EU Funding under Marie Sk\l{}odowska-Curie grant H2020-MSCA-IF-2018-842471. L.G. acknowledges financial support from the Spanish Ministry of Science, Innovation and Universities (MICIU) under the 2019 Ram\'on y Cajal program RYC2019-027683 and from the Spanish MICIU project PID2020-115253GA-I00. MG is supported by the EU Horizon 2020 research and innovation programme under grant agreement No 101004719. T.M.B. was funded by the CONICYT PFCHA / DOCTORADOBECAS CHILE/2017-72180113. This work is based on observations collected at the European Organisation for Astronomical Research in the Southern Hemisphere, Chile, as part of PESSTO/ePESSTO/ePESSTO+ (the extended Public ESO Spectroscopic Survey for Transient Objects Survey). PESSTO/ePESSTO/ePESSTO+ observations were obtained under ESO program IDs 188.D-3003, 191.D-0935, 199.D-0143, 1103.D-0328. This work is based on observations made with the Nordic Optical Telescope, owned in collaboration by the University of Turku and Aarhus University, and operated jointly by Aarhus University, the University of Turku and the University of Oslo, representing Denmark, Finland and Norway, the University of Iceland and Stockholm University at the Observatorio del Roque de los Muchachos, La Palma, Spain, of the Instituto de Astrofisica de Canarias.

\end{acknowledgements}

%-------------------------------------------------------------------
\bibliographystyle{aa}
\bibliography{bib.bib}

%%%%%%%%%%%%%%%%%%%%%%%%%%%%%%%%%%%%%%%%%%%%%%%%%%
%%%%%%%%%%%%%%%%% APPENDICES %%%%%%%%%%%%%%%%%%%%%
\appendix{}
\onecolumn
\centering

\section{Graphs} \label{apdx:graphs}

\begin{figure*}[h]
        \centering
        \begin{subfigure}[b]{1\textwidth}
            \centering
            \includegraphics[trim={0 0cm 0 0},clip,width=\textwidth]{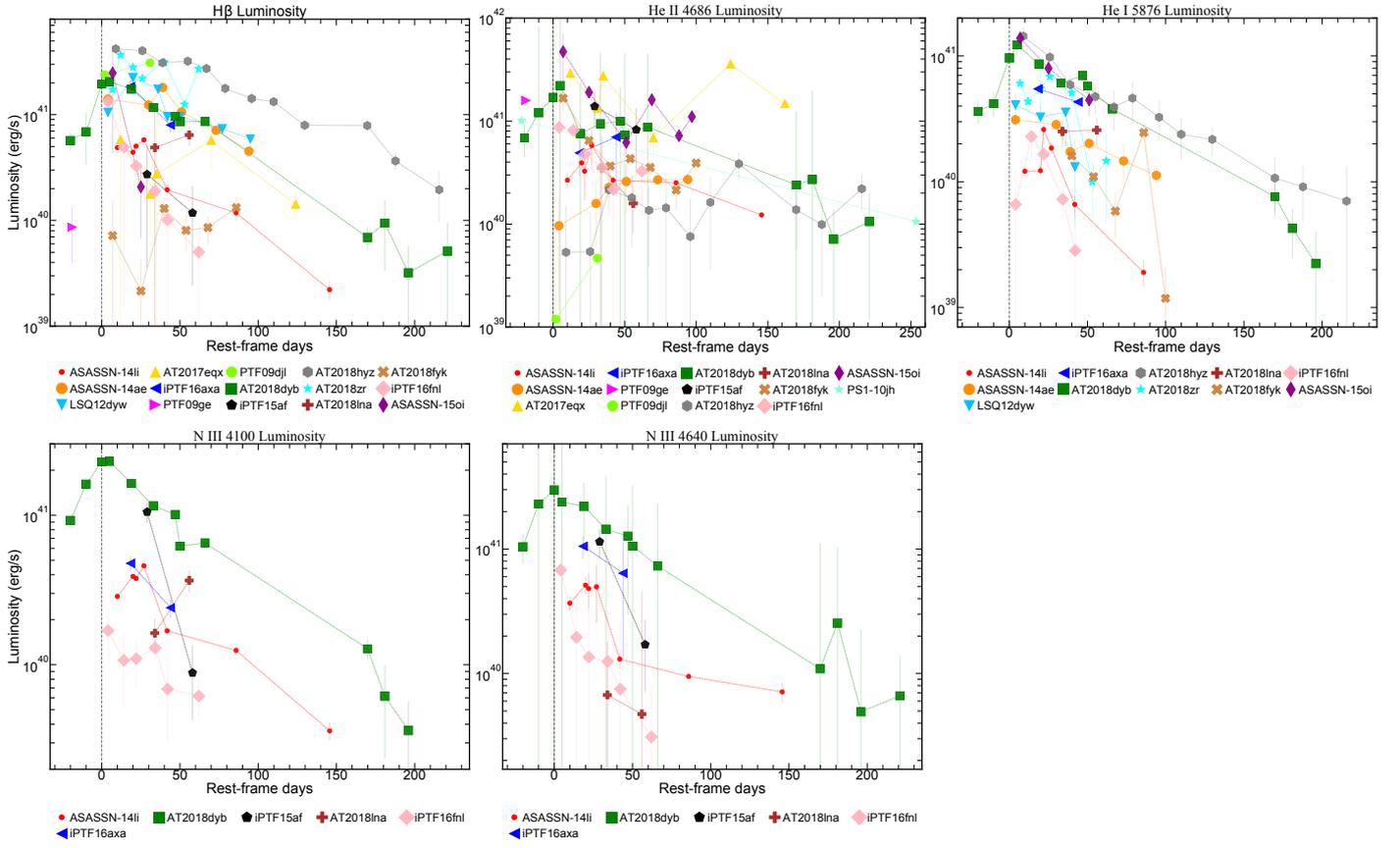}
            % \caption[Network2]%
            % {{\small Network 1}}    
            % \label{fig:off_sub}
        \end{subfigure}
        % \vskip\baselineskip
        \caption
        {From left to right and from top to bottom: H$\beta$, \ion{He}{II} 4686 \AA, \ion{He}{I} 5876 \AA, \ion{N}{III} 4100 \AA, and \ion{N}{III} 46400 \AA\, line luminosities evolution with time.} 
        \label{fig:lums_sub}
    \end{figure*}

\begin{figure}
\centering
\includegraphics[trim={0 0cm 0 0},clip,width=0.47 \textwidth]{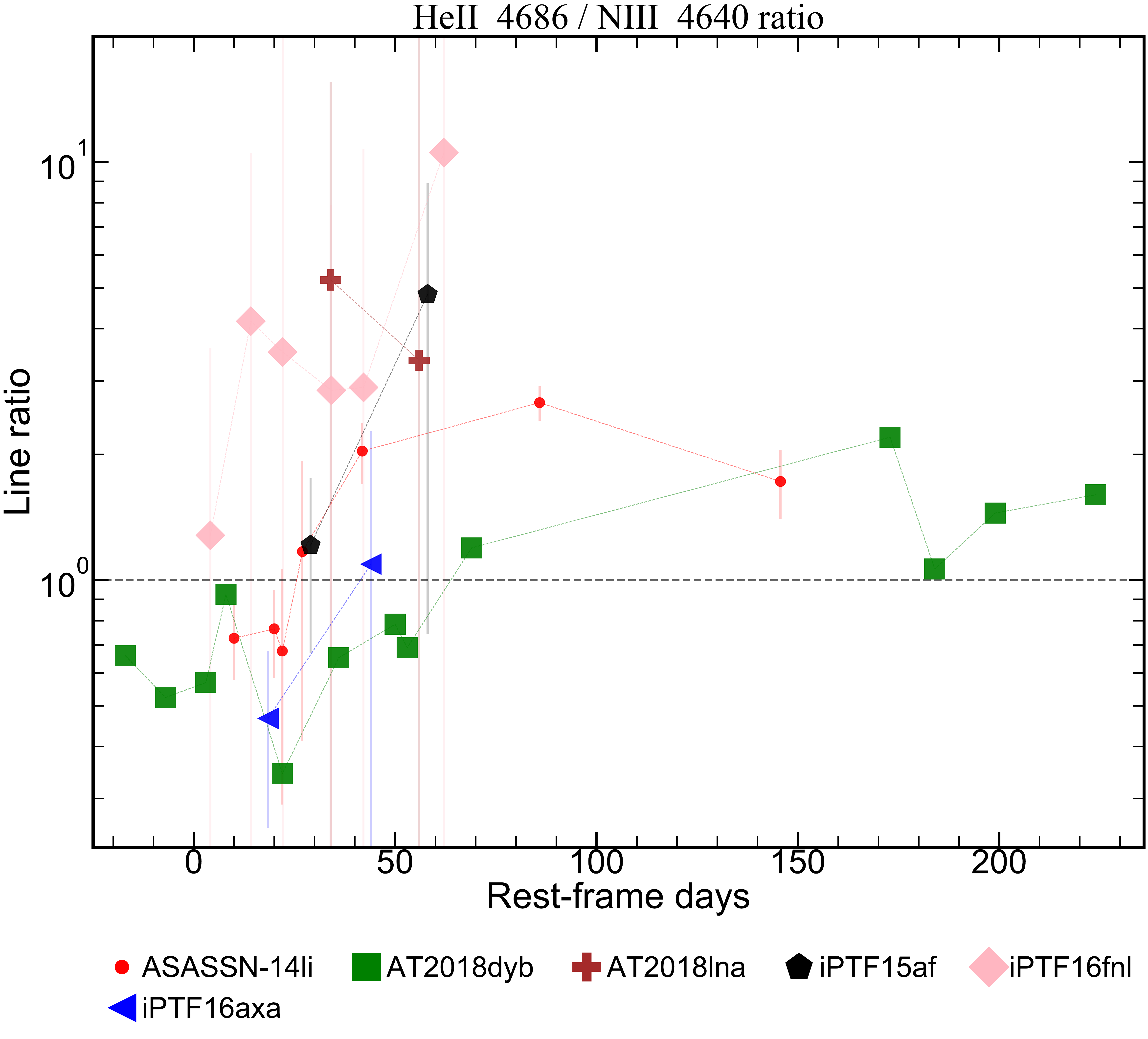}
\includegraphics[trim={0 0cm 0 0},clip,width=0.47 \textwidth]{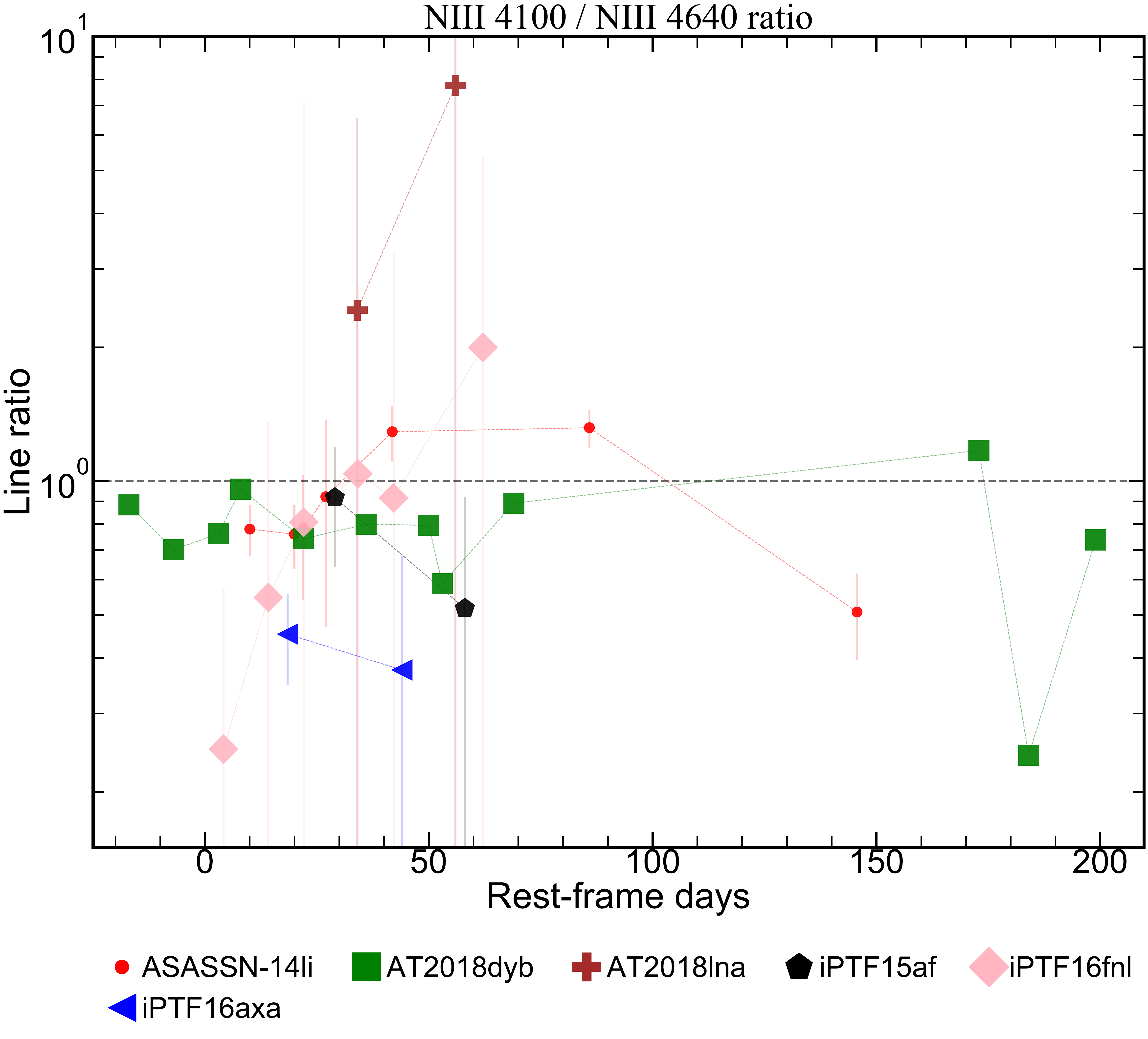}
\caption{\ion{He}{II}/\ion{N}{III} 4640 \AA\, (left) and \ion{N}{III} 4100/4640 \AA\, (right) luminosity ratios' evolution with time.}
\label{fig:HeII_NIIIs}
\end{figure}

\begin{figure*}
        \centering
        \begin{subfigure}[b]{1\textwidth}
            \centering
            \includegraphics[trim={0 0cm 0 0},clip,width=\textwidth]{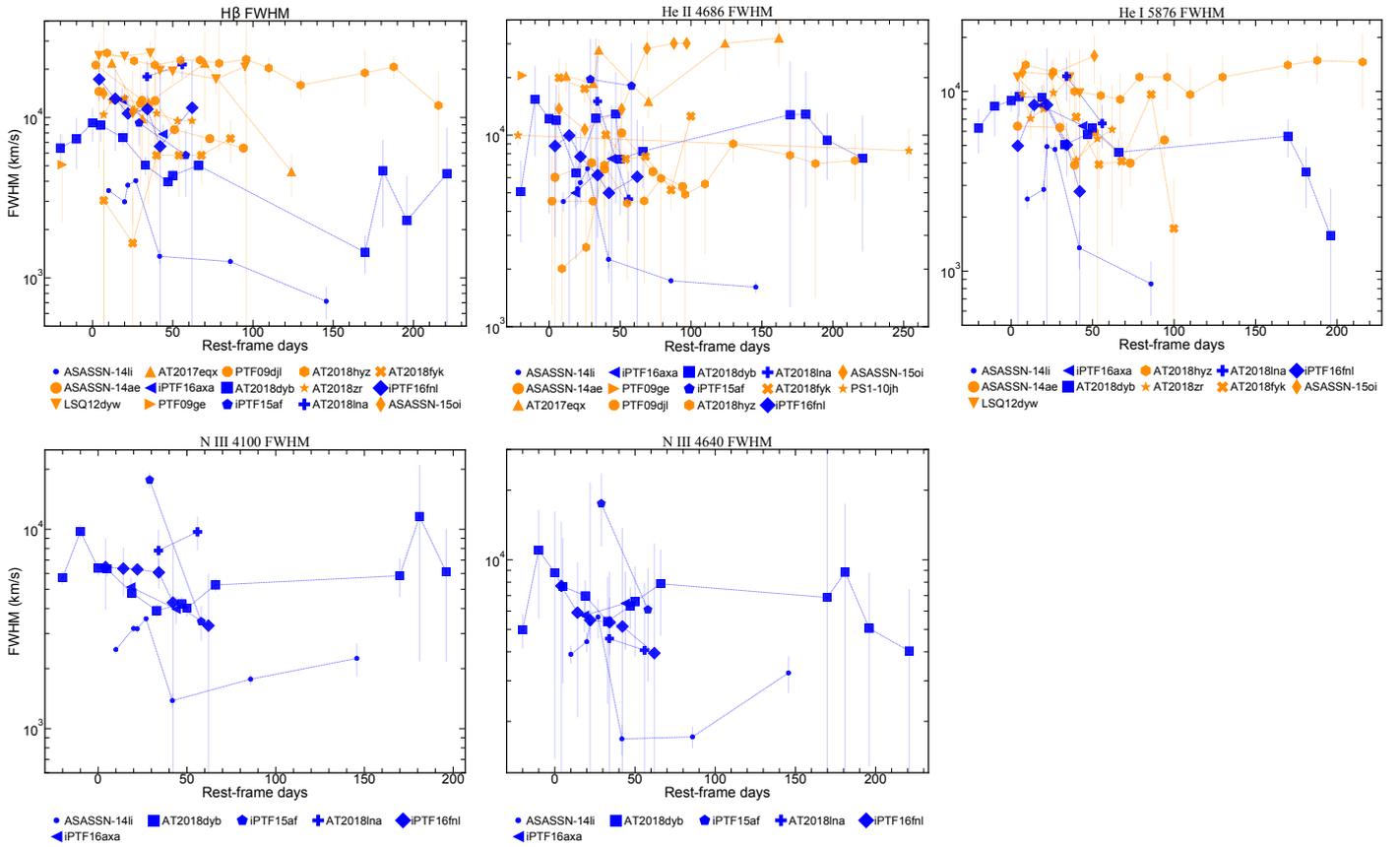}
            % \caption[Network2]%
            % {{\small Network 1}}    
            % \label{fig:off_sub}
        \end{subfigure}
        % \vskip\baselineskip
        \caption
        {From left to right and from top to bottom: H$\beta$, \ion{He}{II} 4686 \AA, \ion{He}{I} 5876 \AA, \ion{N}{III} 4100 \AA, and \ion{N}{III} 46400 \AA\, line widths evolution with time. The graph is color coded for Bowen \ion{N}{III} TDEs (blue) and not Bowen (orange).} 
        \label{fig:fwhms_sub}
    \end{figure*}

\begin{figure*}
        \centering
        \begin{subfigure}[b]{1\textwidth}
            \centering
            \includegraphics[trim={0 0cm 0 0},clip,width=\textwidth]{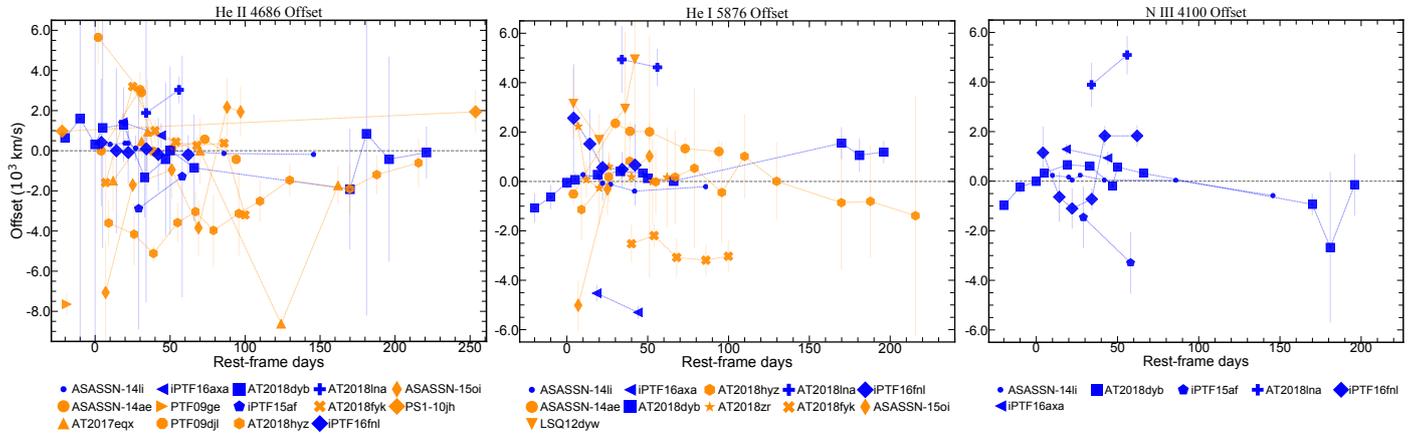}
            % \caption[Network2]%
            % {{\small Network 1}}    
            % \label{fig:off_sub}
        \end{subfigure}
        % \vskip\baselineskip
        \caption
        {From left to right: \ion{He}{II} 4686 \AA, \ion{He}{I} 5876 \AA, and \ion{N}{III} 4100 \AA\, line offsets' evolution with time. The graph is color coded for Bowen \ion{N}{III} TDEs (blue) and not Bowen (orange).} 
        \label{fig:offs_sub}
    \end{figure*}

\clearpage

\onecolumn
\centering
\section{Tables}\label{apdx:tables}
\begin{small}
%%%%%  Sample
\renewcommand{\arraystretch}{1.2}
\setlength\tabcolsep{0.09cm}
\fontsize{9}{11}\selectfont
\begin{longtable}{l l c c | c}
\caption{Phases of our sample's spectra and host galaxies.}
\label{tab:sample2} \\
\hline 
Name & Phase of spectra & Published in: & Notes & Blackbody \\Discovery (IAU)  & (MJD) & & &Ref$^{a}$\\
\noalign{\global\arrayrulewidth=1mm}\hline
\noalign{\global\arrayrulewidth=0.4pt}
PTF09ge  & $-$19 & \citet{Arcavi2014}& &1 \\
\textbf{Host}  & 1549  & \citet{Arcavi2014}& & \\
\arrayrulecolor{gray}\hline
PTF09djl  & 2, 31 & \citet{Arcavi2014}& &2 \\
\textbf{Host}   & 1355 & \citet{Arcavi2014}& & \\
\hline
PS1-10jh  & $-$22, 254 & \citet{Gezari2012}& &1 \\
\textbf{Host}   & 1387 & \citet{Arcavi2014}& & \\
\hline
LSQ12dyw  & 4, 20, 36, 42, 50, 77, 95 & This work & &3 \\
\textbf{Host}   & 1979 & This work & & \\
\hline
ASASSN-14ae  & 4, 30, 39, 51, 73, 94 &\citet{Holoien2014} &The spectra were already host subtracted &4 \\
\textbf{Host}   & Archival SDSS &\citet{Holoien2014}&  in \citet{Holoien2014} & \\
\hline
ASASSN-14li   & 10, 20, 22, 27, 42, 86, 146 & \citet{Holoien2015}&The spectra were already host subtracted &4\\
\textbf{Host}   & Archival SDSS &\citet{Holoien2015} & in \citet{Holoien2015} &  \\
\hline
iPTF15af  & 29, 60 & \citet{Blagorodnova2018}& &5 \\
\textbf{Host}   & 1035 &\citet{Blagorodnova2018} & & \\
\hline
ASASSN-15oi  & 7, 25, 51, 69, 88, 97 & This work & &4 \\
\textbf{Host}   & 359 & This work & & \\
\hline
iPTF16axa  & 12, 38 &\citet{Hung2017} & &1 \\
\textbf{Host}   & 1423 &This work & & \\
\hline
iPTF16fnl  & 2, 12, 20, 32, 40, 60 &\citet{Onori2019} & &4 \\
\textbf{Host}   & 289 & \citet{Onori2019}& & \\
\hline
PS17dhz (AT2017eqx) & 12, 31, 35, 70, 124, 162 &\citet{Nicholl2019} & The spectra were already host subtracted &6\\
\textbf{Host}   & 321 &\citet{Nicholl2019} & in \citet{Nicholl2019} &\\
\hline
PS18kh (AT2018zr) & 7, 12, 20, 26, 40, 52, 61 & \citet{Holoien2018}& &7 \\
\textbf{Host}   & 776 &This work & & \\
\hline
ASASSN-18pg (AT2018dyb) & $-$20, $-$10, 0, 5, 19, 33, 47, 50, & \citet{Leloudas2019}& &8 \\
& 66, 165, 181, 196, 216 & & \\
\textbf{Host}   & 554 &This work & & \\
\hline
ASASSN-18ul (AT2018fyk) & 7, 25, 40, 54, 68, 86, 100 & \citet{Wevers2019}& &9 \\
\textbf{Host}   & 122 &\citet{Wevers2019} & & \\
\hline
ASASSN-18zj (AT2018hyz) & 9, 26, 39, 55, 67, 79, 96, 110,  & \citet{Short2020a}& The spectra were already host subtracted &7 \\
& 130, 170, 188, 216 & & in \citet{Short2020a} \\
\textbf{Host}   & Archival SDSS &\citet{Short2020a} & & \\
\hline
ZTF19aabbnzo (AT2018lna) & 34, 56 & This work& &7 \\
\textbf{Host}   & 394 & This work&  &\\
\hline
\end{longtable}
% \vspace{-0.5cm}
$^{a}$The paper from which we retrieved the blackbody temperatures and radii for each event: $^1$\citet{Hung2017}, $^2$\citet{Arcavi2014}, $^3$This work, $^4$\citet{Holoien2018}, 
 $^5$\citet{Blagorodnova2018},
$^{6}$\citet{Nicholl2019},$^{7}$\citet{vanvelzen2021}, $^{8}$\citet{Leloudas2019}, $^{9}$\citet{Wevers2019}
\end{small}

\vspace{3.5cm}

\begin{small}
%%%%%  Sample
\renewcommand{\arraystretch}{1.2}
\setlength\tabcolsep{0.25cm}
\fontsize{9.5}{11}\selectfont
\begin{longtable}{c c c c c c c}
\caption{Emission line luminosities$^{a}$}
\label{tab:linelum} \\
\hline 
Phase$^{b}$ &H$\delta$/\ion{N}{III} $\lambda$4100 & \ion{N}{III} $\lambda$4640 & \ion{He}{II} & H$\beta$ & \ion{He}{I} $\lambda$5876 & H$\alpha$ \\ (days)     &  ($10^{41}$~erg~s$^{-1}$)      &     ($10^{41}$~erg~s$^{-1}$)      &   ($10^{41}$~erg~s$^{-1}$)     &   ($10^{41}$~erg~s$^{-1}$)       &   ($10^{41}$~erg~s$^{-1}$)       & ($10^{41}$~erg~s$^{-1}$) \\
\noalign{\global\arrayrulewidth=1mm}\hline
\noalign{\global\arrayrulewidth=0.4pt}
	&		&		&	\textbf{PTF09ge}	&		&		&	\\ \hline
$-$19	&	-	&	-	&	1.59 (0.08)	&	0.09 (0.05)	&	-	&	1.19 (0.09)  \\ \hline
	&		&		&	\textbf{PTF09djl}	&		&		&	\\ \hline
2	&	-	&	-	&	0.01 (0.01)	&	2.39 (0.08)	&	-	&	5.25 (0.27)  \\
31	&	-	&	-	&	0.05 (0.01)	&	3.09 (0.02)	&	-	&	3.53 (0.18)  \\ \hline
	&		&		&	\textbf{PS1-10jh}	&		&		&	\\ \hline
$-$22	&	-	&	-	&	1.01 (0.19)	&	-	&	-	&	-                  \\
254	&	-	&	-	&	0.11 (0.03)	&	-	&	-	&	-                  \\ \hline
	&		&		&	\textbf{LSQ12dyw}	&		&		&	\\ \hline
4	&	-	&	-	&	-	&	1.05 (0.09)	&	0.41 (0.10)	&	1.42 (0.14)  \\
20	&	-	&	-	&	-	&	2.24 (0.15)	&	0.33 (0.08)	&	2.21 (0.13)  \\
36	&	-	&	-	&	-	&	1.75 (0.03)	&	0.36 (0.20)	&	2.13 (0.30)  \\
42	&	-	&	-	&	-	&	0.96 (0.08)	&	0.13 (0.09)	&	2.36 (0.14)  \\
50	&	-	&	-	&	-	&	0.96 (0.13)	&	-	&	2.40 (0.13)  \\
77	&	-	&	-	&	-	&	0.74 (0.08)	&	-	&	2.18 (0.14)  \\
95	&	-	&	-	&	-	&	0.59 (0.11)	&	-	&	1.11 (0.15)  \\ \hline
	&		&		&	\textbf{ASASSN-14ae}	&		&		&	\\ \hline
4	&	-	&	-	&	0.10 (0.08)	&	1.41 (0.24)	&	0.31 (0.03)	&	1.13 (0.10)  \\
30	&	-	&	-	&	0.16 (0.10)	&	1.23 (0.19)	&	0.28 (0.03)	&	2.47 (0.05)  \\
39	&	-	&	-	&	0.23 (0.05)	&	1.81 (0.05)	&	0.17 (0.02)	&	2.35 (0.05)  \\
51	&	-	&	-	&	0.26 (0.13)	&	1.06 (0.17)	&	0.20 (0.02)	&	2.08 (0.04)  \\
73	&	-	&	-	&	0.27 (0.02)	&	0.71 (0.03)	&	0.15 (0.06)	&	1.35 (0.05)  \\
94	&	-	&	-	&	0.27 (0.02)	&	0.45 (0.02)	&	0.11 (0.02)	&	1.03 (0.02)  \\ \hline
	&		&		&	\textbf{ASASSN-14li}	&		&		&	\\ \hline
10	&	0.29 (0.01)	&	0.37 (0.05)	&	0.27 (0.04)	&	0.49 (0.01)	&	0.12 (0.01)	&	0.86 (0.02)  \\
20	&	0.39 (0.02)	&	0.51 (0.08)	&	0.39 (0.07)	&	0.44 (0.01)	&	0.12 (0.01)	&	0.89 (0.02)  \\
22	&	0.38 (0.01)	&	0.48 (0.15)	&	0.33 (0.16)	&	0.50 (0.02)	&	0.26 (0.02)	&	0.81 (0.03)  \\
27	&	0.46 (0.01)	&	0.50 (0.24)	&	0.58 (0.25)	&	0.58 (0.03)	&	0.19 (0.02)	&	0.68 (0.02)  \\
42	&	0.17 (0.01)	&	0.13 (0.02)	&	0.27 (0.03)	&	0.19 (0.02)	&	0.07 (0.01)	&	0.52 (0.02)  \\
86	&	0.12 (0.01)	&	0.09 (0.01)	&	0.25 (0.01)	&	0.12 (0.01)	&	0.02 (0.00)	&	0.30 (0.01)  \\
146	&	0.04 (0.00)	&	0.07 (0.01)	&	0.12 (0.01)	&	0.02 (0.00)	&	-	&	0.17 (0.01)  \\ \hline
	&		&		&	\textbf{iPTF15af}	&		&		&	\\ \hline
29	&	1.05 (0.16)	&	1.15 (0.30)	&	1.39 (0.50)	&	0.27 (0.24)	&	-	&	0.18 (0.02)  \\
58	&	0.09 (0.05)	&	0.17 (0.10)	&	0.82 (0.50)	&	0.12 (0.09)	&	-	&	0.47 (0.08)  \\ \hline
	&		&		&	\textbf{ASASSN-15oi}	&		&		&	\\ \hline
7	&	-	&	-	&	4.72 (2.29)	&	2.48 (2.05)	&	1.39 (0.23)	&	1.11 (0.23)  \\
25	&	-	&	-	&	1.90 (0.16)	&	0.21 (0.14)	&	0.80 (0.14)	&	2.17 (0.17)  \\
51	&	-	&	-	&	0.62 (0.15)	&	-	&	0.45 (0.12)	&	1.19 (0.14)  \\
69	&	-	&	-	&	1.61 (0.01)	&	-	&	-	&	0.70 (0.20)  \\
88	&	-	&	-	&	0.72 (0.07)	&	-	&	-	&	0.21 (0.08)  \\
97	&	-	&	-	&	1.10 (0.09)	&	-	&	-	&	-                  \\ \hline
	&		&		&	\textbf{iPTF16axa}	&		&		&	\\ \hline
18	&	0.48 (0.06)	&	1.05 (0.21)	&	0.49 (0.20)	&	1.85 (0.23)	&	0.55 (0.05)	&	1.75 (0.07)  \\
44	&	0.24 (0.03)	&	0.64 (0.51)	&	0.70 (0.51)	&	0.80 (0.12)	&	0.43 (0.04)	&	1.16 (0.05)  \\ \hline
	&		&		&	\textbf{iPTF16fnl}	&		&		&	\\ \hline
4	&	0.17 (0.06)	&	0.68 (0.86)	&	0.87 (1.13)	&	1.33 (0.52)	&	0.07 (0.06)	&	0.30 (0.10)  \\
14	&	0.11 (0.05)	&	0.20 (0.27)	&	0.81 (0.51)	&	0.49 (0.48)	&	0.23 (0.08)	&	0.55 (0.09)  \\
22	&	0.11 (0.04)	&	0.14 (1.06)	&	0.48 (0.27)	&	0.33 (0.08)	&	0.17 (0.05)	&	0.37 (0.06)  \\
34	&	0.13 (0.02)	&	0.12 (0.21)	&	0.36 (0.22)	&	0.19 (0.04)	&	0.07 (0.02)	&	0.21 (0.02)  \\
42	&	0.07 (0.04)	&	0.08 (0.19)	&	0.22 (0.23)	&	0.10 (0.16)	&	0.03 (0.03)	&	0.12 (0.04)  \\
62	&	0.06 (0.02)	&	0.03 (0.05)	&	0.33 (0.20)	&	0.05 (0.04)	&	-	&	0.02 (0.01)  \\ \hline
	&		&		&	\textbf{AT2017eqx}	&		&		&	\\ \hline
12	&	-	&	-	&	2.91 (0.77)	&	0.58 (0.67)	&	-		&	2.30 (0.13)  \\
31	&	-	&	-	&	1.33 (0.12)	&	0.18 (0.11)	&	-		&	1.28 (0.07)  \\
35	&	-	&	-	&	2.75 (0.53)	&	0.27 (0.49)	&	-		&	1.75 (0.11)  \\
70	&	-	&	-	&	0.69 (0.27)	&	0.57 (0.28)	&	-		&	0.69 (0.05)  \\
124	&	-	&	-	&	3.58 (0.10)	&	0.14 (0.04)	&	-		&	0.92 (0.07)  \\
162	&	-	&	-	&	1.47 (0.10)	&	-			&	-       &   -            \\ \hline
	&		&		&	\textbf{AT2018zr}	&		&		&	\\ \hline
7	&	-	&	-	&	-	&	1.73 (0.13)	&	0.61 (0.06)	&	4.37 (0.09)  \\
12	&	-	&	-	&	-	&	3.67 (0.22)	&	0.44 (0.09)	&	3.71 (0.14)  \\
20	&	-	&	-	&	-	&	2.81 (0.47)	&	0.55 (0.21)	&	3.75 (0.28)  \\
26	&	-	&	-	&	-	&	2.20 (0.13)	&	0.68 (0.06)	&	4.85 (0.09)  \\
40	&	-	&	-	&	-	&	3.14 (0.18)	&	0.51 (0.09)	&	6.70 (0.29)  \\
53	&	-	&	-	&	-	&	1.25 (0.06)	&	0.10 (0.05)	&	4.92 (0.08)  \\
62	&	-	&	-	&	-	&	2.71 (0.18)	&	0.15 (0.10)	&	5.17 (0.15)  \\ \hline
	&		&		&	\textbf{AT2018dyb}	&		&		&	\\ \hline
$-$20	&	0.92 (0.05)	&	1.04 (0.27)	&	0.69 (0.24)	&	0.57 (0.11)	&	0.36 (0.08)	&	0.73 (0.07)  \\
$-$10	&	1.61 (0.10)	&	2.30 (4.59)	&	1.20 (7.07)	&	0.69 (0.36)	&	0.42 (0.09)	&	1.72 (0.11)  \\
0	&	2.27 (0.14)	&	2.99 (10.26)	&	1.70 (16.95)	&	1.95 (0.67)	&	0.96 (0.09)	&	3.39 (0.20)  \\
5	&	2.29 (0.14)	&	2.39 (4.78)	&	2.21 (9.36)	&	2.05 (0.65)	&	1.22 (0.20)	&	4.32 (0.19)  \\
19	&	1.63 (0.07)	&	2.20 (1.19)	&	0.76 (1.65)	&	1.74 (0.18)	&	0.86 (0.06)	&	3.45 (0.12)  \\
33	&	1.16 (0.06)	&	1.45 (2.38)	&	0.94 (3.66)	&	1.17 (0.16)	&	0.61 (0.03)	&	3.02 (0.10)  \\
47	&	1.01 (0.06)	&	1.27 (0.97)	&	1.00 (1.11)	&	0.96 (0.08)	&	0.70 (0.02)	&	2.80 (0.09)  \\
50	&	0.62 (0.04)	&	1.06 (2.20)	&	0.73 (3.82)	&	0.86 (0.08)	&	0.58 (0.07)	&	2.40 (0.06)  \\
66	&	0.65 (0.03)	&	0.73 (1.59)	&	0.87 (3.62)	&	0.86 (0.06)	&	0.38 (0.04)	&	2.14 (0.04)  \\
170	&	0.13 (0.02)	&	0.11 (1.01)	&	0.24 (1.00)	&	0.07 (0.02)	&	0.08 (0.02)	&	0.38 (0.03)  \\
181	&	0.06 (0.04)	&	0.25 (0.78)	&	0.27 (1.72)	&	0.09 (0.06)	&	0.04 (0.02)	&	0.31 (0.03)  \\
196	&	0.04 (0.02)	&	0.05 (0.18)	&	0.07 (0.07)	&	0.03 (0.03)	&	0.02 (0.02)	&	0.23 (0.03)  \\
221	&	-	&	0.07 (0.07)	&	0.11 (0.09)	&	0.05 (0.04)	&	-	&	0.22 (0.04)  \\ \hline
	&		&		&	\textbf{AT2018fyk}	&		&		&	\\ \hline
7	&	-	&	-	&	1.67 (0.42)	&	0.07 (0.08)	&	-	&	0.97 (0.15)  \\
25	&	-	&	-	&	0.65 (0.09)	&	0.02 (0.02)	&	-	&	0.94 (0.06)  \\
40	&	-	&	-	&	0.37 (0.04)	&	0.13 (0.04)	&	0.16 (0.03)	&	0.10 (0.03)  \\
54	&	-	&	-	&	0.43 (0.03)	&	0.08 (0.03)	&	0.11 (0.03)	&	0.34 (0.05)  \\
68	&	-	&	-	&	0.35 (0.06)	&	0.09 (0.03)	&	0.06 (0.02)	&	0.70 (0.05)  \\
86	&	-	&	-	&	0.21 (0.08)	&	0.13 (0.03)	&	0.25 (0.03)	&	0.02 (0.02)  \\
100	&	-	&	-	&	0.39 (0.02)	&	-	&	0.01 (0.01)	&	0.25 (0.02)  \\ \hline
	&		&		&	\textbf{AT2018hyz}	&		&		&	\\ \hline
9	&	-	&	-	&	0.05 (0.10)	&	4.19 (0.33)	&	1.44 (0.26)	&	4.27 (0.30)  \\
26	&	-	&	-	&	0.05 (0.08)	&	4.02 (0.21)	&	0.98 (0.16)	&	4.00 (0.19)  \\
39	&	-	&	-	&	0.21 (0.09)	&	3.09 (0.20)	&	0.59 (0.12)	&	3.92 (0.15)  \\
55	&	-	&	-	&	0.18 (0.12)	&	3.21 (0.23)	&	0.48 (0.13)	&	3.31 (0.18)  \\
67	&	-	&	-	&	0.14 (0.10)	&	2.73 (0.18)	&	0.39 (0.14)	&	2.88 (0.14)  \\
79	&	-	&	-	&	0.14 (0.17)	&	1.77 (0.26)	&	0.46 (0.17)	&	2.32 (0.16)  \\
96	&	-	&	-	&	0.08 (0.10)	&	1.42 (0.20)	&	0.33 (0.11)	&	1.64 (0.14)  \\
110	&	-	&	-	&	0.16 (0.13)	&	1.33 (0.19)	&	0.24 (0.09)	&	1.75 (0.11)  \\
130	&	-	&	-	&	0.39 (0.11)	&	0.79 (0.10)	&	0.22 (0.06)	&	1.02 (0.06)  \\
170	&	-	&	-	&	0.14 (0.08)	&	0.79 (0.09)	&	0.11 (0.05)	&	0.46 (0.09)  \\
188	&	-	&	-	&	0.10 (0.08)	&	0.37 (0.07)	&	0.09 (0.04)	&	0.35 (0.05)  \\
216	&	-	&	-	&	0.22 (0.08)	&	0.20 (0.10)	&	0.07 (0.06)	&	0.40 (0.10)  \\ \hline
	&		&		&	\textbf{AT2018lna}	&		&		&	\\ \hline
34	&	0.16 (0.04)	&	0.07 (0.11)	&	0.35 (0.36)	&	0.49 (0.75)	&	0.25 (0.06)	&	1.11 (0.19)  \\
56	&	0.37 (0.06)	&	0.05 (0.41)	&	0.16 (0.08)	&	0.64 (0.13)	&	0.26 (0.06)	&	0.94 (0.23)  \\  \hline

\end{longtable}
%  \vspace{0.5cm}
$^{a}$De-reddened for Galactic extinction. The $A_{V}$ of each TDE is provided in Table  \ref{tab:sample}. \\
$^{b}$With respect to the MJD of the t$_{\rm peak/max}$ of each TDE provided in Table \ref{tab:sample}. %\\
\end{small}

\begin{small}
%%%%%  Sample
\renewcommand{\arraystretch}{1.2}
\setlength\tabcolsep{0.25cm}
\fontsize{9.5}{11}\selectfont
\begin{longtable}{c c c c c c c}
\caption{Emission line widths}
\label{tab:linewidths} \\
\hline 
Phase$^{a}$ &H$\delta$/\ion{N}{III} $\lambda$4100 & \ion{N}{III} $\lambda$4640 & \ion{He}{II} & H$\beta$ & \ion{He}{I} $\lambda$5876 & H$\alpha$ \\ (days)     &  ($10^{3}$~km~s$^{-1}$)      &     ($10^{3}$~km~s$^{-1}$)      &   ($10^{3}$~km~s$^{-1}$)     &   ($10^{3}$~km~s$^{-1}$)       &   ($10^{3}$~km~s$^{-1}$)        & ($10^{3}$~km~s$^{-1}$)  \\
\noalign{\global\arrayrulewidth=1mm}\hline
\noalign{\global\arrayrulewidth=0.4pt}
	&		&		&	\textbf{PTF09ge}	&		&		&	\\ \hline
$-$19	&	-	&	-	&	20.49 (1.34)	&	5.08 (2.84)	&	-	&	21.51 (0.17)  \\ \hline
	&		&		&	\textbf{PTF09djl}	&		&		&	\\ \hline
2	&	-	&	-	&	4.52 (3.23)	&	21.25 (2.80)	&	-	&	24.18 (1.41)  \\
31	&	-	&	-	&	4.52 (0.48)	&	12.81 (1.28)	&	-	&	12.68 (0.74)  \\ \hline
	&		&		&	\textbf{PS1-10jh}	&		&		&	\\ \hline
$-$22	&	-	&	-	&	9.98 (1.19)	&	-	&	-	&	- \\
254	&	-	&	-	&	8.30 (2.50)	&	-	&	-	&	-  \\ \hline
	&		&		&	\textbf{LSQ12dyw}	&		&		&	\\ \hline
4	&	-	&	-	&	-	&	24.32 (4.93)	&	12.01 (3.35)	&	19.25 (2.27)  \\
20	&	-	&	-	&	-	&	24.21 (3.05)	&	8.37 (2.48)	&	16.00 (1.30)  \\
36	&	-	&	-	&	-	&	25.29 (4.50)	&	12.01 (4.75)	&	15.51 (3.45)  \\
42	&	-	&	-	&	-	&	19.68 (2.51)	&	9.81 (7.47)	&	21.95 (1.57)  \\
50	&	-	&	-	&	-	&	19.44 (3.19)	&	-	&	20.76 (1.49)  \\
77	&	-	&	-	&	-	&	17.39 (3.34)	&	-	&	22.45 (1.43)  \\
95	&	-	&	-	&	-	&	20.74 (4.82)	&	-	&	20.71 (3.18)  \\ \hline
	&		&		&	\textbf{ASASSN-14ae}	&		&		&	\\ \hline
4	&	-	&	-	&	6.03 (5.74)	&	14.52 (9.69)	&	6.41 (1.06)	&	10.76 (1.00)  \\
30	&	-	&	-	&	7.17 (6.83)	&	12.24 (8.50)	&	6.29 (0.67)	&	12.87 (0.31)  \\
39	&	-	&	-	&	6.64 (3.86)	&	12.72 (8.47)	&	3.89 (0.89)	&	11.75 (0.32)  \\
51	&	-	&	-	&	10.27 (2.39)	&	8.41 (0.73)	&	5.94 (0.81)	&	11.00 (0.26)  \\
73	&	-	&	-	&	6.45 (0.61)	&	7.37 (0.43)	&	3.99 (1.04)	&	9.51 (0.26)  \\
94	&	-	&	-	&	5.39 (0.39)	&	6.44 (0.31)	&	5.36 (1.02)	&	7.93 (0.21)  \\ \hline
	&		&		&	\textbf{ASASSN-14li}	&		&		&	\\ \hline
10	&	2.49 (0.12)	&	3.90 (0.35)	&	4.50 (0.53)	&	3.50 (0.12)	&	2.52 (0.29)	&	4.34 (0.12)  \\
20	&	3.18 (0.14)	&	4.42 (0.44)	&	5.25 (0.92)	&	2.98 (0.11)	&	2.85 (0.35)	&	4.10 (0.10)  \\
22	&	3.17 (0.17)	&	5.30 (0.75)	&	5.66 (1.31)	&	3.78 (0.21)	&	4.93 (0.59)	&	4.11 (0.17)  \\
27	&	3.56 (0.13)	&	5.65 (1.09)	&	6.68 (0.94)	&	4.04 (0.20)	&	4.76 (0.62)	&	3.45 (0.12)  \\
42	&	1.38 (0.11)	&	1.68 (0.25)	&	2.25 (0.24)	&	1.36 (0.14)	&	1.35 (0.32)	&	2.12 (0.09)  \\
86	&	1.77 (0.11)	&	1.71 (0.18)	&	1.74 (0.07)	&	1.26 (0.09)	&	0.85 (0.29)	&	1.74 (0.06)  \\
146	&	2.25 (0.43)	&	3.24 (0.59)	&	1.61 (0.12)	&	0.71 (0.16)	&	-	&	1.68 (0.08)  \\ \hline
	&		&		&	\textbf{iPTF15af}	&		&		&	\\ \hline
29	&	17.67 (1.68)	&	17.51 (6.12)	&	19.57 (12.27)	&	9.25 (2.63)	&	-	&	5.62 (0.87)  \\
58	&	3.44 (0.68)	&	6.09 (3.12)	&	18.08 (12.27)	&	5.81 (2.63)	&	-	&	15.06 (6.57)  \\ \hline
	&		&		&	\textbf{ASASSN-15oi}	&		&		&	\\ \hline
7	&	-	&	-	&	13.72 (2.78)	&	14.14 (134.30)	&	12.79 (2.41)	&	11.89 (2.84)  \\
25	&	-	&	-	&	10.72 (1.07)	&	10.86 (8.74)	&	12.47 (2.47)	&	17.28 (1.57)  \\
51	&	-	&	-	&	13.66 (4.26)	&	-	&	15.78 (4.77)	&	19.46 (2.61)  \\
69	&	-	&	-	&	28.34 (6.53)	&	-	&	-	&	17.86 (5.78)  \\
88	&	-	&	-	&	30.13 (2.29)	&	-	&	-	&	26.89 (9.31)  \\
97	&	-	&	-	&	30.13 (1.56)	&	-	&	-	&	- \\  \hline
	&		&		&	\textbf{iPTF16axa}	&		&		&	\\ \hline
18	&	5.13 (0.57)	&	5.75 (0.89)	&	4.99 (0.81)	&	12.87 (2.37)	&	8.34 (0.83)	&	12.01 (0.49)  \\
44	&	3.97 (0.62)	&	6.48 (2.40)	&	7.53 (6.76)	&	7.87 (1.14)	&	6.43 (0.64)	&	7.78 (0.34)  \\ \hline
	&		&		&	\textbf{iPTF16fnl}	&		&		&	\\ \hline
4	&	6.46 (2.51)	&	7.71 (6.94)	&	8.78 (5.84)	&	17.32 (4.68)	&	4.98 (5.19)	&	11.42 (4.15)  \\
14	&	6.35 (1.74)	&	5.91 (3.83)	&	9.96 (8.94)	&	13.09 (6.90)	&	8.41 (1.05)	&	9.72 (1.81)  \\
22	&	6.29 (0.16)	&	5.49 (16.08)	&	7.72 (2.96)	&	10.59 (3.14)	&	8.41 (9.01)	&	8.44 (1.49)  \\
34	&	6.07 (1.03)	&	5.36 (1.49)	&	6.18 (1.50)	&	11.28 (2.40)	&	5.04 (1.67)	&	6.14 (0.83)  \\
42	&	4.28 (4.03)	&	5.15 (8.59)	&	5.00 (3.31)	&	6.61 (15.11)	&	2.78 (3.86)	&	3.75 (1.53)  \\
62	&	3.28 (2.70)	&	3.95 (7.83)	&	6.07 (5.98)	&	11.51 (11.11)	&	-	&	3.23 (1.09)  \\ \hline
	&		&		&	\textbf{AT2017eqx}	&		&		&	\\ \hline
12	&	-	&	-	&	20.34 (3.62)	&	21.78 (0.41)	&	-	&	23.70 (1.53)  \\
31	&	-	&	-	&	18.62 (1.91)	&	9.80 (4.38)	&	-	&	16.23 (1.07)  \\
35	&	-	&	-	&	27.78 (3.93)	&	12.40 (9.51)	&	-	&	17.45 (1.23)  \\
70	&	-	&	-	&	14.97 (2.74)	&	21.78 (12.25)	&	-	&	8.53 (0.73)  \\
124	&	-	&	-	&	30.13 (8.57)	&	4.57 (1.38)	&	-	&	12.91 (1.04)  \\
162	&	-	&	-	&	32.14 (9.27)	&	-	&	-	&	- \\  \hline
	&		&		&	\textbf{AT2018zr}	&		&		&	\\ \hline
7	&	-	&	-	&	-	&	10.42 (0.64)	&	9.64 (1.14)	&	14.65 (0.56)  \\
12	&	-	&	-	&	-	&	12.76 (0.78)	&	7.07 (1.70)	&	12.56 (0.64)  \\
20	&	-	&	-	&	-	&	13.23 (2.89)	&	7.99 (3.53)	&	11.40 (2.06)  \\
26	&	-	&	-	&	-	&	11.12 (0.59)	&	9.84 (1.03)	&	11.79 (0.27)  \\
40	&	-	&	-	&	-	&	10.62 (0.59)	&	4.15 (0.87)	&	13.63 (0.42)  \\
53	&	-	&	-	&	-	&	9.54 (0.63)	&	5.45 (5.01)	&	10.47 (0.18)  \\
62	&	-	&	-	&	-	&	9.54 (0.95)	&	6.14 (4.78)	&	11.87 (0.39)  \\ \hline
	&		&		&	\textbf{AT2018dyb}	&		&		&	\\ \hline
$-$20	&	5.72 (0.44)	&	4.98 (0.85)	&	5.06 (2.30)	&	6.45 (1.41)	&	6.24 (1.74)	&	6.25 (0.76)  \\
$-$10	&	9.73 (0.81)	&	10.99 (5.44)	&	15.35 (7.62)	&	7.35 (2.61)	&	8.27 (2.63)	&	9.24 (0.82)  \\
0	&	6.38 (0.54)	&	8.80 (7.43)	&	12.20 (8.30)	&	9.26 (2.19)	&	8.88 (1.18)	&	8.92 (0.72)  \\
5	&	6.32 (0.51)	&	7.67 (4.75)	&	11.95 (6.31)	&	8.96 (1.94)	&	9.34 (2.11)	&	9.24 (0.58)  \\
19	&	4.79 (0.30)	&	6.98 (1.18)	&	6.35 (4.10)	&	7.47 (0.81)	&	9.29 (1.53)	&	7.61 (0.36)  \\
33	&	3.91 (0.29)	&	5.41 (3.03)	&	12.25 (19.70)	&	5.05 (0.65)	&	5.05 (0.54)	&	6.36 (0.29)  \\
47	&	4.23 (0.35)	&	6.32 (1.30)	&	12.89 (8.46)	&	3.97 (0.41)	&	5.78 (0.24)	&	5.53 (0.24)  \\
50	&	4.01 (0.37)	&	6.59 (2.77)	&	7.50 (3.68)	&	4.35 (0.43)	&	6.25 (0.54)	&	5.21 (0.18)  \\
66	&	5.25 (0.39)	&	7.87 (3.19)	&	8.24 (2.95)	&	5.02 (0.39)	&	4.59 (0.86)	&	5.57 (0.16)  \\
170	&	5.85 (1.29)	&	6.87 (26.36)	&	12.80 (11.53)	&	1.44 (0.39)	&	5.63 (1.35)	&	5.67 (0.55)  \\
181	&	11.58 (9.40)	&	8.84 (8.64)	&	12.85 (8.65)	&	4.65 (2.58)	&	3.57 (1.32)	&	5.34 (0.67)  \\
196	&	6.11 (3.95)	&	5.06 (3.78)	&	9.42 (3.58)	&	2.28 (2.00)	&	1.58 (1.31)	&	6.43 (1.18)  \\
221	&	-	&	4.03 (3.44)	&	7.58 (5.12)	&	4.46 (4.15)	&	-	&	6.41 (1.46)  \\ \hline
	&		&		&	\textbf{AT2018fyk}	&		&		&	\\ \hline
7	&	-	&	-	&	19.93 (5.33)	&	3.04 (3.18)	&	-	&	16.06 (2.83)  \\
25	&	-	&	-	&	17.48 (2.77)	&	1.65 (1.60)	&	-	&	17.02 (1.28)  \\
40	&	-	&	-	&	10.05 (1.46)	&	5.81 (1.72)	&	7.21 (1.80)	&	3.72 (1.12)  \\
54	&	-	&	-	&	7.50 (0.70)	&	5.81 (2.59)	&	3.92 (1.07)	&	13.43 (2.32)  \\
68	&	-	&	-	&	7.74 (1.02)	&	5.81 (2.02)	&	4.09 (1.85)	&	14.75 (1.12)  \\
86	&	-	&	-	&	5.17 (1.09)	&	7.39 (2.22)	&	9.61 (0.82)	&	2.70 (2.94)  \\
100	&	-	&	-	&	12.55 (0.81)	&	-	&	1.73 (1.52)	&	10.98 (1.24)  \\ \hline
	&		&		&	\textbf{AT2018hyz}	&		&		&	\\ \hline
9	&	-	&	-	&	2.01 (2.77)	&	25.25 (2.00)	&	14.06 (2.90)	&	17.29 (1.48)  \\
26	&	-	&	-	&	2.60 (3.83)	&	22.55 (1.04)	&	12.84 (2.45)	&	14.70 (0.84)  \\
39	&	-	&	-	&	6.98 (2.75)	&	21.23 (14.61)	&	10.00 (2.34)	&	12.82 (0.69)  \\
55	&	-	&	-	&	4.42 (2.67)	&	22.70 (1.98)	&	9.48 (3.06)	&	12.69 (0.98)  \\
67	&	-	&	-	&	4.54 (2.84)	&	22.79 (1.80)	&	9.02 (3.59)	&	12.57 (0.79)  \\
79	&	-	&	-	&	5.95 (5.49)	&	21.76 (15.65)	&	12.02 (0.14)	&	12.83 (1.19)  \\
96	&	-	&	-	&	4.91 (5.53)	&	23.06 (135.03)	&	12.01 (4.22)	&	13.95 (1.73)  \\
110	&	-	&	-	&	5.56 (3.18)	&	20.34 (3.17)	&	9.60 (4.01)	&	12.65 (0.98)  \\
130	&	-	&	-	&	9.01 (1.80)	&	15.90 (2.74)	&	12.01 (3.81)	&	11.96 (1.18)  \\
170	&	-	&	-	&	7.85 (3.74)	&	18.97 (7.29)	&	14.01 (2.07)	&	12.10 (2.56)  \\
188	&	-	&	-	&	7.10 (5.70)	&	20.69 (3.30)	&	14.87 (3.87)	&	16.14 (2.33)  \\
216	&	-	&	-	&	7.37 (2.81)	&	11.89 (7.41)	&	14.58 (6.50)	&	13.82 (3.79)  \\ \hline
	&		&		&	\textbf{AT2018lna}	&		&		&	\\ \hline
34	&	7.81 (2.12)	&	4.56 (4.31)	&	15.01 (12.11)	&	17.97 (4.80)	&	12.14 (3.26)	&	15.28 (3.82)  \\
56	&	9.70 (1.87)	&	4.06 (3.91)	&	4.63 (1.86)	&	21.42 (3.66)	&	6.63 (1.79)	&	16.66 (4.82)  \\ \hline

\end{longtable}
%  \vspace{0.5cm}
$^{a}$With respect to the MJD of the t$_{\rm peak/max}$ of each TDE provided in Table \ref{tab:sample}. %\\
\end{small}

\begin{small}
%%%%%  Sample
\renewcommand{\arraystretch}{1.2}
\setlength\tabcolsep{0.2cm}
\fontsize{9}{11}\selectfont
\begin{longtable}{c c c c c c c}
\caption{Emission line velocity offsets}
\label{tab:lineoffsets} \\
\hline 
Phase$^{a}$ &H$\delta$/\ion{N}{III} $\lambda$4100 & \ion{N}{III} $\lambda$4640 & \ion{He}{II} & H$\beta$ & \ion{He}{I} $\lambda$5876 & H$\alpha$ \\  (days)     &  (km~s$^{-1}$)      &     (km~s$^{-1}$)      &   (km~s$^{-1}$)     &   (km~s$^{-1}$)       &   (km~s$^{-1}$)        & (km~s$^{-1}$) \\
\noalign{\global\arrayrulewidth=1mm}\hline
\noalign{\global\arrayrulewidth=0.4pt}
	&		&		&	\textbf{PTF09ge}	&		&		&	\\ \hline
$-$19	&	-	&	-	&	$-$7647.19 (518.49)	&	383.52 (1156.79)	&	-	&	$-$8751.49 (708.59)  \\ \hline 
	&		&		&	\textbf{PTF09djl}	&		&		&	\\ \hline
2	&	-	&	-	&	5649.44 (1309.94)	&	2443.70 (1236.05)	&	-	&	2284.02 (255.49)  \\ \hline
31	&	-	&	-	&	2890.93 (350.98)	&	4293.90 (499.47)	&	-	&	305.02 (316.12)  \\ \hline
	&		&		&	\textbf{PS1-10jh}	&		&		&	\\ \hline
$-$22	&	-	&	-	&	974.52 (624.37)	&	-	&	-	&	- \\
254	&	-	&	-	&	1942.34 (1057.18)	&	-	&	-	&	- \\ \hline
	&		&		&	\textbf{LSQ12dyw}	&		&		&	\\ \hline
4	&	-	&	-	&	-	&	$-$2421.01 (1387.33)	&	3169.27 (1402.66)	&	-1008.24 (920.18)  \\
20	&	-	&	-	&	-	&	$-$6166.78 (1872.74)	&	1689.13 (1049.17)	&	$-$3240.83 (69.46)  \\
36	&	-	&	-	&	-	&	1480.34 (754.69)	&	2953.73 (3113.03)	&	$-$2152.45 (1204.55)  \\
42	&	-	&	-	&	-	&	-1099.58 (1005.06)	&	4950.63 (3131.93)	&	1013.10 (642.24)  \\
50	&	-	&	-	&	-	&	$-$1905.85 (1246.23)	&	-	&	628.57 (633.15)  \\
77	&	-	&	-	&	-	&	$-$1374.21 (801.41)	&	-	&	1356.88 (605.48)  \\
95	&	-	&	-	&	-	&	$-$4068.31 (1896.24)	&	-	&	$-$1101.02 (1149.10)  \\ \hline
	&		&		&	\textbf{ASASSN-14ae}	&		&	-	&	\\ \hline
4	&	-	&	-	&	$-$127.952 (322.14)	&	1233.37 (314.35)	&	$-$508.00 (477.28)	&	$-$2284.03 (445.07)  \\
30	&	-	&	-	&	-104.921 (481.79)	&	2374.77 (172.03)	&	2352.17 (280.32)	&	1373.36 (113.97)  \\
39	&	-	&	-	&	$-$40.25 (305.00)	&	2541.13 (155.50)	&	2031.07 (373.67)	&	1991.47 (120.39)  \\
51	&	-	&	-	&	90.11 (565.46)	&	1274.37 (236.18)	&	2009.33 (340.38)	&	1785.14 (97.79)  \\
73	&	-	&	-	&	570.48 (244.06)	&	916.04 (145.03)	&	1328.25 (438.55)	&	1153.81 (115.59)  \\
94	&	-	&	-	&	$-$425.96 (161.10)	&	651.17 (128.64)	&	1211.50 (429.35)	&	1065.82 (89.02)  \\ \hline
	&		&		&	\textbf{ASASSN-14li}	&		&		&	\\ \hline
10	&	232.87 (34.56)	&	$-$373.80 (101.03)	&	320.38 (199.72)	&	$-$24.85 (36.78)	&	273.59 (98.16)	&	$-$286.56 (37.11)  \\
20	&	162.67 (38.12)	&	$-$208.08 (116.95)	&	371.83 (247.84)	&	$-$56.70 (33.07)	&	164.70 (119.54)	&	$-$154.86 (31.68)  \\
22	&	46.52 (58.04)	&	-109.58 (349.97)	&	373.96 (633.63)	&	5.78 (63.65)	&	$-$77.13 (195.96)	&	$-$185.41 (52.56)  \\
27	&	238.59 (44.88)	&	$-$82.07 (439.07)	&	128.15 (626.61)	&	$-$70.71 (51.82)	&	$-$119.28 (205.70)	&	$-$136.87 (37.48)  \\
42	&	44.38 (38.47)	&	$-$34.62 (77.38)	&	14.56 (61.94)	&	75.61 (36.24)	&	$-$393.75 (113.23)	&	$-$115.77 (29.97)  \\
86	&	35.29 (38.36)	&	$-$29.02 (55.11)	&	$-$134.71 (21.32)	&	12.07 (26.72)	&	$-$210.83 (100.22)	&	$-$54.22 (19.84)  \\
146	&	$-$581.66 (150.73)	&	$-$58.18 (197.98)	&	$-$183.92 (32.67)	&	$-$158.06 (48.38)	&	-	&	$-$118.97 (26.09)  \\ \hline
	&		&		&	\textbf{iPTF15af}	&		&	-	&	\\ \hline
29	&	$-$1457.19 (1247.91)	&	$-$5831.79 (6872.78)	&	$-$2884.59 (6021.04)	&	$-$416.26 (791.90)	&	-	&	$-$4617.78 (451.23)  \\
58	&	$-$3290.41 (1247.91)	&	$-$1292.21 (6872.78)	&	$-$1279.52 (6021.04)	&	$-$2466.74 (791.90)	&	-	&	$-$92.67 (591.23)  \\ \hline
	&		&		&	\textbf{ASASSN-15oi}	&		&		&	\\ \hline
7	&	-	&	-	&	$-$7060.59 (2546.38)	&	$-$4283.77 (5283.63)	&	$-$5014.67 (1023.99)	&	$-$4568.05 (1224.75)  \\
25	&	-	&	-	&	$-$1704.58 (373.73)	&	3700.11 (99.28)	&	$-$310.75 (1048.22)	&	2664.92 (666.18)  \\
51	&	-	&	-	&	$-$947.87 (1441.30)	&	-	&	1020.39 (4898.18)	&	856.03 (1073.05)  \\
69	&	-	&	-	&	$-$3838.57 (1384.78)	&	-	&	-	&	620.67 (2452.38)  \\
88	&	-	&	-	&	2170.21 (1437.09)	&	-	&	-	&	2496.81 (4614.11)  \\
97	&	-	&	-	&	1938.05 (1256.42)	&	-	&	-	&	- \\ \hline
	&		&		&	\textbf{iPTF16axa}	&		&		&	\\ \hline
18	&	1289.07 (220.12)	&	$-$343.67 (460.01)	&	1418.01 (445.50)	&	$-$1721.55 (1068.97)	&	$-$4525.57 (337.58)	&	335.96 (194.48)  \\
44	&	932.80 (261.74)	&	554.17 (517.61)	&	764.68 (603.10)	&	525.64 (852.41)	&	$-$5301.20 (263.03)	&	365.69 (138.74)  \\ \hline
	&		&		&	\textbf{iPTF16fnl}	&		&		&	\\ \hline
4	&	1148.98 (1064.73)	&	$-$3876.62 (2306.45)	&	412.01 (3179.91)	&	496.88 (3314.78)	&	2562.59 (2204.28)	&	$-$3904.20 (1761.06)  \\
14	&	$-$636.47 (1003.42)	&	2192.13 (1097.81)	&	0.44 (4113.10)	&	3370.77 (4882.07)	&	1519.00 (1407.23)	&	418.23 (769.66)  \\
22	&	$-$1101.00 (816.20)	&	$-$4977.33 (6363.56)	&	$-$84.82 (1532.78)	&	1580.64 (1257.82)	&	563.13 (1345.10)	&	577.12 (633.78)  \\
34	&	$-$722.95 (437.83)	&	$-$2638.79 (5324.63)	&	96.08 (1139.41)	&	1212.02 (926.14)	&	486.92 (708.30)	&	532.46 (351.74)  \\
42	&	1828.00 (402.83)	&	$-$2295.53 (5815.22)	&	$-$182.93 (1831.40)	&	-1044.72 (6929.07)	&	662.33 (1640.74)	&	$-$189.40 (650.43)  \\
62	&	1828.00 (427.83)	&	$-$2031.34 (4591.23)	&	$-$199.29 (983.32)	&	3700.11 (5692.56)	&	-	&	$-$817.05 (601.20)  \\ \hline
	&		&		&	\textbf{AT2017eqx}	&		&		&	\\ \hline
12	&	-	&	-	&	$-$1501.81 (1523.15)	&	6166.84 (7107.93)	&	-	&	6852.08 (242.40)  \\
31	&	-	&	-	&	464.46 (814.43)	&	5499.39 (1981.71)	&	-	&	961.58 (455.14)  \\
35	&	-	&	-	&	924.88 (2641.20)	&	4797.78 (2329.31)	&	-	&	2042.93 (521.23)  \\
70	&	-	&	-	&	-10.97 (1637.70)	&	6166.84 (728.10)	&	-	&	$-$37.81 (309.48)  \\
124	&	-	&	-	&	$-$8622.61 (387.70)	&	2950.51 (535.37)	&	-	&	$-$3649.80 (453.35)  \\
162	&	-	&	-	&	$-$1728.64 (955.90)	&	-	&	-	&	- \\ \hline
	&		&		&	\textbf{AT2018zr}			&		&	\\ \hline
7	&	-	&	-	&	-	&	400.58 (314.07)	&	2229.51 (483.18)	&	$-$575.14 (193.49)  \\
12	&	-	&	-	&	-	&	148.12 (374.12)	&	77.87 (722.37)	&	$-$453.18 (270.37)  \\
20	&	-	&	-	&	-	&	1054.17 (1339.45)	&	$-$261.46 (1499.68)	&	$-$593.56 (450.83)  \\
26	&	-	&	-	&	-	&	1547.78 (283.48)	&	594.74 (438.91)	&	$-$1994.85 (116.06)  \\
40	&	-	&	-	&	-	&	$-$230.33 (203.19)	&	179.94 (368.05)	&	$-$892.30 (180.19)  \\
53	&	-	&	-	&	-	&	-107.89 (251.58)	&	$-$5.61 (1705.71)	&	$-$1438.49 (76.80)  \\
62	&	-	&	-	&	-	&	$-$620.80 (378.31)	&	169.77 (2028.61)	&	$-$1853.60 (164.69)  \\ \hline
	&		&		&	\textbf{AT2018dyb}	&		&		&	\\ \hline
$-$20	&	$-$975.61 (151.50)	&	$-$565.36 (271.22)	&	636.11 (201.19)	&	$-$614.36 (417.09)	&	-1072.93 (596.87)	&	$-$406.23 (257.17)  \\
$-$10	&	$-$242.25 (258.76)	&	569.10 (2122.40)	&	1599.37 (12066.40)	&	784.21 (668.85)	&	$-$631.09 (491.84)	&	$-$18.84 (282.25)  \\
0	&	$-$6.27 (177.48)	&	293.85 (2486.09)	&	316.65 (13695.10)	&	666.74 (613.52)	&	$-$47.62 (267.24)	&	166.42 (247.44)  \\
5	&	336.16 (170.44)	&	392.78 (1260.59)	&	1137.68 (5997.61)	&	1058.21 (533.08)	&	65.14 (292.11)	&	438.38 (201.32)  \\
19	&	662.48 (104.11)	&	821.73 (801.19)	&	1298.12 (1870.42)	&	831.03 (253.45)	&	275.05 (121.10)	&	657.07 (124.71)  \\
33	&	600.39 (100.07)	&	1330.30 (240.16)	&	$-$1331.75 (1985.83)	&	674.13 (167.62)	&	408.02 (83.58)	&	478.67 (100.69)  \\
47	&	$-$186.54 (123.81)	&	682.25 (1188.87)	&	$-$423.23 (3820.98)	&	246.52 (131.18)	&	341.39 (77.43)	&	107.61 (84.50)  \\
50	&	575.79 (127.97)	&	503.08 (1763.18)	&	11.17 (4188.71)	&	577.93 (120.29)	&	130.38 (140.13)	&	384.53 (62.14)  \\
66	&	317.19 (134.28)	&	$-$790.53 (2517.10)	&	$-$852.37 (2643.77)	&	495.41 (110.61)	&	13.97 (163.68)	&	208.79 (56.47)  \\
170	&	$-$934.87 (446.18)	&	$-$46.30 (3489.14)	&	$-$1919.29 (3014.40)	&	$-$3152.05 (110.61)	&	1549.37 (651.96)	&	$-$306.97 (180.67)  \\
181	&	$-$2671.59 (3031.80)	&	$-$358.68 (2907.20)	&	846.05 (9060.25)	&	$-$119.77 (599.77)	&	1061.90 (683.19)	&	$-$398.12 (235.17)  \\
196	&	$-$142.35 (1245.60)	&	$-$977.28 (696.58)	&	$-$420.77 (5107.98)	&	$-$191.85 (701.80)	&	1189.01 (458.00)	&	$-$467.81 (397.24)  \\
221	&	-	&	720.74 (1135.26)	&	$-$89.51 (1292.11)	&	$-$1946.03 (1269.69)	&	-	&	$-$807.47 (516.78)  \\ \hline
	&		&		&	\textbf{AT2018fyk}	&		&		&	\\ \hline
7	&	-	&	-	&	$-$1585.25 (2005.19)	&	1850.05 (1149.48)	&	-	&	$-$1310.58 (1198.35)  \\
25	&	-	&	-	&	3198.81 (113.52)	&	1206.50 (615.85)	&	-	&	$-$1726.51 (542.50)  \\
40	&	-	&	-	&	991.19 (562.45)	&	1541.82 (722.51)	&	$-$2522.40 (766.53)	&	778.38 (477.70)  \\
54	&	-	&	-	&	440.19 (283.20)	&	866.68 (1060.15)	&	$-$2192.57 (456.18)	&	$-$318.83 (986.95)  \\
68	&	-	&	-	&	256.51 (429.92)	&	1850.05 (837.55)	&	$-$3075.81 (783.74)	&	$-$2528.83 (474.31)  \\
86	&	-	&	-	&	377.25 (340.94)	&	$-$1566.16 (879.92)	&	$-$3187.17 (625.51)	&	$-$548.50 (1248.04)  \\
100	&	-	&	-	&	$-$3198.81 (341.70)	&	-	&	$-$3028.25 (644.94)	&	1975.18 (526.20)  \\ \hline
	&		&		&	\textbf{AT2018hyz}			&		&	\\ \hline
9	&	-	&	-	&	$-$3600.09 (1146.19)	&	2424.40 (962.63)	&	$-$1139.28 (1226.98)	&	$-$304.55 (623.69)  \\
26	&	-	&	-	&	$-$4159.37 (1536.07)	&	2561.67 (565.83)	&	179.21 (1039.95)	&	169.65 (255.61)  \\
39	&	-	&	-	&	$-$5118.09 (388.40)	&	2378.67 (642.42)	&	822.32 (991.99)	&	785.35 (294.85)  \\
55	&	-	&	-	&	$-$3582.03 (985.46)	&	1586.07 (783.85)	&	$-$17.09 (1299.68)	&	587.71 (416.86)  \\
67	&	-	&	-	&	$-$3035.32 (1014.56)	&	978.58 (739.42)	&	177.36 (1903.37)	&	501.57 (326.85)  \\
79	&	-	&	-	&	$-$3969.65 (1822.51)	&	$-$415.19 (1557.03)	&	533.13 (3244.81)	&	809.32 (502.31)  \\
96	&	-	&	-	&	$-$3126.65 (2092.23)	&	$-$3173.98 (1775.46)	&	$-$445.85 (2056.07)	&	989.63 (733.34)  \\
110	&	-	&	-	&	$-$2511.79 (1017.07)	&	$-$783.66 (1412.31)	&	1015.43 (1704.25)	&	790.95 (417.11)  \\
130	&	-	&	-	&	$-$1466.32 (828.01)	&	1138.39 (1233.27)	&	6.06 (1582.69)	&	769.27 (375.65)  \\
170	&	-	&	-	&	$-$1897.16 (1266.90)	&	$-$3700.10 (1382.48)	&	$-$859.70 (2703.44)	&	768.89 (1085.47)  \\
188	&	-	&	-	&	$-$1192.61 (979.04)	&	$-$3700.09 (2013.05)	&	$-$806.26 (2291.70)	&	59.39 (1303.07)  \\
216	&	-	&	-	&	$-$606.29 (1209.89)	&	1569.49 (2857.38)	&	$-$1388.33 (4854.42)	&	358.24 (1613.23)  \\ \hline
	&		&		&	\textbf{AT2018lna}	&		&		&	\\ \hline
34	&	3889.42 (885.51)	&	3876.63 (445.77)	&	1881.88 (9439.60)	&	3700.11 (1065.19)	&	4939.60 (1349.97)	&	893.05 (510.05)  \\
56	&	5091.91 (779.75)	&	4976.63 (445.77)	&	3031.95 (667.98)	&	$-$2657.00 (2070.11)	&	4621.26 (751.86)	&	1954.83 (889.79)  \\
\end{longtable}
%  \vspace{0.5cm}
$^{a}$With respect to the MJD of the t$_{\rm peak/max}$ of each TDE provided in Table \ref{tab:sample}. %\\
\end{small}

\end{document}